\documentclass[%
preprint,
amsmath,amssymb,
aps,
prx,
floatfix,
]{revtex4-2}

\usepackage[dvipsnames]{xcolor}

\def\MODE{SUBMIT}

\usepackage[normalem]{ulem}
\usepackage{xr}

\usepackage{dsfont}

\usepackage{footmisc}
\usepackage{xparse}
\usepackage{lipsum}
\usepackage[base]{babel}
\usepackage{physics}

\makeatletter
\def\mathcolor#1#{\@mathcolor{#1}}
\def\@mathcolor#1#2#3{%
  \protect\leavevmode
  \begingroup
    \color#1{#2}#3%
  \endgroup
}
\makeatother

\usepackage{graphicx}

\usepackage{tikz}

\usepackage{placeins}

\usepackage{xspace}

\usepackage{multirow}
\usepackage{array}

\usepackage{ltablex}

\usepackage{booktabs}
\usepackage{tabularray}

\newcolumntype{Y}{>{\centering\arraybackslash}X}

\usepackage{amsmath,amsthm,amssymb}
\usepackage{mathtools}

\usepackage{graphicx}% Include figure files
\usepackage{dcolumn}% Align table columns on decimal point
\usepackage{bm}% bold math
\usepackage{booktabs}
\usepackage{IEEEtrantools}

\usepackage{tikz}
\usetikzlibrary{graphs}
\usepackage{amsthm}
\usepackage{amsfonts}
\usepackage{mathtools}
\usepackage{mathrsfs}
\usepackage{cancel}
\usepackage{xargs}
\usepackage{xparse}
\usepackage{xcolor}
\usepackage[at]{easylist}
\usepackage{enumitem}
\usepackage{algpseudocode}% http://ctan.org/pkg/algorithmicx
\usepackage[caption=false]{subfig}
\usepackage{xifthen}
\usepackage{pict2e,picture}
\usepackage{enumitem}
\usepackage{physics}

\newcommand{\setmin}{\setminus}

\DeclareMathSymbol{\mrq}{\mathord}{operators}{`'}
\DeclareMathSymbol{\mlq}{\mathord}{operators}{``}

\newcommand{\Sum}[2]{\sum_{#1}^{#2}}

\newcommand{\dashh}{^{\prime}}
\newcommand{\prm}[0]{\dashh}

\newcommand{\langlee}[0]{\left\langle}
\newcommand{\ranglee}[0]{\right\rangle}

\DeclarePairedDelimiterX{\infdivx}[2]{(}{)}{%
  #1\;\delimsize|\delimsize|\;#2%
}
\DeclareDocumentCommand \expectation { o m } {%
  \ensuremath{\left\langle%
  \IfValueTF {#1} {%
    #2\right\rangle_{#1}%
  }{%
    #2\right\rangle%
  }%
  }\xspace%
}

\newcommand{\MutualInfoSymbol}{I}
\newcommand{\MutualInfo}[3]{\MutualInfoSymbol\left(#1;#2\right)_{#3}}

\newcommand{\Temperature}{T}
\newcommand{\invTemp}{\beta}

\newcommand{\shaentSymbol}{S}
\newcommand{\shaent}[1]{\shaentSymbol\left(#1\right)}

\def\|#1{\ensuremath{\mathtt{#1}}}
\def\!#1{\ensuremath{\mathbf{#1}}}
\def\*#1{\ensuremath{\mathcal{#1}}}

\newcommand{\pprop}[0]{Proposition}
\newtheorem{theorem}{Theorem}%[section]
\newtheorem{corollary}{Corollary}%[theorem]

\newtheorem{proposition}{\pprop}%[section]
\newtheorem{lemma}{Lemma}%[section]
\usepackage{xargs}
\newcommandx{\unsure}[2][1=]{\todo[linecolor=red,backgroundcolor=red!25,bordercolor=red,#1]{#2}}
\newcommandx{\change}[2][1=]{\todo[linecolor=blue,backgroundcolor=blue!25,bordercolor=blue,#1]{#2}}
\newcommandx{\info}[2][1=]{\todo[linecolor=OliveGreen,backgroundcolor=OliveGreen!25,bordercolor=OliveGreen,#1]{#2}}
\newcommandx{\improvement}[2][1=]{\todo[linecolor=Plum,backgroundcolor=Plum!25,bordercolor=Plum,#1]{#2}}
\newcommandx{\thiswillnotshow}[2][1=]{\todo[disable,#1]{#2}}

\newcommand{\QQ}{\quad\quad}

\newcommand{\NonumberNewline}{\nonumber\\}
\newcommand{\NonumberNewlineQQ}{\NonumberNewline\QQ}
\newcommand{\EqualText}[1]{\stackrel{\text{#1}}{=}}

\makeatletter
\DeclareRobustCommand{\AArrow}[1][]{%
  \check@mathfonts
  \if\relax\detokenize{#1}\relax
    \settowidth{\dimen@}{$\m@th\rightarrow$}%
  \else
    \setlength{\dimen@}{#1}%
  \fi
  \sbox\z@{\usefont{U}{lasy}{m}{n}\symbol{41}}%
  \begin{picture}(\dimen@,\ht\z@)
  \roundcap
  \put(\dimexpr\dimen@-.7\wd\z@,0){\usebox\z@}
  \put(0,\fontdimen22\textfont2){\line(1,0){\dimen@}}
  \end{picture}%
}
\makeatother

\newcommand{\boltz}{k_{\operatorname{B}}}

\newcommand{\sSys}{\operatorname{tot}}

\newcommand{\DepX}[1]{\operatorname{pa}({#1})}
\newcommand{\DepXJ}{\DepX{j}}

\newcommand{\Xs}[1]{X_{#1}}

\newcommand{\Xtot}{{{\Xs{}}_{\sSys}}}

\newcommand{\Xtotprm}{X\prm_{\sSys}}
\newcommand{\Qtot}[0]{Q_{\sSys}}

\newcommand{\Qs}[1]{Q_{{#1}}}

\newcommand{\Htot}[0]{\shaent{\Xtot}}

\newcommand{\EntropyProduction}{\DeltaStot}

\newcommand{\ChiIndex}[1]{1:#1-1}
\newcommand{\MyChi}[1]{\Xs{\ChiIndex{#1}}}

\newcommand{\MyV}[1]{\Xs{#1}}

\newcommand{\IonePlus}[1]{I(\MyV{#1}\prm;\MyChi{#1}\prm\mid \DepX{1:#1})}
\newcommand{\IoneMinus}[1]{I(\MyV{#1};\MyChi{#1}\mid \DepX{1:#1})}
\newcommand{\Ione}[1]{\IonePlus{k}-\IoneMinus{k}}

\newcommand{\ItwoPlus}[1]{I(\MyChi{#1}\prm;\DepX{#1}\mid\DepX{\ChiIndex{#1}})}
\newcommand{\ItwoMinus}[1]{I(\MyChi{#1};\DepX{#1}\mid\DepX{\ChiIndex{#1}})}
\newcommand{\Itwo}[1]{\ItwoPlus{k}-\ItwoMinus{k}}

\newcommand{\DelI}{D}
\newcommand{\Dissipation}{\DelI}
\newcommand{\DissipationI}{\DelI_i}

\newcommand{\Wext}{W_{\mathrm{ext}}}
\newcommand{\Floc}{F_{\mathrm{loc}}}

\newcommand{\DelIstocha}{d}

\newcommand{\Prb}[2]{
\ifthenelse{\isempty{#2}}%
    {P_{#1}}% if #1 is empty
    {P_{#1}(#2)}% if #1 is not empty
}

\newcommand{\DelIinter}[1]{\widebar{\DelI}}

\newcommand{\Nx}{N}

\newcommand{\Xj}{\Xs{j}}

\newcommand{\Xk}{\Xs{k}}
\newcommand{\Xl}{\Xs{l}}

\newcommand{\RVa}{A}
\newcommand{\Ra}{\RVa}
\newcommand{\RVb}{B}
\newcommand{\Rb}{\RVb}

\newcommand{\RVc}{C}
\newcommand{\Rc}{\RVc}

\newcommand{\RVz}{Z}
\newcommand{\Rz}{\RVz}

\newcommand{\FB}{\operatorname{FBC}}

\newcommand{\SumIN}{\sum_{i=2}^{\Nx}}

\newcommand{\SumKinMu}{\sum_{k\in\minusI}}

\newcommand{\RESn}{2:N}

\newcommand{\Xresn}{\Xs{\RESn}}

\newcommand{\sInCor}{\mathcal{C}}

\newcommand{\InCor}[1]{
\ifthenelse{\isempty{#1}}%
    {\sInCor_{\Xresn}}
    {\sInCor_{\Xresn\mid#1}}
}

\newcommand{\Ngen}{n}

\newcommand{\Xsub}{\Xs{j}}

\newcommand{\Zd}{\DepXJ}
\newcommand{\cj}{^{\dag}}
\newcommand{\Xtotmin}{{\bar{X}_{j}}}

\newcommand{\OpenLoopWithoutMeasure}{NFC\xspace}

\newcommand{\set}[1]{\{#1\}}

\newcommand{\sstar}{\star}
\newcommand{\M}{^{\sstar}}
\newcommand{\symTotalCorrelation}{\mathcal{C}}
\newcommand{\TotalCorrelation}{\symTotalCorrelation_{\sSys}}
\newcommand{\delitot}{\Delta \TotalCorrelation}

\newcommand{\totalCorrI}{\symTotalCorrelation_{i}}
\newcommand{\usable}{\gamma}

\newcommand{\B}{^B}

\newcommand{\influTo}[1]{\operatorname{pa}(#1)}
\newcommand{\infPa}[1]{\influTo{#1}}
\newcommand{\infPaI}{\infPa{i}}

\newcommand{\infPaK}{\infPa{k}}

\newcommand{\minusX}[1]{{{\prec}#1}}
\newcommand{\minusI}{\minusX{i}}
\newcommand{\minusJ}{\minusX{j}}
\newcommand{\minusK}{\minusX{k}}

\newcommand{\plusX}[1]{{{\succ}#1}}
\newcommand{\plusI}{\plusX{i}}

\newcommand{\SAHEN}{\Delta\Htot + \invTemp\Qtot}

\newcommand{\atamaSymbol}{\mathcal{L}^{\mathrm{both}}}
\newcommand{\centerSymbol}{\mathcal{L}^{\mathrm{self}}}
\newcommand{\bottomSymbol}{\mathcal{L}^{\mathrm{other}}}
\newcommand{\memotouSymbol}{\mathcal{V}}

\newcommand{\bottomObsSymbol}{\bar{\mathcal{L}}}% unrealized variant of L^other (notation redesign)

\newcommand{\MuPrimeSeibunI}{i\scl k\M\scl\minusI'\mid\infPaK}

\newcommand{\centerSeibun}[1]{\centerSymbol_{#1}}

\newcommand{\centerSeibunKMu}{\centerSeibun{k,\minusK}}

\newcommand{\centerSeibunIK}{\centerSeibun{i,k}}

\newcommand{\centerSeibunIMu}{\centerSeibun{i}}% abbreviated subscript (notation redesign)

\newcommand{\bottomSeibun}[1]{\bottomSymbol_{#1}}

\newcommand{\bottomSeibunIKMu}{\bottomSeibun{i, k, \minusI}}
\newcommand{\bottomSeibunIMuMu}{\bottomSeibun{i}}% abbreviated subscript (notation redesign)
\newcommand{\bottomSeibunKMuMu}{\bottomSeibun{k,\minusK,\minusK}}

\newcommand{\memotouSeibun}[1]{\memotouSymbol_{#1}}
\newcommand{\memotouSeibunIMu}[0]{\memotouSymbol_{i}}% abbreviated subscript (notation redesign)
\newcommand{\memotouSeibunL}[1]{\memotouSymbol^L_{#1}}
\newcommand{\memotouSeibunLL}[1]{\mathcal{W}_{#1}}
\newcommand{\memotouSeibunR}[1]{\memotouSymbol^R_{#1}}
\newcommand{\memotouSeibunLIMu}[0]{\memotouSeibunL{i}}% abbreviated subscript (notation redesign)
\newcommand{\memotouSeibunRIMu}[0]{\memotouSeibunR{i}}% abbreviated subscript (notation redesign)

\newcommand{\botObsSeibun}[3]{\bottomObsSymbol_{#1, #2}^{#3}}

\newcommand{\botObsSeibunIKMuk}{\botObsSeibun{i}{k}{\minusK}}

\newcommand{\botObsSeibunUVW}{\botObsSeibun{u}{v}{w}}
\newcommand{\botObsSeibunIMukK}{\botObsSeibun{i}{\minusK}{k}}

\newcommand{\atamaOmega}[3]{\atamaSymbol_{#1,#2,#3}}
\newcommand{\atamaOmegaIMuMu}{\atamaSymbol_{i}}% abbreviated subscript (notation redesign)

\newcommand{\atamaOmegaIKMu}{\atamaOmega{i}{k}{\minusI}}
\newcommand{\atamaOmegaIKK}{\atamaOmega{i}{k}{k}}

\newcommand{\Yeng}[1]{\mathcal{Y}\left(#1\right)}

\newcommand{\DeltaStot}{\sigma}
\newcommand{\relright}[1]{\mathrel{}\right#1\mathrel{}}

\newcommand{\semicolon}{\,\mathpunct{;}\,}
\newcommand{\scl}{\semicolon}

\NewDocumentCommand{\muI}{m m o o o}{%
    \MutualInfoSymbol\left(#1\semicolon#2%
    \IfValueT{#3}{\semicolon#3}%
    \IfValueT{#4}{\semicolon#4}%
    \IfValueT{#5}{\semicolon#5}%
    \right)
}

\NewDocumentCommand{\muInfo}{m m o o}{%
    #1\semicolon#2%
    \IfValueT{#3}{\semicolon#3}%
    \IfValueT{#4}{\semicolon#4}%
}

\newcommand{\UsableNew}[1]{\usable_{#1}}
\newcommand{\UsableNewAll}{\UsableNew{}}

\newcommand{\sTuriB}{b}
\newcommand{\turiB}[1]{\sTuriB_{#1}}
\newcommand{\turiBfull}[2]{\sTuriB({#1,#2})}

\newcommand{\delIP}[1]{\Delta \symTotalCorrelation_{#1}}

\NewDocumentCommand{\infPaMulti}{o}{%
    \IfNoValueTF{#1}{%
        \hat{\Pi}_{\minusX{i}}%
    }{%
        \hat{\Pi}_{\minusX{#1}}%
    }%
}
\NewDocumentCommand{\blind}{o}{%
    \IfNoValueTF{#1}{%
        \mathbb{B}_{\minusX{i}}%
    }{%
        \mathbb{B}_{\minusX{#1}}%
    }%
}
\NewDocumentCommand{\effPa}{o}{%
    \IfNoValueTF{#1}{%
        \widehat{\operatorname{pa}}({\minusX{i}})%
    }{%
        \widehat{\operatorname{pa}}({\minusX{#1}})%
    }%
}

\ifthenelse{\equal{\MODE}{EDIT}}{
}{
    \definecolor{myPageColor}{gray}{1}
}
\ifdefined\ShowVennDiagram
    \newcommand{\VennDiagramThree}[4]{%
        #4
        \begin{center}
            \begin{tikzpicture}[scale=0.8]
                \draw[thick] (0,0) circle (1.2cm);
                \draw[thick] (2,0) circle (1.2cm);
                \draw[thick] (1,1.6) circle (1.2cm);

                \node[fill=myPageColor] at (1,2.8)  {#1};
                \node[fill=myPageColor] at (-1.2,0)    {#2};
                \node[fill=myPageColor] at (3.1,0)    {#3};
            \end{tikzpicture}
        \end{center}
    }
    \newcommand{\VennDiagramTwo}[3]{%
        #3
        \begin{center}
            \begin{tikzpicture}[scale=0.8]
            \draw[thick] (0,0) circle (1.2cm);
            \draw[thick] (2,0) circle (1.2cm);

            \node[fill=myPageColor] at (-1.2,0) {#1};
            \node[fill=myPageColor] at (3.1,0) {#2};
            \end{tikzpicture}
        \end{center}
    }
\else
    \newcommand{\VennDiagramThree}[4]{}
    \newcommand{\VennDiagramTwo}[3]{}
\fi

\newcommand{\syDecI}{\mathcal{C}}
\newcommand{\decI}[0]{\syDecI^{dec}}
\newcommand{\decIX}[2]{\decI_{#1,#2}}
\newcommand{\decIij}[0]{\decI_{i,j}}
\newcommand{\decIiMu}[0]{\decI_{i,\minusI}}

\newcommand{\incIX}[2]{\syDecI^{inc}_{#1,#2}}
\newcommand{\incIij}[0]{\syDecI^{inc}_{i,j}}
\newcommand{\incIiMu}[0]{\syDecI^{inc}_{i,\minusI}}

\newcommand{\accIiMu}[0]{\syDecI^{acc}_{i,\minusI}}

\newtheorem{assumption}{Assumption}
\newcommand{\Dmiss}[1][i]{D^{\mathrm{miss}}_{#1}}
\newcommand{\Ucost}[1][i]{U_{#1}}
\newcommand{\Rreb}[1][i]{R_{#1}}
\renewcommand{\Wext}{W_{\mathrm{ext}}}
\RenewDocumentCommand{\Floc}{g}{F^{\mathrm{loc}}\IfNoValueF{#1}{_{#1}}}

\newcommand{\decIk}[1][k]{\syDecI^{dec,(#1)}}
\newcommand{\incIk}[1][k]{\syDecI^{inc,(#1)}}
\newcommand{\accIk}[1][k]{\syDecI^{acc,(#1)}}
\newcommand{\Fcart}{F_{\mathrm{ctg}}}
\usepackage{color_text}
\usepackage{stmaryrd}

\usepackage[colorlinks=true]{hyperref}
\begin{document}

\title{
    Mechanism-resolved second law for multipartite systems:\texorpdfstring{\\}{ } An entropy-production ledger for correlation loss
}% Force line breaks with \\

\author{Akihito Sudo}%
\email{sudo.akihito@shizuoka.ac.jp}
\affiliation{%
    ZeroStruct Inc.
}%

\begin{abstract}
Internal correlation among the subsystems of a many-body system with additive bare energies stores free energy at $\boltz\Temperature$ per nat. Endpoint accounts admit its release into the work budget but do not determine which mechanism can retain that release instead of surrendering it to dissipation. Accounting based on realized statistics cannot resolve this question. Three one-step processes on three bits can share the same initial--final joint distribution yet differ in their minimum total entropy production, zero or $\ln 2$, depending on whether the mechanism causing the correlation loss itself reads the variable that carries it. Sharing statistics is not the end of the degeneracy. Two such processes can further share a valid declaration of their reading patterns and the heat released on every trajectory, so that every per-stage balance and every fixed-block modularity value computed under that declaration coincides; the minimum total entropy production over implementations of their kernels still differs by $\ln 2$, the kernels remaining distinguishable only at the level of their full-state-space structure. We consider one synchronized step of a classical multipartite system in contact with a single heat bath. Each subsystem is updated by its own local mechanism, which reads a beginning-of-step snapshot of a fixed subset of the others under local detailed balance. We derive an exact entropy-production ledger for correlation loss that is valid for arbitrary reading patterns, including reciprocal ones, and is saturated by explicit processes. Under acyclic reading and for protocols that build no correlation erased within the same step, the ledger collapses to a mechanism-resolved second law. The total entropy production is bounded below by the destroyed correlation hidden from the mechanisms that caused each loss. Blind dynamics dissipate destroyed correlation in full, whereas complete informational coverage recovers the conventional second law. Destroyed correlation is credited, in aggregate, to extractable work only through its part shared with the responsible mechanisms. For a fixed conversion task, the informed--blind gap along conversion chains is capped at $\boltz\Temperature$ times the initial total correlation, and the cap is exact. Under a per-run work budget, information decides access rather than price. In an explicit correlation-keyed device, informed protocols collect an independent free-energy resource at zero injected work, while blind protocols in the stated class, for budgets below a stated linear threshold, succeed with probability exponentially small in the number of correlated bits. This unmeasured contrast identifies a candidate single-electron test.
\end{abstract}
\maketitle

\section{Introduction}

Internal correlation among the subsystems of a many-body system is a free-energy resource \cite{Brillouin_1951,Landauer_1961,Bennett_1973,Bennett_1982,Jarzynski_1997,Crooks_1999,sagawa2008second,sagawa2010generalized,Toyabe_2010,del2011thermodynamic,Mandal_2012,B_rut_2012,seifert2012stochastic,horowitz2014thermodynamics,Koski_2014,Parrondo_2015}. For additive bare energies, the nonequilibrium free energy separates into a local part and a correlation part,
$$
F=\Floc+\boltz\Temperature\,\TotalCorrelation,
$$
where $\Floc$ is the free energy that the system would possess if its subsystems were uncorrelated, while the total correlation $\TotalCorrelation$ \cite{Watanabe1960} measures the statistical dependence carried by the joint state over and above its parts. This decomposition establishes both the existence of the resource and its conventional exchange value: correlation stores free energy at $\boltz\Temperature$ per nat \cite{sagawa2008second,sagawa2010generalized,sagawa2012fluctuation}, and an endpoint account admits any release of correlation into the work budget on top of the local free-energy changes \cite{Mandal_2012,Kolchinsky_20219,Song_2021}. The decomposition does not, however, determine which local mechanism can retain that release in the work balance rather than surrender it to dissipation. We first examine this mechanism-sensitive informational requirement for correlation conversion. We will show that what matters is not whether the relevant information exists somewhere in the system, but whether it enters the readout of the mechanism whose action causes the correlation loss.

Section~\ref{sect:ex_bit} demonstrates this dependence through a minimal construction of three one-step processes on three bits. Under the specified initial ensemble, the processes induce the same realized initial--final joint statistics $P(\Xtot,\Xtotprm)$ but have different transition kernels on the full state space. By a mechanism's reading pattern we mean here the initial variables on which its local kernel actually depends. Every quantity computed from the shared statistics is identical across the three processes, but the minimum total entropy production attainable in the step differs. It is zero when the mechanism causing the correlation loss itself reads the variable carrying the destroyed correlation, and $\ln 2$ when that variable is present in the system but read by no mechanism, or read only by a mechanism not causing the loss; each minimum is attained by an explicit protocol, exactly or in the quasistatic limit. Realized statistics alone therefore do not fix the thermodynamic verdicts of this paper: the reading structure must enter the account.

We consider one synchronized step of a classical many-body system with finite state spaces coupled to a single heat bath at temperature $\Temperature$. Each subsystem is updated by its own local mechanism, a stochastic channel with its own independent thermal noise. The channel is conditioned on a beginning-of-step snapshot of its own state and a fixed subset of the others; the joint transition kernel therefore factorizes into local channels \cite{horowitz2014thermodynamics,hartich2014stochastic,ito2013information}. The informational parents of mechanism $i$ are the unique minimal set of initial variables, beyond its own state, on which its channel actually depends. The channel is defined on the whole state space, not merely on the support of the initial distribution. Parents are a property of the kernel, not of the statistics. Variables merely correlated with the update under the chosen initial distribution, or sets inferred from realized statistics, do not qualify. Standard local detailed balance \cite{Evans_1994,Jarzynski_1997,Sekimoto_1998,Crooks_1998,Crooks_1999,Hatano_2001,Seifert_2005,sekimoto2010stochastic,seifert2012stochastic} ties each channel to its dissipated heat, thereby attaching thermodynamics to the step. This definition separates the three processes above: they share realized statistics but not informational parents.

Our problem sits among several established accounts, each successful on its own question. Feedback thermodynamics with a held memory bounds extra work from consumed mutual information at $\boltz\Temperature$ per nat \cite{sagawa2008second,sagawa2010generalized,sagawa2012fluctuation}. Constrained-protocol theories characterize free energy accessible under a prescribed generator class \cite{Kolchinsky_20219}. Predictive-information analyses distinguish task-relevant predictive memory from retained information irrelevant to that task \cite{Still_2020,Still_2022}. Modularity accounts quantify dissipation imposed by localized control or specified dependency and circuit structure \cite{Boyd_2018,wolpert2019stochastic,Wolpert_2020a,WolpertMinEP2020}. Information-flow theories on causal networks relate entropy production to information exchanged along a specified interaction structure \cite{horowitz2014thermodynamics,hartich2014stochastic,ito2013information,barato2014efficiency,Horowitz_2015,Wolpert_2020,WolpertCoevolving2020}. Endpoint formulations bound work through initial-to-final nonequilibrium-free-energy differences \cite{Kolchinsky_20219}. Our problem calls for a more specialized object: an account that takes full-state-space local transition kernels, and thus their unique minimal informational parents, as input and returns a one-step aggregate dissipation bound for correlation loss in a synchronous snapshot update, resolved by the state change causing each loss and by the variables read by that mechanism. The formulations above do not, as stated, directly supply this kernel-minimal, synchronous bound. The gap is structural rather than presentational. Fix any valid acyclic declaration of the reading patterns. The stage quantities and fixed-block modularity values these accounts then generate are functionals of the declaration, the realized statistics, and the trajectory heats alone (Appendix~\ref{ap:declaration_measurability}). Yet Sec.~\ref{sect:ex_bit} exhibits two processes that share all three but whose minimal attainable dissipation differs by $\ln 2$. What separates such processes is a judgement these quantities do not carry: which declarations are valid for which kernel. That judgement is the kernel-level input constructed in Sec.~\ref{sect:setting}; read off realized statistics instead, it cannot be sound, because the processes share them. We therefore ask: in one synchronized step of local mechanisms reading partial beginning-of-step snapshots, how does the reading structure set the minimum aggregate dissipation accompanying internal-correlation loss?

Our first result (Theorem~\ref{thm:ledger}) answers this question at the level of arbitrary reading patterns. Assuming only local mechanisms and local detailed balance (the setting above, with no restriction on the parent structure), it establishes an exact entropy-production ledger for the step: a lower bound on the total entropy production that remains valid under reciprocal reading, in which mechanisms read each other within the same step. The ledger comprises three families of signed terms: hidden-loss contributions, which collect the destroyed correlation invisible to the controllers of the side or sides whose state change caused the loss; unrealized-correlation costs, which charge for correlation built toward a partner whose own move erases it within the step; and rebate terms, which enter with a negative sign and are present in general. These are signed bookkeeping quantities, not independently nonnegative costs. The bound is sharp. Equality holds exactly when every local entropy production vanishes, and explicit processes, including reciprocal ones, attain it.

Two structural conditions make this ledger transparent. The first, acyclicity of the reading pattern (Assumption~\ref{asm:acyclic}), makes every rebate term vanish identically. This condition cannot simply be dropped. An explicit two-subsystem process with reciprocal reading falsifies the rebate-free reduction (Appendix~\ref{ap:cex_refresh}); we claim neither that every cyclic architecture fails nor that acyclicity is necessary. The second condition (Assumption~\ref{asm:no_unrealized}) removes the unrealized-correlation costs by forbidding their source: correlation built toward a partner whose own move erases it within the same step. It is a strong condition. Fixed-memory feedback satisfies it by construction, but determinism alone does not imply it. Under Assumptions~\ref{asm:local_mech}--\ref{asm:no_unrealized} the ledger collapses to our main result (Theorem~\ref{thm:main}),
$$
\DeltaStot \geq \Dissipation,
$$
where $\DeltaStot$ is the total entropy production of the step and $\Dissipation$ is the one-step aggregate of destroyed internal correlation hidden from the controllers of the side or sides whose state change caused each loss. The inequality concerns the aggregate alone. The subsystem-resolved terms are signed chain-rule contributions to this aggregate and need not be nonnegative. Thus, Theorem~\ref{thm:main} establishes a mechanism-sensitive aggregate penalty, not a decomposition into independently nonnegative pieces. At the endpoints, fully blind dynamics are charged the entire destroyed correlation, while complete parental coverage removes the correlation-specific penalty and returns the conventional second law without, by itself, guaranteeing that the conventional limit is attained. The work-balance consequences of this penalty are developed next.

Combined with the first law, Theorem~\ref{thm:main} yields a work bound (Corollary~\ref{cor:work_bound}). The aggregate hidden loss cancels from the extractable work. Every nat of the step total $\Dissipation$ releases $\boltz\Temperature$ of correlation free energy and simultaneously incurs at least the same dissipation. Beyond the local free-energy change, the destroyed correlation enters the work balance only through the aggregate reading-pattern credit, the portion shared with the responsible controllers through their parent variables, while newly created correlation enters as a cost. Here ``credit'' means a term admitted by the work bound; its attainment requires a separate protocol construction. Under additive bare energies, steps of this class compose into conversion chains (Proposition~\ref{prop:chain_work}). A fully blind chain can never credit internal correlations to its work balance. For a fixed conversion task, with final marginals and bare energies prescribed, and with work injected freely along the way and repaid from later proceeds, the optimal informed and blind work balances differ by at most $\boltz\Temperature\,\TotalCorrelation(t_0)$, however long the chain and however large the final proceeds; the blind optimum is approached quasistatically subsystem by subsystem, and the cap is exact (Sec.~\ref{sect:ex_cartridge}). The single-step bound, too, is attained at macroscopic scale by an explicit quasistatic protocol.

Figure~\ref{fig:ex} supplies that protocol. A disposable container holds $\Nx-1$ gas particles, initially confined together to one of two equally likely halves. They form an extensive correlation reservoir, $\TotalCorrelation=(\Nx-1)\ln 2$, while a one-bit memory records which half. Supplied as informational parent to every particle's update mechanism, the bit covers the aggregate loss, and the stated quasistatic feedback protocol attains the conventional free-energy limit: a one-bit readout controls the entire extensive correlation budget. When no mechanism reads the memory, the same correlation decrease generates no reading-pattern credit and is dissipated in full. Both poles are standard fixed-memory feedback \cite{sagawa2008second,Parrondo_2015} and calibrate the bound. The verdicts that separate this account from distribution-level accounting are those of Sec.~\ref{sect:ex_bit}.

A per-run budget changes what is at stake. When the work injectable before a resource is unlocked is capped at $b$ on every run, information no longer prices an exchange: it decides whether the resource is reached at all. Section~\ref{sect:ex_cartridge} proves this separation for an explicit correlation-keyed cartridge. The construction has $n$ working bits, each perfectly correlated with a frozen reference bit; a cartridge whose free energy exists prior to and independently of those bits; a fixed, tamper-proof gate releasing the cartridge only from a pass state of exponentially small prior weight; and, on the blind side, channels acting on each working bit separately, which renders the final bits independent. Within this stated per-bit local class, the informed protocol, whose channels read the reference bits, reaches the pass state with certainty at zero injected work, collecting the full cartridge, whereas the blind passing probability is at most $\exp(-n/2+\sqrt{n\invTemp b/2})$, exponentially small whenever the budget stays below the extensive preparation cost. The required correlation grows only logarithmically in the size of the locked resource and linearly in the budget: information does not amplify its own exchange value but switches reachability. This yields a budget-conditioned indispensability for access to an independent resource, a tightening of the general kind that state-function formulations were noted to leave open (Sec.~IX in \cite{Kolchinsky_20219}). Extending the probability bound to joint blind operations with fully priced auxiliary systems remains open. Several ingredients of the proposed test have related single-electron precedents \cite{Koski_2014,Scandi_2022}, but an integrated device measuring the informed--blind unlocking contrast under a declared per-run budget remains unrealized.

The three examples have distinct evidentiary roles: Sec.~\ref{sect:ex_bit} provides the operational separation, Sec.~\ref{sect:ex_container} the attainment and scaling, and Sec.~\ref{sect:ex_cartridge} the budget-conditioned reachability construction. Throughout, the scope is classical and limited to finite-state many-body systems in contact with a single heat bath, evolving by synchronized snapshot steps and their compositions. The synchronized step is a design choice, not an approximation. It is the update pattern of clocked architectures, in which every mechanism latches a beginning-of-step snapshot and the updates are released together. This pattern makes the attribution well defined, since every mechanism's input is fixed before any output moves. Nor is the class a special case of continuous-time dynamics. The acyclic two-bit keyed bijection $\Xs{1}\prm=\Xs{1}\oplus\Xs{2}$ with $\Xs{2}$ static already realizes a joint kernel of determinant $-1$; the mutual-refresh kernel of Appendix~\ref{ap:cex_refresh} one of determinant $-(2a-1)^{4}$; and every propagator of a continuous-time Markov process on the same state space has positive determinant. Continuous-time dynamics, where mechanisms read continuously rather than through a declared snapshot, and quantum systems lie beyond the present framework. Section~\ref{sect:setting} formalizes the setting and the informational parents; Sec.~\ref{sect:generalized-ineq} establishes the general ledger (Theorem~\ref{thm:ledger}) and its transparent reduction (Theorem~\ref{thm:main}); Sec.~\ref{sect:indispensability} develops the work consequences and conversion chains; Sec.~\ref{sect:ex} presents the three examples; Sec.~\ref{sect:conclusion} concludes.\section{Setting}
\label{sect:setting}
We set out the theoretical framework for our analysis. We consider the single-step stochastic evolution of a classical system in contact with a single heat bath maintained at a constant temperature $T$.
Let $\invTemp$ denote the inverse temperature $(\boltz \Temperature)^{-1}$, where $\boltz$ is the Boltzmann constant.
The total system is partitioned into $\Nx$ subsystems. Its state space is the product of the subsystem state spaces, each of which we assume to be finite\footnote{Finiteness is assumed for definiteness of the Shannon quantities. The countable case carries over verbatim wherever the entropies and mutual informations involved are finite; continuous state spaces require, beyond the replacement of sums by integrals (integral notation in the appendices is read in this sense), the usual regularity of conditional kernels and finiteness of the differential quantities, and are not treated formally here. All examples and counterexamples in this paper have finite state spaces.}. The state of the $k$-th subsystem is represented by a random variable, denoted by $\Xs{k}$ at the beginning of the process and by $\Xs{k}'$ at its conclusion. The initial states $(\Xs{1},\dots,\Xs{\Nx})$ are drawn from an arbitrary joint distribution, which may include correlations among the subsystems.
We use a colon-based notation for contiguous groups of subsystems. For example, $\Xs{j:k}$ denotes the joint state of subsystems $j$ through $k$, i.e., $(\Xs{j},\Xs{j+1},\dots,\Xs{k})$, and similarly for the final states $\Xs{j:k}'$. Under this convention, the state of the total system is $\Xtot=\Xs{1:N}$ initially and $\Xtotprm=\Xs{1:N}'$ finally.
Thus, $\Xtot$ and $\Xtotprm$ denote the initial and final states, respectively, of the entire composite system comprising subsystems 1 through $\Nx$.

For collections of subsystems defined relative to a given index, we introduce, for each $j\in\{1,\cdots,N\}$, the index ranges
\begin{IEEEeqnarray}{rCl}
    \minusJ := 1\mathbin{:}(j-1),
    \\
    \plusX{j} := (j+1):\Nx,
\end{IEEEeqnarray}
so that $\Xs{\minusJ}=\Xs{1:j-1}$ denotes the joint state of all subsystems preceding $j$, whereas $\Xs{\plusX{j}}=\Xs{j+1:N}$ denotes that of all subsystems succeeding $j$. The symbols $\prec$ and $\succ$, which recall the precedence order of the indices, are part of these range tokens; they are never used as binary relations in this paper.

To specify the causal structure of the evolution, we assume that the composite system evolves in one step through \textit{local mechanisms}. Each subsystem is updated by its own stochastic channel, driven by its own source of thermal noise. Throughout, these mechanisms specify the transition kernel $P(\Xtotprm\mid\Xtot)$ on the whole state space, not merely on the support of the initial distribution.

\begin{assumption}[Local mechanisms]
    \label{asm:local_mech}
    The joint transition probability of the composite system factorizes into local channels,
    \begin{IEEEeqnarray}{lCl}
        P(\Xtotprm\mid \Xtot)
        = \prod_{i=1}^{\Nx} P\big(\Xs{i}\prm\mid \Xs{i}, \DepX{i}\big),
        \label{eq:asm_local_mech}
    \end{IEEEeqnarray}
    for some collections $\DepX{i}\subseteq \Xtot\setminus\Xs{i}$ of initial-state variables.
\end{assumption}

Assumption~\ref{asm:local_mech} leaves the collections $\DepX{i}$ underdetermined because enlarging a valid collection produces another valid one. We therefore fix them by minimality. A factorization of the form \eqref{eq:asm_local_mech} exists if and only if the final states are conditionally independent given $\Xtot$. In that case, each factor necessarily equals the conditional law of $\Xs{i}\prm$ given $\Xtot$, so the collections can be minimized separately for each $i$. Because the kernel is defined on the whole state space, each channel has a \emph{unique} smallest dependence set\footnote{Write $f$ for the conditional law of $\Xs{i}\prm$ given $\Xtot$, and suppose $f$ depends on $\Xtot$ only through $(\Xs{i},A)$ and also only through $(\Xs{i},B)$. Given two configurations of $\Xtot$ agreeing on $(\Xs{i},A\cap B)$, alter the first outside $A$ to match the second: $f$ is unchanged, and the result agrees with the second configuration on $(\Xs{i},B)$, so $f$ is unchanged again. Hence $f$ depends on $\Xtot$ only through $(\Xs{i},A\cap B)$, and the minimal dependence set is unique. The argument evaluates $f$ on every configuration of the product state space; this is why the kernel, rather than the joint statistics, must be the primitive object.}. We adopt this minimal choice throughout and call $\DepX{i}$ the \textit{informational parents} of subsystem $i$: the initial-state variables, other than its own state, on which the update of $i$ depends. This definition is mechanism-based rather than distributional in two respects. First, a variable that a mechanism physically probes but that leaves its channel unchanged is not a parent; minimality discards it. Second, if the parents were defined through the joint statistics $P(\Xtot,\Xtotprm)$ rather than through the kernel, the smallest sets would not generally be unique. Under a degenerate initial distribution, two perfectly correlated initial variables can screen for one another. The bounds derived below would then depend on the chosen representative. Specifying the kernel on the whole state space removes this ambiguity. The parents are, finally, resolved at the granularity of the declared subsystem partition: a channel that depends on any part of a composite subsystem state has, in this accounting, the whole subsystem as a parent, so the operational reading of the bounds below presupposes a partition chosen at the granularity of the signals the mechanisms physically read.

Assumption~\ref{asm:local_mech} has two consequences that recur throughout the derivations.

\begin{lemma}[Screening]
    \label{lem:screening}
    Under Assumption~\ref{asm:local_mech}, for every initial distribution: (i) the final states $\Xs{1}\prm,\dots,\Xs{\Nx}\prm$ are conditionally mutually independent given the initial state $\Xtot$; (ii) conditioned on $(\Xs{i},\DepX{i})$, the final state $\Xs{i}\prm$ is statistically independent of all remaining initial and final variables.
\end{lemma}
\begin{proof}
    Part (i) is immediate from \eqref{eq:asm_local_mech}: given $\Xtot$, the joint law of the final states is a product in which each factor involves a single final variable. For part (ii), write $Z$ for the tuple of initial variables outside $(\Xs{i},\DepX{i})$, so that $\Xtot=(\Xs{i},\DepX{i},Z)$ up to reordering, and $W$ for the tuple of final variables other than $\Xs{i}\prm$. Multiplying \eqref{eq:asm_local_mech} by the conditional law of $Z$ given $(\Xs{i},\DepX{i})$ yields, for any initial distribution,
    \begin{IEEEeqnarray}{lCl}
        P\big(\Xs{i}\prm, W, Z\mid \Xs{i},\DepX{i}\big)
        = P\big(\Xs{i}\prm\mid \Xs{i},\DepX{i}\big)\, g(W,Z),
        \nonumber
    \end{IEEEeqnarray}
    where $g(W,Z):=P(Z\mid \Xs{i},\DepX{i})\prod_{j\neq i}P\big(\Xs{j}\prm\mid \Xs{j},\DepX{j}\big)$ collects every factor other than the channel of $i$; note that each $(\Xs{j},\DepX{j})$ is a function of $(\Xs{i},\DepX{i},Z)$. Conditional on $(\Xs{i},\DepX{i})$, the joint law thus splits into a factor in $\Xs{i}\prm$ alone and a factor in $(W,Z)$ alone, which is the asserted independence.
\end{proof}

Part (i) states that correlations between final states can only be inherited from the initial state and cannot be introduced by shared noise. Part (ii), the \textit{screening property}, is strictly stronger than requiring $\DepX{i}$ to contain every initial variable marginally correlated with $\Xs{i}\prm$. Appendix~\ref{ap:cex_xor} gives a process for which a marginal-dependence definition of the parents invalidates the identity \eqref{eq:199DD} underlying our main results, whereas the parents defined above preserve it. The factorization \eqref{eq:asm_local_mech} is the discrete-time, single-step analogue of the multipartite assumption used in the continuous-time thermodynamics of information \cite{horowitz2014thermodynamics,hartich2014stochastic,Horowitz_2015}. It holds by construction in the causal-network framework of Ref.~\cite{ito2013information}.

We now add thermodynamic assumptions to this probabilistic setting. All mechanisms exchange heat with the common bath, but each couples to the bath through its own degrees of freedom. This is the same idealization that makes the noise sources driving different subsystems statistically independent, as expressed by the factorization \eqref{eq:asm_local_mech}. In particular, the heat dissipated during the update of subsystem $i$, denoted by $\Qs{i}$, is separately well defined. Each channel is related to its dissipated heat through a local detailed balance condition.

\begin{assumption}[Local detailed balance]
    \label{asm:ldb}
    For every subsystem $i$,
    \begin{IEEEeqnarray}{lCr}
        \invTemp\Qs{i}
        =
        \langlee \ln\frac{P(X_i\dashh\mid X_i,\DepX{i})}{P\B(X_i\mid X_i\dashh,\DepX{i})}\ranglee,
        \label{eq:local_detail_subsystem}
    \end{IEEEeqnarray}
    where $P\B$ is the transition probability of the time-reversed process.
\end{assumption}

The integrand of \eqref{eq:local_detail_subsystem}, evaluated on a realized transition, is the \textit{trajectory heat} of that transition; its average is $\invTemp\Qs{i}$, and the comparisons of Sec.~\ref{sect:ex_bit} match processes at the level of these trajectory values. Throughout, the reversed channel is supported on the reversals of the forward transitions: $P\B(X_i\mid X_i\dashh,\DepX{i})>0$ only where $P(X_i\dashh\mid X_i,\DepX{i})>0$. Every implementation in this paper satisfies this condition, read on the forward-reachable outputs; in the ideal limits below it fixes the trajectory heats of bijective limits, while many-to-one resets retain the freedom of distributing reversed weight among their forward preimages. The exact form of the fluctuation theorem of Appendix~\ref{ap:fluctuation_subsystem} requires, in addition, that the conjugate distribution built there charge no event of zero forward probability---automatic when the relevant initial conditionals have full support, and accounted for explicitly in Appendix~\ref{ap:fluctuation_subsystem} otherwise; every averaged statement holds regardless. Degenerate channels---deterministic maps and resets whose targets lack full support---are admitted as ideal limits of full-support channels within this class, the forward channel, the reversed channel, and the trajectory heats converging jointly to a tuple that itself obeys the support convention: entropic quantities are then read in the extended-real sense with $0\ln0:=0$, the backward kernel is defined on the support of the forward path measure, and every statement involving such channels (the keyed bijections and quasistatic resets below) is understood in the corresponding limit. Assumption~\ref{asm:ldb} holds for a broad class of non-equilibrium dynamics, including discretized Langevin systems \cite{ito2013information}. Appendix~\ref{ap:fluctuation_subsystem} converts this condition into the per-subsystem entropy-production inequality from which the derivation in Sec.~\ref{sect:generalized-ineq} begins. With the total dissipated heat $\Qtot:=\sum_{i=1}^{\Nx}\Qs{i}$, the \textit{total entropy production} of the composite system is
\begin{IEEEeqnarray}{lCl}
\DeltaStot := \SAHEN,
\label{eq:def_sigma}
\end{IEEEeqnarray}
where $\Htot$ is the Shannon entropy of the total system.

\begin{assumption}[Acyclic architecture]
    \label{asm:acyclic}
    The directed graph on $\{1,\dots,\Nx\}$ obtained by drawing an edge $j\to i$ whenever $\Xs{j}\in\DepX{i}$ is acyclic.
\end{assumption}

Acyclicity is the substantive assumption; the associated labeling is only bookkeeping. Relabeling the subsystems in reverse topological order, we adopt throughout the \textit{parents-later convention}
\begin{IEEEeqnarray}{lCl}
    \DepX{i} \subseteq \Xs{\plusI}
    \qquad (i=1,\dots,\Nx).
    \label{eq:parents_later}
\end{IEEEeqnarray}
When several topological orders exist, any fixed choice suffices because every result below holds for each such labeling. In the multistage chains of Sec.~\ref{sect:indispensability}, where the parent structure may change between steps, the labeling is chosen anew at each step.

Assumption~\ref{asm:acyclic} states that the reading pattern contains no closed loop: there is no sequence of subsystems, each read by the next, that returns to its starting point. In particular, although this is not the only restriction, no two subsystems read each other within a single step. Acyclicity is equivalent to the existence of an update schedule in which every mechanism fires while its parents still retain their initial values. Under the parents-later convention, updating the subsystems in the order $i=1,2,\dots,\Nx$ realizes the dynamics \eqref{eq:asm_local_mech} exactly without copying an initial state into an external record. No such schedule exists for cyclic reading. In every update order, some mechanism fires after a variable that it reads has already been overwritten. An in-place sequential realization of the declared snapshot kernel, one in which each declared mechanism fires exactly once, then requires retaining that variable's initial value past its overwriting, effectively creating a frozen record. The cost of erasing such records is not captured by the entropy bookkeeping of the composite system alone. Appendix~\ref{ap:counterexamples} quantifies this loophole for the shortest cycle: explicit two-subsystem processes with reciprocal reading falsify the strengthened bounds derived below. Assumption~\ref{asm:acyclic} therefore cannot be omitted. We regard it as a physical restriction on the architecture rather than a technical convenience. Whether cycles of length three or more are equally fatal is not settled by these examples, so we impose acyclicity as the structural hypothesis under which the strengthened bounds are proved. The ledger inequality of Theorem~\ref{thm:ledger} below does not use Assumption~\ref{asm:acyclic} and remains valid for arbitrary reading patterns, including cyclic ones.

We first derive an exact lower bound on the total entropy production \eqref{eq:def_sigma}, namely the \textit{ledger} inequality \eqref{eq:158FF} of Theorem~\ref{thm:ledger}. This inequality holds under Assumptions~\ref{asm:local_mech} and \ref{asm:ldb} for an arbitrary parent structure and is saturated by explicit processes. Adding Assumption~\ref{asm:acyclic} makes its rebate terms vanish identically (Corollary~\ref{cor:no_rebate}). Restricting further to protocols that generate no \textit{unrealized correlation} yields the readily interpretable main result \eqref{eq:main_no_new_obs} of Theorem~\ref{thm:main}.

\section{Generalized Second-Law Inequality}
\label{sect:generalized-ineq}

We first specify the information-theoretic framework and notational conventions used to derive our main result.

\subsection{Information-Theoretic Preliminaries}

We begin with the standard information-theoretic quantities. The conditional Shannon entropy of a random variable $\Xj$ given another variable $\Xk$ is defined as
\begin{IEEEeqnarray}{c}
\shaent{\Xj\mid\Xk}:=\langlee -\ln P(\Xj\mid\Xk)\ranglee,
\end{IEEEeqnarray}
where $\langlee\cdot\ranglee$ denotes the expectation over the joint probability distribution. The conditional mutual information between $\Xj$ and $\Xk$ given $\Xl$ is defined as
\begin{IEEEeqnarray}{cCc}
\muI{\Xj}{\Xk\mid\Xl}:=\langlee\ln \left[\frac{P(\Xj,\Xk\mid\Xl)}{P(\Xj\mid\Xl)P(\Xk\mid\Xl)}\right]\ranglee.
\end{IEEEeqnarray}
To characterize dependencies among more than two variables, we use interaction information, which generalizes mutual information and is defined in Eq.~(\ref{eq:def_interaction_cond}).

To describe how these quantities change during the system's evolution, we adopt the following difference convention. The prefix $\Delta$ denotes the final value of a quantity minus its initial value when \emph{all} of its arguments evolve. For example,
\begin{IEEEeqnarray}{rCl}
\Delta \muI{A}{B\mid C} &:=& \muI{A\prm}{B\prm\mid C\prm} - \muI{A}{B\mid C}.
\label{eq:conv_mui}
\end{IEEEeqnarray}
When only some arguments evolve, we do not use $\Delta$. Instead, we write the difference explicitly, as in $\muI{A\prm}{B}-\muI{A}{B}$.

We now introduce the two quantities that enter our generalized inequality. The first is the change in the system's \textit{total correlation} \cite{Watanabe1960}, which measures the total statistical dependence among all subsystems. We denote this change by $\delitot$. As detailed in Appendix~\ref{ap:delitotTotalIC}, $\delitot$ represents the total increase in correlations within the system and is defined as the sum of the changes in mutual information between each subsystem and all preceding subsystems:
\begin{IEEEeqnarray}{rCl}
\delitot := \Sum{i=2}{N} \Delta \totalCorrI, \label{eq:defIJ}\label{eq:def_delitot}
\end{IEEEeqnarray}
where $\totalCorrI := \muI{\Xs{i}}{\Xs{\minusI}}$. The change in total correlation is related to the change in the system's total Shannon entropy, $\Delta\Htot$. Equation~(\ref{eq:F18YY}) gives the following decomposition, whose correlation sum equals $\delitot$ by the total-correlation decomposition in Appendix~\ref{ap:delitotTotalIC}:
\begin{IEEEeqnarray}{lCl}
\Delta \Htot = \Sum{i=1}{\Nx}\Delta\shaent{\Xs{i}} - \delitot. \label{eq:split_entropy} \IEEEeqnarraynumspace
\end{IEEEeqnarray}
The quantity $\delitot$ will be used throughout the subsequent discussion.

The second quantity, denoted $\UsableNew{}$, describes the change in mutual information between each subsystem and the other subsystems that causally influence its evolution. For each subsystem $i$, we define the explicit difference
\begin{IEEEeqnarray}{lCl}
    \UsableNew{i}:= \muI{\Xs{i}\prm}{\DepX{i}} - \muI{\Xs{i}}{\DepX{i}},
 \label{eq:def_eta}
\end{IEEEeqnarray}
Thus, $\UsableNew{i}$ is the change in mutual information between subsystem $i$ and its informational parents $\DepX{i}$. Only $\Xs{i}$ evolves, while $\DepX{i}$ remains evaluated at the initial time. The total contribution is
\begin{IEEEeqnarray}{lCl}
 \UsableNew{}:= \Sum{i=1}{N}\UsableNew{i}.
 \label{eq:defUsableNew}
\end{IEEEeqnarray}
This quantity is the information gain that the bounds derived below credit toward extractable work---the portion whose conversion is not ruled out; whether it is actually harvested depends on the protocol, and Sec.~\ref{sect:ex} exhibits explicit processes attaining the credit.

\subsection{A Preliminary Lower Bound on Entropy Production under Internal Correlations}
Appendix~\ref{ap:fluctuation_subsystem} shows that the local detailed balance condition in Assumption~\ref{asm:ldb} yields an entropy production inequality for each subsystem:
\begin{IEEEeqnarray}{rCl}
\shaent{\Xs{i}\prm\mid \DepX{i}} - \shaent{\Xs{i}\mid \DepX{i}}
 + \invTemp\Qs{i} \geq 0.
\label{eq:yuragi_k}
\end{IEEEeqnarray}
Applying \eqref{eq:SRARBSHARB_D1} to the left-hand side and summing over $i$ gives
\begin{IEEEeqnarray}{lCl}
    \Sum{i=1}{\Nx}\left[
        \Delta\shaent{\Xs{i}}
        - \UsableNew{i}
        + \invTemp\Qs{i}
    \right]
    =
    \Sum{i=1}{\Nx} \Delta\shaent{\Xs{i}}
    - \UsableNewAll +\invTemp\Qtot
    \geq 0.
    \label{eq:21FR}
\end{IEEEeqnarray}
Using \eqref{eq:F18YY}, we have
\begin{IEEEeqnarray}{lCl}
    \Sum{i=1}{\Nx} \Delta\shaent{\Xs{i}}
    = \Delta \shaent{\Xtot} + \delitot.
\end{IEEEeqnarray}
Substitution into \eqref{eq:21FR} yields
\begin{IEEEeqnarray}{lCl}
    \DeltaStot \geq \UsableNewAll - \delitot,
    \label{eq:2ndlaw1}
\end{IEEEeqnarray}
where $\DeltaStot$ is the total entropy production defined in \eqref{eq:def_sigma}.
If the initial distribution also factorizes along the parent structure and has full support, Appendix~\ref{ap:fluctuation_internal} strengthens this average inequality to an exact integral fluctuation theorem; without full support, its $1-\Lambda$ form yields the same average inequality.
Both conditions are essential for the exact form, whereas the derivation above imposes no condition on the initial distribution.
Under Assumptions~\ref{asm:local_mech} and \ref{asm:acyclic} alone, the right-hand side of \eqref{eq:2ndlaw1} is nonnegative (Lemma~\ref{lem:step_sum} of Appendix~\ref{ap:multistage}).
Thus, the generalized inequality is at least as strong as the conventional second law $\DeltaStot \geq 0$.

This inequality has a direct physical interpretation.
Under fully blind dynamics, $\DepX{i}=\emptyset$ for all $i$, so every information gain $\UsableNew{i}$ vanishes.
The inequality then reduces to $\DeltaStot \geq -\delitot$: any reduction in the system's total internal correlation must be dissipated entirely as entropy rather than converted into usable free energy\footnote{The condition $\DepX{i}=\emptyset$ cannot be relaxed to the mere absence of initial knowledge: a mechanism that reads $\DepX{i}$ without holding initial correlation with its own subsystem, $\muI{\Xs{i}}{\DepX{i}}=0$, can still build up $\muI{\Xs{i}\prm}{\DepX{i}}>0$ during the update (e.g., by copying), so that $\UsableNew{i}>0$.}.
Conversely, when $\UsableNewAll$ is non-zero, it can offset the $-\delitot$ term.
This suggests that the corresponding portion of the released internal correlation can be converted into work rather than dissipated.

The following sections reformulate the inequality to clarify the implications of the main claim stated in the Introduction.

\subsection{Information Components for the Main Result}
\label{sect:components}

Our main inequality uses a small number of \textit{information components}. These are all standard (conditional, possibly multivariate) Shannon information quantities. In addition to conditional mutual information, we use the conditional interaction information $\muI{A}{B;C\mid Z}$, defined in Eq.~(\ref{eq:def_interaction_cond}). Conditioning on a set of variables, such as $\DepX{i}$, means conditioning on the corresponding joint random variable. Table~\ref{tbl:notation} summarizes all symbols used in the main text. The derivation of the main result involves lengthy intermediate manipulations, which we defer to Appendix~\ref{ap:derivation}. That appendix uses the set-theoretic shorthand introduced in Appendix~\ref{ap:shorthand}; neither is needed to read the results below.

The first three components concern the pair formed by subsystem $i$ and the preceding block $\Xs{\minusI}$. As shown in Sec.~\ref{sect:indispensability}, they partition the portion of the initial correlation between $\Xs{i}$ and $\Xs{\minusI}$ that is lost during the evolution, according to which side's state change causes the loss (see Fig.~\ref{fig:venn_haita_seibun}):
\begin{IEEEeqnarray}{lCl}
    \atamaOmegaIMuMu := \muI{\Xs{i}}{\Xs{\minusI}\mid \Xs{i}\prm,\Xs{\minusI}\prm},
    \label{eq:atamaOmega}
    \\
    \centerSeibunIMu := \muI{\Xs{i}}{\Xs{\minusI};\Xs{\minusI}\prm\mid \Xs{i}\prm},
    \label{eq:centerSeibun}
    \\
    \bottomSeibunIMuMu := \muI{\Xs{i}}{\Xs{\minusI};\Xs{i}\prm\mid \Xs{\minusI}\prm}.
    \label{eq:bottomSeibun3}
\end{IEEEeqnarray}
Here, $\centerSeibunIMu$ is the loss caused solely by the state change of subsystem $i$, $\bottomSeibunIMuMu$ is the loss caused solely by the state changes within $\Xs{\minusI}$, and $\atamaOmegaIMuMu$ is the loss that requires the simultaneous change of both sides.

The remaining components quantify correlations that one subsystem attempts to establish with another during the evolution but that do not survive in the final state (\textit{unrealized correlations}; see Sec.~\ref{sect:mainresult}). For subsystems or blocks $u$, $v$, and $w$, we define
\begin{IEEEeqnarray}{lCl}
    \memotouSeibunL{u,v} := \muI{\Xs{u}}{\Xs{v}\prm\mid \Xs{u}\prm,\Xs{v}},
    \label{eq:memotouSeibunL}
    \\
    \memotouSeibunR{u,v} := \muI{\Xs{u}\prm}{\Xs{v}\mid \Xs{u},\Xs{v}\prm},
    \label{eq:memotouSeibunR}
    \\
    \memotouSeibunIMu := \memotouSeibunL{i,\minusI} + \memotouSeibunR{i,\minusI},
    \label{eq:memotouSeibunSum}
    \\
    \memotouSeibunLL{u,v,w} := \muI{\Xs{u}}{\Xs{v};\Xs{w}\prm\mid \Xs{u}\prm,\Xs{v}\prm,\Xs{w}},
    \label{eq:memotouSeibunLL}
    \\
    \botObsSeibun{u}{v}{w} := \muI{\Xs{u}}{\Xs{v};\Xs{u}\prm;\Xs{w}\prm\mid \Xs{v}\prm,\Xs{w}}.
    \label{eq:botobsSeibun}
\end{IEEEeqnarray}

\paragraph*{Conditioning rule.}
For any component $\mathcal{K}$ defined above and any collection $V$ of random variables, $\mathcal{K}\mid V$ denotes the quantity obtained by appending $V$ to the conditioning side of its defining expression. For example,
\begin{IEEEeqnarray}{lCl}
    \centerSeibunIMu\mid\infPaI := \muI{\Xs{i}}{\Xs{\minusI};\Xs{\minusI}\prm\mid \Xs{i}\prm,\infPaI}.
\end{IEEEeqnarray}
For a sum of components, we define $(\mathcal{K}_1+\mathcal{K}_2)\mid V := \mathcal{K}_1\mid V + \mathcal{K}_2\mid V$.

\paragraph*{The effective collective parent region $\effPa$.}
The only non-standard symbol in our main result is $\effPa$. Whereas $\infPaI$ is an ordinary collection of random variables, $\effPa$ denotes a \emph{region} of the information diagram \cite{yeung1991new,yeung2012first}. This region represents the informational parents of the block $\minusI$ as a whole. Writing $\Yeng{Z}$ for the region associated with a random variable $Z$, we define
\begin{IEEEeqnarray}{lCl}
    \effPa := \Yeng{\Xs{\minusI}}\setminus
    \bigcup_{n\in\minusI}\bigl(\Yeng{\Xs{n}}\setminus\Yeng{\infPa{n}}\bigr),
    \label{eq:103YY}
\end{IEEEeqnarray}
i.e., the part of the initial state of the block $\minusI$ that remains after removing the per-subsystem \emph{blind spots}. For each subsystem $n$, the difference $\Yeng{\Xs{n}}\setminus\Yeng{\infPa{n}}$ is the region of the conditional entropy $\shaent{\Xs{n}\mid\infPa{n}}$. It represents the part of the initial state of subsystem $n$ that is unknown to the variables controlling $n$ itself. A piece of the block's state therefore belongs to $\effPa$ exactly when every block member carrying it is controlled with knowledge of it. Thus, $\effPa$ is the \emph{effective} counterpart of $\infPa{n}$ for the block: information about a member that is held only by the controllers of other subsystems is not credited. Conditioning a component on $\effPa$, as in $\bottomSeibunIMuMu\mid\effPa$, removes the region $\effPa$ from the region of that component, in direct analogy with the conditioning rule above. When variables and the region $\effPa$ occur together in a conditioning list, as in $\atamaOmegaIMuMu\mid\infPaI,\effPa$, the removed region is the union of the regions of all listed entries. The same construction applies to any preceding block. For a member $k$ of the block, $\effPa[k]:=\Yeng{\Xs{\minusK}}\setminus\bigcup_{n\in\minusK}\bigl(\Yeng{\Xs{n}}\setminus\Yeng{\infPa{n}}\bigr)$ denotes the effective collective parent region of the sub-block $\minusK$; it enters $\Ucost$ below and the derivations of Appendix~\ref{ap:derivation}.

\begin{table*}[t]
    \caption{\label{tbl:notation}%
        Summary of the notation used in the main text. Apart from the region $\effPa$, every entry is a standard Shannon information quantity or a named combination thereof.
    }
    \resizebox{\textwidth}{!}{%
    \begin{tabular}{lll}
        \toprule
        Symbol & Meaning & Defined in \\
        \colrule
        $\Xs{i}$,\ $\Xs{i}\prm$ & initial / final state of subsystem $i$ & Sec.~\ref{sect:setting} \\
        $\Xs{j:k}$ & joint state $(\Xs{j},\dots,\Xs{k})$ & Sec.~\ref{sect:setting} \\
        $\Xs{\minusJ}$,\ $\Xs{\plusX{j}}$ & joint state of subsystems $1{:}(j{-}1)$ / $(j{+}1){:}N$ & Sec.~\ref{sect:setting} \\
        $\DepX{j}$ & informational parents (mechanism inputs) & Eq.~(\ref{eq:asm_local_mech}) \\
        $\muI{A}{B;C\mid Z}$ & conditional interaction information & Eq.~(\ref{eq:def_interaction_cond}) \\
        $\Delta$ & time difference in which all arguments evolve & Eq.~(\ref{eq:conv_mui}) \\
        $\delitot$ & change in the total correlation & Eq.~(\ref{eq:defIJ}) \\
        $\UsableNew{i}$,\ $\UsableNew{}$ & information gain of subsystem $i$ / its total & Eqs.~(\ref{eq:def_eta}) and (\ref{eq:defUsableNew}) \\
        $\DeltaStot$ & total entropy production & Eq.~(\ref{eq:def_sigma}) \\
        $\atamaOmegaIMuMu$,\ $\centerSeibunIMu$,\ $\bottomSeibunIMuMu$ & correlation-loss components & Eqs.~(\ref{eq:atamaOmega})--(\ref{eq:bottomSeibun3}) \\
        $\memotouSeibunL{u,v}$,\ $\memotouSeibunR{u,v}$,\ $\memotouSeibunIMu$ & unrealized-correlation components & Eqs.~(\ref{eq:memotouSeibunL})--(\ref{eq:memotouSeibunSum}) \\
        $\memotouSeibunLL{u,v,w}$,\ $\botObsSeibun{u}{v}{w}$ & higher-order unrealized components & Eqs.~(\ref{eq:memotouSeibunLL}) and (\ref{eq:botobsSeibun}) \\
        $\mathcal{K}\mid V$ & component $\mathcal{K}$ additionally conditioned on $V$ & Sec.~\ref{sect:components} \\
        $\effPa$ & effective collective parent region of the block $\minusI$ & Eq.~(\ref{eq:103YY}) \\
        $\Dmiss$,\ $\Ucost$,\ $\Rreb$ & ledger aggregates & Eqs.~(\ref{eq:def_Dmiss})--(\ref{eq:def_Rreb}) \\
        $\decIij$,\ $\incIij$ & destroyed / newly created correlation & Eqs.~(\ref{eq:def_decIij}) and (\ref{eq:def_incIij}) \\
        $\DissipationI$,\ $\Dissipation$ & hidden part of the destroyed correlation ($\DissipationI=\Dmiss$) / its total & Eqs.~(\ref{eq:def_DissipationI}) and (\ref{eq:sigma_geq_D}) \\
        $\accIiMu$ & controller-shared part of the destroyed correlation & Eq.~(\ref{eq:dec_split}) \\
        $\Wext$,\ $F$,\ $\Floc$ & extracted work; free energy and its local part & Eq.~(\ref{eq:def_Floc}) \\
        $\TotalCorrelation$ & total correlation of the composite system & Appendix~\ref{ap:delitotTotalIC} \\
        \botrule
    \end{tabular}
    }
\end{table*}

 \subsection{Main Result}
\label{sect:mainresult}

Starting from inequality \eqref{eq:2ndlaw1}, elementary identities reorganize the difference $\UsableNewAll - \delitot$ into the information components introduced in Sec.~\ref{sect:components}. Appendix~\ref{ap:derivation} gives the derivation, which culminates in the exact identity \eqref{eq:199DD}. This identity recasts the right-hand side of \eqref{eq:2ndlaw1} as a sum of these components. We state the result in a form that separates the physically distinct contributions.

One family of contributions consists of \textit{unrealized correlations}. We call the conditional mutual information
$\muI{\Xs{j}\prm}{\Xs{k}\mid \Xs{j},\Xs{k}\prm}$
the unrealized correlation from $j$ to $k$. It quantifies the correlation that subsystem $j$ attempts to establish with subsystem $k$ during its evolution but that does not survive in the final state because $k$ also evolves.

For each $i$, we group the components in the identity \eqref{eq:199DD} into three aggregates. The first,
\begin{IEEEeqnarray}{lCl}
    \Dmiss &:=&
    \atamaOmegaIMuMu\mid\infPaI,\effPa
    + \centerSeibunIMu\mid\infPaI
    + \bottomSeibunIMuMu\mid\effPa,
    \IEEEeqnarraynumspace
    \label{eq:def_Dmiss}
\end{IEEEeqnarray}
collects the loss of internal correlation that is invisible to the controllers of the side or sides whose state changes caused the loss. The second,
\begin{IEEEeqnarray}{rCl}
    \Ucost &:=& \Biggl[
    \memotouSeibunIMu
    + \sum_{k=2}^{i-1}\Bigl(
    \botObsSeibunIKMuk\mid\infPaK
    + \botObsSeibunIMukK\mid\effPa[k]
    \NonumberNewline
    &&\quad
    + \memotouSeibunLL{i,k,\minusK}\mid\infPaK
    + \memotouSeibunLL{i,\minusK,k}\mid\effPa[k]
    \Bigr)
    \Biggr]\Bigm|\Xs{\plusI},
    \IEEEeqnarraynumspace
    \label{eq:def_Ucost}
\end{IEEEeqnarray}
collects the cost of unrealized correlations. Each constituent has the form
\begin{IEEEeqnarray*}{c}
    \muI{\Xs{j}\prm}{\Xs{k};\cdots\mid \Xs{j},\Xs{k}\prm,\cdots},
\end{IEEEeqnarray*}
i.e., an unrealized correlation with additional variables in its interaction and conditioning slots. The third,
\begin{IEEEeqnarray}{lCl}
    \Rreb &:=& \left[
    \atamaOmegaIMuMu\scl\infPaI\scl\effPa
    + \muI{\Xs{i}\prm}{\Xs{\minusI}\prm\mid \Xs{i},\Xs{\minusI}}
    \middle]\relright|\Xs{\plusI},
    \IEEEeqnarraynumspace
    \label{eq:def_Rreb}
\end{IEEEeqnarray}
is a \textit{rebate} that enters the bound with a negative sign. In the first term, the semicolons denote the restriction of the region of $\atamaOmegaIMuMu$ to its common part with $\Yeng{\infPaI}$ and $\effPa$. This is the shorthand of Appendix~\ref{ap:shorthand}, dual to the conditioning rule of Sec.~\ref{sect:components}. The term is therefore the part of the jointly destroyed correlation that is simultaneously visible to the controller of $i$ \emph{and} to the effective controllers of the block. The second term is the correlation among final states that is not inherited from the initial state.

\begin{theorem}[Entropy-production ledger]
    \label{thm:ledger}
    Under Assumptions~\ref{asm:local_mech} and \ref{asm:ldb},
    \begin{IEEEeqnarray}{lCl}
        \DeltaStot \;\geq\;
        \Sum{i=2}{\Nx}\left(
        \Dmiss + \Ucost - \Rreb
        \right),
        \label{eq:158FF}
    \end{IEEEeqnarray}
    with equality if and only if the local entropy production \eqref{eq:yuragi_k} vanishes for every subsystem.
\end{theorem}
\begin{proof}
    By Proposition~\ref{prop:rigorous_uhen} of Appendix~\ref{ap:derivation}, together with Eq.~\eqref{eq:154CC}, the right-hand side of \eqref{eq:2ndlaw1} is identically equal to the right-hand side of \eqref{eq:158FF}. The inequality and its equality condition follow from \eqref{eq:2ndlaw1}, whose slack is the sum over $i$ of the nonnegative local entropy productions \eqref{eq:yuragi_k}.
\end{proof}

Three remarks are in order. First, Theorem~\ref{thm:ledger} does not use Assumption~\ref{asm:acyclic} and therefore remains valid for reciprocal parent structures. Second, the bound is tight: Appendix~\ref{ap:counterexamples} presents reciprocal processes that saturate \eqref{eq:158FF}. For reciprocal patterns, both the right-hand side and $\DeltaStot$ itself can be negative. This contradicts no second law. The quantity $\DeltaStot$ in Eq.~\eqref{eq:def_sigma} is the balance of the declared accounting boundary: the partitioned subsystems, with the heats that Assumption~\ref{asm:ldb} assigns to their mechanisms. When apparatus is left undeclared, it is a partial balance rather than the entropy production of a closed device. Unless declared as subsystems, clocks and controllers lie outside this boundary, as do, for example, the frozen records that a single-firing sequential realization of a reciprocal step must keep (Sec.~\ref{sect:setting}). The mutual-overwriting process of Appendix~\ref{ap:cex_xor}, with $\DeltaStot=-\ln 2$, illustrates both points; tightness there means that the bound is attained, not that it sharpens a bound on the closed apparatus. Third, the rebate $\Rreb$ cannot simply be dropped. For reciprocal architectures, the subtraction is necessary because the mutually visible part of a jointly destroyed correlation offsets part of the unrealized-correlation cost $\Ucost$ \footnote{An earlier version of this work argued heuristically that the term $\atamaOmegaIMuMu\scl\infPaI\scl\effPa$ could be discarded on the grounds that a shared resource must not be counted twice. That argument is refuted by the mutual-refresh process of Appendix~\ref{ap:cex_refresh}, which saturates \eqref{eq:158FF} with equality while the rebate is strictly positive; any strengthening that omits $\Rreb$ without further assumptions is therefore false.}.

For acyclic architectures, however, the rebate vanishes identically, and the ledger simplifies without heuristic input.

\begin{corollary}
    \label{cor:no_rebate}
    Under Assumptions~\ref{asm:local_mech}--\ref{asm:acyclic}, $\Rreb = 0$ for every $i$, and hence
    \begin{IEEEeqnarray}{lCl}
        \DeltaStot \;\geq\;
        \Sum{i=2}{\Nx}\left(
        \Dmiss + \Ucost
        \right).
        \label{eq:204EE}
    \end{IEEEeqnarray}
\end{corollary}
\begin{proof}
    Consider the second term of $\Rreb$. Appending $\Xs{\plusI}$ to its conditioning side yields $\muI{\Xs{i}\prm}{\Xs{\minusI}\prm\mid \Xtot}$, which vanishes because the final states are conditionally mutually independent given $\Xtot$ (Lemma~\ref{lem:screening}). For the first term, the parents-later convention \eqref{eq:parents_later} gives $\Yeng{\infPaI}\subseteq\Yeng{\Xs{\plusI}}$, so the region of $\atamaOmegaIMuMu\scl\infPaI\scl\effPa$, being contained in $\Yeng{\infPaI}$, is removed entirely by the conditioning on $\Xs{\plusI}$; an empty region carries zero signed measure.
\end{proof}

The assumptions have distinct roles. Assumptions~\ref{asm:local_mech} and \ref{asm:acyclic} eliminate the two rebate terms separately, without any appeal to physical plausibility. Outside the acyclic class, the strengthening \eqref{eq:204EE} is false (Appendix~\ref{ap:cex_refresh}).

To reduce inequality \eqref{eq:204EE} to a form that directly expresses the central claim of this work, we restrict attention to protocols that generate no unrealized correlation.

\begin{assumption}[No unrealized correlation]
    \label{asm:no_unrealized}
    For all $j\neq k$ and every collection $Z$ of initial and final subsystem states,
    \begin{IEEEeqnarray}{lCl}
        \muI{\Xs{j}\prm}{\Xs{k}\mid \Xs{j},\Xs{k}\prm, Z} \;=\; 0.
        \label{eq:asm_no_unrealized}
    \end{IEEEeqnarray}
\end{assumption}

The conditional form of \eqref{eq:asm_no_unrealized}, with arbitrary $Z$, is required because the derivation below produces such terms with additional variables in the conditioning slot. Assumption~\ref{asm:no_unrealized} restricts how correlations evolve, not their initial strength. Two classes of protocols satisfy it by construction. The first is the standard feedback setting, in which a single subsystem is updated while every other variable, including the memories being read, remains unchanged during the step. For a static target $\Xs{k}\prm=\Xs{k}$, the conditioning in \eqref{eq:asm_no_unrealized} determines $\Xs{k}$. For a static source $\Xs{j}\prm=\Xs{j}$, it determines $\Xs{j}\prm$. Thus, every term vanishes. The second class consists of dynamics for which the pair $(\Xs{j},\Xs{k}\prm)$ determines $\Xs{k}$ for all $j\neq k$, as in the mutual-overwriting process of Appendix~\ref{ap:cex_xor}. Determinism alone, however, does \emph{not} suffice. A deterministic mechanism can build correlation toward a partner whose own move erases it within the same step \footnote{For instance, take $\Nx=4$ with $\Xs{1}\prm=\Xs{1}\oplus\Xs{3}$, $\Xs{2}\prm=\Xs{2}\oplus\Xs{4}$, $\Xs{3}\prm=\Xs{3}$, $\Xs{4}\prm=\Xs{4}$, the initial state maximally correlated as $\Xs{2}=\Xs{3}$ with $\Xs{1},\Xs{4}$ independent and uniform. Every local update is a bijection of the subsystem's state space, yet $\muI{\Xs{1}\prm}{\Xs{2}\mid \Xs{1},\Xs{2}\prm}=\ln 2$: subsystem 1 correlates itself with $\Xs{2}$ (through $\Xs{3}$), while the simultaneous move of subsystem 2 hides that correlation from the final state.}.

\begin{lemma}
    \label{lem:U_vanish}
    Under Assumptions~\ref{asm:local_mech} and \ref{asm:no_unrealized}, $\Ucost = 0$ for every $i$.
\end{lemma}
\begin{proof}
    Each constituent of $\Ucost$ reduces to a finite signed combination of terms of the form \eqref{eq:asm_no_unrealized}, and hence vanishes: the block components split into pairwise unrealized correlations by the chain rule; the extra interaction slots of the higher-order components are removed one at a time via the recursion \eqref{eq:def_interaction_cond}; and the conditionings on the regions $\effPa[k]$ expand, by inclusion--exclusion over the blind spots in \eqref{eq:103YY}, into conditionings on collections of variables, which are absorbed into $Z$. The full reduction is carried out in Appendix~\ref{ap:proof_U_vanish}.
\end{proof}

Combining Corollary~\ref{cor:no_rebate} with Lemma~\ref{lem:U_vanish} yields our main result.

\begin{theorem}[Main result]
    \label{thm:main}
    Under Assumptions~\ref{asm:local_mech}--\ref{asm:no_unrealized},
    \begin{IEEEeqnarray}{rCl}
        \DeltaStot&\geq&
        \Sum{i=2}{\Nx}\left(
        \atamaOmegaIMuMu\mid\infPaI,\effPa
        + \centerSeibunIMu\mid\infPaI
        + \bottomSeibunIMuMu\mid\effPa
        \right).
        \IEEEeqnarraynumspace
        \label{eq:main_no_new_obs}
    \end{IEEEeqnarray}
\end{theorem}

The right-hand side of \eqref{eq:main_no_new_obs} has a compact interpretation in the language of the information diagram. For each $i$, let
\begin{IEEEeqnarray}{lCl}
    \mathcal{D}_i := \big(\Yeng{\Xs{i}}\cap\Yeng{\Xs{\minusI}}\big)\setminus\big(\Yeng{\Xs{i}\prm}\cap\Yeng{\Xs{\minusI}\prm}\big)
    \label{eq:def_destroyed_region}
\end{IEEEeqnarray}
denote the region of the initial correlation between $i$ and the preceding block that is destroyed during the step. We assign each atom in this region a \textit{rescue region}: $\Yeng{\infPaI}$ for atoms of $\centerSeibunIMu$ destroyed by the move of $i$, and $\effPa$ for atoms of $\bottomSeibunIMuMu$ destroyed by the move of the block. For atoms of $\atamaOmegaIMuMu$ destroyed jointly, it is $\Yeng{\infPaI}\cup\effPa$, because the conditioning may come from either side. The term \textit{rescue region} names the conditioning attached to a destroyed atom; it implies neither that the corresponding contribution is nonnegative nor that the correlation is operationally recovered. The right-hand side of \eqref{eq:main_no_new_obs} is precisely the measure of the destroyed correlation lying outside the rescue regions: \emph{the ledger conditions each destroyed component on the information read by the controllers of the side that caused the loss}---information located elsewhere does not enter the conditioning. The resulting subsystem-resolved terms are signed, and nonnegativity is guaranteed only for the aggregate over the step (Sec.~\ref{sect:indispensability}). The three components $\atamaOmegaIMuMu$, $\centerSeibunIMu$, and $\bottomSeibunIMuMu$ enumerate the possible destroying sides.

Two limiting cases illustrate the bound. For fully blind dynamics, $\DepX{i}=\emptyset$ for all $i$, the screening property (Lemma~\ref{lem:screening}) forces every unrealized-correlation component and both rebate terms to vanish identically. Theorem~\ref{thm:ledger} then reduces, without invoking Assumption~\ref{asm:no_unrealized}, to $\DeltaStot\geq\Sum{i=2}{\Nx}(\atamaOmegaIMuMu + \centerSeibunIMu + \bottomSeibunIMuMu)$: in the absence of observation, the destroyed internal correlation is dissipated in full. Conversely, when the parents cover all destroyed correlations, the right-hand side of \eqref{eq:main_no_new_obs} vanishes. The bound then reduces to the conventional second law, leaving room for the released correlation to be converted into work.

How much of this simplification survives without Assumption~\ref{asm:no_unrealized}? The ledger \eqref{eq:204EE} remains valid, so the question reduces to the sign of the aggregate unrealized-correlation cost separating \eqref{eq:204EE} from \eqref{eq:main_no_new_obs}. At the subsystem level, the sign can be negative. Take $\Nx=3$ with $\Xs{1}\prm=\Xs{2}\oplus\Xs{3}$, $\Xs{2}\prm=0$, and $\Xs{3}\prm=\Xs{3}$, so that $\DepX{1}=(\Xs{2},\Xs{3})$ and $\DepX{2}=\DepX{3}=\emptyset$, with the three initial states independent and uniform. Then $\Ucost[2]=\ln2$ while $\Ucost[3]=-\ln2$. The update of subsystem $1$ records the parity of $(\Xs{2},\Xs{3})$ in its own state. The simultaneous erasure of subsystem $2$ turns the resulting correlation into a negative interaction contribution to $\Ucost[3]$. The step total, by contrast, appears to be universally nonnegative. We conjecture that
\begin{IEEEeqnarray}{lCl}
    \Sum{i=2}{\Nx}\Ucost \;\geq\; 0
    \label{eq:Ucost_sum}
\end{IEEEeqnarray}
under Assumptions~\ref{asm:local_mech} and \ref{asm:acyclic} alone. Equality holds under Assumption~\ref{asm:no_unrealized} (Lemma~\ref{lem:U_vanish}), but not only under that assumption: the example above saturates \eqref{eq:Ucost_sum} while violating it. We have verified \eqref{eq:Ucost_sum} for every parent structure with $\Nx\leq5$ using exact linear-programming certificates over the Shannon cone constrained by the screening property. Each certificate decomposes the step total into $\Nx(\Nx-1)$ conditional mutual informations. We have also verified the inequality on extensive random ensembles of deterministic and stochastic processes; a structure-independent proof remains open. Whenever \eqref{eq:Ucost_sum} holds, Corollary~\ref{cor:no_rebate} alone yields $\DeltaStot\geq\Sum{i=2}{\Nx}\Dmiss$, the inequality of Theorem~\ref{thm:main}, without Assumption~\ref{asm:no_unrealized}. The role of that assumption is then to localize the bound: every $\Ucost$ vanishes individually, and the aggregate ledger reduces to the subsystem-resolved form.

The resolution by parents also locates the bound relative to accounts that price the step by sequential or modular bookkeeping \cite{ito2013information,Wolpert_2020,WolpertMinEP2020,Boyd_2018,wolpert2019stochastic}. Those accounts take a declared reading pattern as input. Once a valid acyclic declaration is fixed, the stage quantities of every parents-later schedule and the values of the fixed-block modularity difference, evaluated as algebra on every bipartition, admissible or not, are functionals of the declaration, the realized statistics, and the trajectory heats (Lemma~\ref{lem:declaration_measurability} of Appendix~\ref{ap:declaration_measurability}). The right-hand side of \eqref{eq:main_no_new_obs}, however, is not such a functional. Sec.~\ref{sect:ex_bit} exhibits two processes sharing all three whose bounds and implementation-optimal costs differ by $\ln 2$. In the next section, we show that \eqref{eq:main_no_new_obs} implies the central claim of this work.

\section{Indispensability of Information}
\label{sect:indispensability}

The inequality \eqref{eq:main_no_new_obs} implies the central claim of this work:
\textit{internal correlations can be converted into work only through the part that is informationally shared with the controllers; whatever correlation is destroyed while hidden from them is, in aggregate, entirely dissipated.}
The argument rests on one identity and one inequality. We first establish the identity, which splits the correlation destroyed during the evolution into the right-hand side of \eqref{eq:main_no_new_obs} and a part shared with the controllers (Eqs.~\eqref{eq:decIiMu_decomp} and \eqref{eq:dec_split}). Calling the former the hidden part is then a matter of definition, Eq.~\eqref{eq:def_DissipationI}, motivated by the conditioning structure of its three terms. The inequality is \eqref{eq:main_no_new_obs} itself. Combining the two with energy conservation yields a bound on the extractable work, Corollary~\ref{cor:work_bound}, from which the hidden part has dropped out. This establishes the claim.

\paragraph*{Destroyed correlation.}
The change of correlation between subsystems $i$ and $j$ combines two distinct processes: the destruction of correlation that existed initially and the creation of new correlation in the final state. Both amounts are obtained by subtracting from the initial and the final correlation, respectively, the surviving part of the initial correlation, namely the interaction information $\muI{\Xs{i}}{\Xs{j};\Xs{i}\prm;\Xs{j}\prm}$:
\begin{IEEEeqnarray}{lCl}
    \decIij &:=& \muI{\Xs{i}}{\Xs{j}} - \muI{\Xs{i}}{\Xs{j};\Xs{i}\prm;\Xs{j}\prm},
    \label{eq:def_decIij}
    \\
    \incIij &:=& \muI{\Xs{i}\prm}{\Xs{j}\prm} - \muI{\Xs{i}}{\Xs{j};\Xs{i}\prm;\Xs{j}\prm},
    \label{eq:def_incIij}
\end{IEEEeqnarray}
so that $-\Delta\muI{\Xs{i}}{\Xs{j}} = \decIij - \incIij$. Under Assumption~\ref{asm:no_unrealized} both amounts are nonnegative (Lemma~\ref{lem:dec_inc_signs} of Appendix~\ref{ap:subsystem_signs}). The surviving part itself, however, is an interaction information and carries no definite sign. Where it is negative, the destroyed correlation \eqref{eq:def_decIij} exceeds the correlation initially present (see the two-bit example below Lemma~\ref{lem:dec_inc_signs}). The same split applies to the pair formed by subsystem $i$ and the block $\Xs{\minusI}$, and summing over the decomposition \eqref{eq:defIJ} of the total correlation yields
\begin{IEEEeqnarray}{lCl}
    -\delitot = \Sum{i=2}{\Nx} \left( \decIiMu - \incIiMu \right).
    \label{eq:release_split}
\end{IEEEeqnarray}
The destroyed correlation $\decIiMu$ is precisely the quantity that the three components of Sec.~\ref{sect:components} partition (see Appendix~\ref{ap:shorthand} for the elementary computation):
\begin{IEEEeqnarray}{lCl}
\decIiMu
&=& \atamaOmegaIMuMu + \centerSeibunIMu + \bottomSeibunIMuMu,
\label{eq:decIiMu_decomp}
\end{IEEEeqnarray}
The three terms sort the lost correlation according to the state change that causes the loss: that of $i$ alone ($\centerSeibunIMu$), those inside $\minusI$ alone ($\bottomSeibunIMuMu$), or only both together ($\atamaOmegaIMuMu$). These three cases are visualized in Fig.~\ref{fig:venn_haita_seibun}.

\begin{figure}[t]
    \centering
    % TikZ replacement for venn_haita_seibun_1.eps (notation redesign, issue #2).
% Region topology:
%   T_i = i & <i, outside both primed curves          (top of lens)
%   M_i = i & <i & <i', outside i'                    (middle of lens)
%   B_i = i & <i & i', outside <i'                    (bottom of lens)
%   V^L = i & <i', outside i' and <i                  (left)
%   V^R = i' & <i, outside i and <i'                  (right)
\begin{tikzpicture}[scale=0.78, every node/.style={font=\small}]
    % initial states: solid circles
    \draw[line width=1.1pt] (-1.1,0) circle [radius=2.2];
    \draw[line width=1.1pt] (1.1,0) circle [radius=2.2];
    % final state of X_i: dashed ellipse (reaches right into X_{<i})
    \draw[line width=1.1pt, dashed]
        ([shift={(0.7,-1.0)}] 0,0) ellipse [x radius=2.6, y radius=0.8, rotate=-15];
    % final state of X_{<i}: dotted ellipse (reaches left into X_i)
    \draw[line width=1.1pt, dotted]
        ([shift={(-0.7,0.5)}] 0,0) ellipse [x radius=2.6, y radius=0.8, rotate=-15];
    % curve labels
    \node at (-2.75,2.05) {$\Xs{i}$};
    \node at (2.75,2.05) {$\Xs{\minusI}$};
    \node at (3.85,-1.85) {$\Xs{i}\prm$};
    \node at (-3.85,1.45) {$\Xs{\minusI}\prm$};
    % region labels
    \fill (0,1.5) circle [radius=1.4pt];
    \draw[gray] (0,1.5) -- (1.7,2.6);
    \node[right] at (1.7,2.6) {$\atamaOmegaIMuMu$};
    \node at (0,0.42) {$\centerSeibunIMu$};
    \node at (0,-1.05) {$\bottomSeibunIMuMu$};
    \node at (-2.25,0.78) {$\memotouSeibunL{i,\minusI}$};
    \node at (2.3,-0.82) {$\memotouSeibunR{i,\minusI}$};
\end{tikzpicture}
    \caption{
        Information diagram illustrating the evolution of the correlation between subsystem $i$ and the preceding subsystems $\Xs{\minusI}$. The quantities $\atamaOmegaIMuMu$, $\centerSeibunIMu$, and $\bottomSeibunIMuMu$ represent an exclusive partitioning of the initial correlation $\muI{\Xs{i}}{\Xs{\minusI}}$ that is lost during the time evolution.
        For simplicity, the region corresponding to $\muI{\Xs{i}\prm}{\Xs{\minusI}\prm\mid \Xs{i},\Xs{\minusI}}$ is omitted.
    }
    \label{fig:venn_haita_seibun}
\end{figure}

\paragraph*{Hidden and shared parts of the loss.}
A lost component of correlation could be preserved, and its free energy harvested, only by acting on the very state changes that cause the loss. A controller can act only on what it knows. The evolution of subsystem $i$ is conditioned on its informational parents $\infPaI$. For the block $\minusI$ as a whole, the role of the parents is played by the region $\effPa$ of Eq.~\eqref{eq:103YY}. This is the region that the right-hand side of \eqref{eq:main_no_new_obs} itself attaches to the block components, rather than a conditioning derived from the dynamics. Conditioning each component on exactly these regions isolates the part of the loss about which the responsible controllers hold no information. The three cases are $\centerSeibunIMu\mid\infPaI$ for the loss caused by $i$ alone, $\bottomSeibunIMuMu\mid\effPa$ for the loss caused inside $\minusI$, and $\atamaOmegaIMuMu\mid\infPaI,\effPa$ for the loss requiring both changes. The last loss can be averted from either side and is beyond control only when hidden from both. These are precisely the three summands of \eqref{eq:main_no_new_obs}. We write
\begin{IEEEeqnarray}{lCl}
\DissipationI :=
\atamaOmegaIMuMu\mid\infPaI,\effPa
+ \centerSeibunIMu\mid\infPaI
+ \bottomSeibunIMuMu\mid\effPa
\IEEEeqnarraynumspace
\label{eq:def_DissipationI}
\end{IEEEeqnarray}
for this hidden part. Term for term, this is the ledger aggregate $\Dmiss$ of \eqref{eq:def_Dmiss}. The two symbols denote the same quantity, and we write $\DissipationI$ from here on. With this notation, the decomposition \eqref{eq:decIiMu_decomp} splits the destroyed correlation into hidden and shared parts,
\begin{IEEEeqnarray}{lCl}
\decIiMu = \DissipationI + \accIiMu,
\qquad
\accIiMu := \decIiMu - \DissipationI,
\IEEEeqnarraynumspace
\label{eq:dec_split}
\end{IEEEeqnarray}
where the shared part $\accIiMu$ collects, from each component, the portion shared with the information available to its controllers. For instance, the contribution of $\centerSeibunIMu$ to $\accIiMu$ is $\centerSeibunIMu - \centerSeibunIMu\mid\infPaI = \muI{\Xs{i}}{\Xs{\minusI};\Xs{\minusI}\prm;\infPaI\mid \Xs{i}\prm}$, the part of the loss actually known to the controller of $i$. Two remarks temper this reading. First, the split \eqref{eq:dec_split} is an exact identity, and the quantity it splits is itself nonnegative ($\decIiMu\geq0$ for every $i$ under Assumption~\ref{asm:no_unrealized}, as is $\incIiMu$; Lemma~\ref{lem:dec_inc_signs} of Appendix~\ref{ap:subsystem_signs}), but it is \emph{not} a decomposition into nonnegative parts: appending variables to the conditioning side can \emph{increase} an information component, so $\accIiMu$ (and, for $\Nx\geq 5$, even $\DissipationI$ itself) can be negative under Assumptions~\ref{asm:local_mech}--\ref{asm:no_unrealized}. Appendix~\ref{ap:subsystem_signs} delimits the failure: $\DissipationI\geq0$ is proved there for a class of parent structures including every acyclic process with $\Nx\leq4$ (Corollary~\ref{cor:safe_classes}), and an explicit five-subsystem counterexample shows the restriction to be essential. Only the subsystem-resolved signs fail, however: the step totals obey $\Sum{i=2}{\Nx}\DissipationI\geq0$ and $\Sum{i=2}{\Nx}\accIiMu\geq0$ (Lemmas~\ref{lem:step_sum} and \ref{lem:acc_sum} of Appendix~\ref{ap:multistage}). Second, one conditioned component is nonnegative unconditionally: Assumption~\ref{asm:no_unrealized} reduces it to an ordinary conditional mutual information,
\begin{IEEEeqnarray}{lCl}
\centerSeibunIMu\mid\infPaI = \muI{\Xs{i}}{\Xs{\minusI}\prm\mid \Xs{i}\prm,\infPaI} \;\geq\; 0
\label{eq:cpa_closed_form}
\end{IEEEeqnarray}
\footnote{By the recursion \eqref{eq:def_interaction_cond} and the symmetry of the interaction information, $\centerSeibunIMu\mid\infPaI = \muI{\Xs{i}}{\Xs{\minusI}\prm\mid \Xs{i}\prm,\infPaI} - \muI{\Xs{i}}{\Xs{\minusI}\prm\mid \Xs{\minusI},\Xs{i}\prm,\infPaI}$. Expanding $\Xs{\minusI}\prm$ in the subtrahend by the chain rule yields a sum of terms of the form $\muI{\Xs{k}\prm}{\Xs{i}\mid \Xs{k},\Xs{i}\prm,Z}$ with $k\in\minusI$, each of which vanishes by \eqref{eq:asm_no_unrealized}.}. In this notation, the main result \eqref{eq:main_no_new_obs} takes the compact form
\begin{IEEEeqnarray}{lCl}
\DeltaStot \geq \Dissipation := \Sum{i=2}{\Nx}\DissipationI.
\label{eq:sigma_geq_D}
\end{IEEEeqnarray}
Only the hidden parts appear in the bound; the shared parts $\accIiMu$ are absent. Because the mechanisms read only part of the state, this correlation remains hidden from the controllers and is charged, in full, to entropy production. The charge applies to the step total $\Dissipation$, not to each $\DissipationI$ separately (see the remarks below \eqref{eq:dec_split}).

\paragraph*{Work bound.}
To state the thermodynamic consequence, let $E$ denote the internal energy of the composite system, $\Wext$ the work extracted during the process, and
\begin{IEEEeqnarray}{lCl}
F := E - \boltz\Temperature\,\shaent{\Xtot},
\quad
\Floc := E - \boltz\Temperature\Sum{i=1}{\Nx}\shaent{\Xs{i}}
\IEEEeqnarraynumspace
\label{eq:def_Floc}
\end{IEEEeqnarray}
the nonequilibrium free energy of the system and its \textit{local} part. For additive bare energies, the latter is the free energy the system would possess if the subsystems were uncorrelated. For interacting bare energies, $\Floc$ is the formal local part evaluated at the actual mean energy, and the resource interpretations attached to the bounds below assume additivity. The first law, $\Wext = -\Delta E - \Qtot$, combined with the definition \eqref{eq:def_sigma} of $\DeltaStot$, gives the exact identity
\begin{IEEEeqnarray}{lCl}
\Wext = -\Delta F - \boltz\Temperature\,\DeltaStot,
\label{eq:work_identity}
\end{IEEEeqnarray}
while the entropy decomposition \eqref{eq:split_entropy} separates the free energy into its local and correlation parts, $\Delta F = \Delta\Floc + \boltz\Temperature\,\delitot$.

\begin{corollary}
\label{cor:work_bound}
Under Assumptions~\ref{asm:local_mech}--\ref{asm:no_unrealized},
\begin{IEEEeqnarray}{lCl}
\Wext \leq -\Delta\Floc
+ \boltz\Temperature\Sum{i=2}{\Nx}\left( \accIiMu - \incIiMu \right).
\IEEEeqnarraynumspace
\label{eq:work_bound}
\end{IEEEeqnarray}
\end{corollary}
\begin{proof}
Substituting $\Delta F = \Delta\Floc + \boltz\Temperature\,\delitot$ into \eqref{eq:work_identity} gives
\begin{IEEEeqnarray}{rCl}
\Wext
&=& -\Delta\Floc
- \boltz\Temperature\left(\delitot + \DeltaStot\right).
\nonumber
\end{IEEEeqnarray}
By \eqref{eq:sigma_geq_D} together with Eqs.~\eqref{eq:release_split} and \eqref{eq:dec_split},
\begin{IEEEeqnarray}{rCl}
-\delitot - \DeltaStot
&\leq& -\delitot - \Dissipation
\NonumberNewline
&=& \Sum{i=2}{\Nx}\left(\accIiMu - \incIiMu\right).
\nonumber
\end{IEEEeqnarray}
\end{proof}

The bound \eqref{eq:work_bound} exhibits the indispensability of information at the level of step totals. The work extractable beyond the local free-energy change is fed by the destroyed correlation, but only through its shared part $\accIiMu$. The hidden part $\DissipationI$ has cancelled between the released correlation and the guaranteed dissipation. This cancellation occurs because every nat of the step total $\Dissipation$ of correlation destroyed while hidden from its controllers releases $\boltz\Temperature$ of free energy while simultaneously incurring at least the same amount of dissipation. (The newly created correlation $\incIiMu$ enters with a negative sign because building up correlation consumes free energy. The aggregate wording is essential: the individual $\accIiMu$ and $\DissipationI$ are not sign-definite, so a particular destroyed correlation may even reduce the budget. Only the step totals are guaranteed, $\Sum{i=2}{\Nx}\accIiMu\geq0$ and $\Dissipation\geq0$, by Lemmas~\ref{lem:acc_sum} and \ref{lem:step_sum} of Appendix~\ref{ap:multistage}; cf.\ the remarks below \eqref{eq:dec_split}.) In the limit of no information, $\infPa{n}=\emptyset$ for all $n$, the region $\effPa$ is empty by \eqref{eq:103YY}, so $\DissipationI = \decIiMu$ and $\accIiMu = 0$. The released correlations then contribute nothing to the work budget, consistent with the discussion below Eq.~\eqref{eq:2ndlaw1}. Conversely, when the parents cover all destroyed correlations, every atom of the destroyed region $\mathcal{D}_i$ of \eqref{eq:def_destroyed_region} lies inside its rescue region. All three conditioned components then vanish because the conditioning in \eqref{eq:def_DissipationI} removes exactly those rescue regions \footnote{The covering condition is realizable within the acyclic class. In the standard feedback setting---subsystem $1$ alone moves, reading all others, $\infPa{1}=\Xs{\plusX{1}}$, while $\Xs{n}\prm=\Xs{n}$ and $\infPa{n}=\emptyset$ for $n\geq2$---the regions of $\centerSeibunIMu$ and $\atamaOmegaIMuMu$ are empty for every $i$, being contained in $\Yeng{\Xs{i}}\setminus\Yeng{\Xs{i}\prm}=\emptyset$, and every atom of the region of $\bottomSeibunIMuMu$ lies in $\Yeng{\Xs{i}}\cap\Yeng{\Xs{1}}$ outside $\Yeng{\Xs{\minusI}\prm}$; such an atom belongs to $\effPa$, because the blind spot $\Yeng{\Xs{n}}$ of each static member of the block is contained in $\Yeng{\Xs{\minusI}\prm}$, while the blind spot $\Yeng{\Xs{1}}\setminus\Yeng{\Xs{\plusX{1}}}$ of the mover is disjoint from $\Yeng{\Xs{i}}$.}. Then $\accIiMu = \decIiMu$, and \eqref{eq:work_bound} reduces to the conventional bound $\Wext \leq -\Delta F$, in which the entire released correlation is available to the conventional work balance, subject to ordinary attainability conditions. Between these extremes, the bound \eqref{eq:work_bound} states that the information held by the controllers, and nothing else, delimits the portion of the internal correlation that can act as a thermodynamic resource: whatever correlation is lost outside this information is, in aggregate, irretrievably dissipated. The bound itself asserts only the impossibility of exceeding this portion; that it can indeed be harvested is shown by explicit processes, such as the feedback protocol of Sec.~\ref{sect:ex_container}, which saturates \eqref{eq:work_bound} with equality.

\paragraph*{Multistage conversion chains.}
In applications, the quantity an agent ultimately cares about is often separated from the correlation resource by a chain of intermediate conversions. The final resource, say an externally supplied free-energy unit, bears no visible relation to any internal correlation, yet reaching it may require steps in which internal correlations are consumed. To cover such situations, we consider a process composed of $K$ successive steps $t_0\to t_1\to\cdots\to t_K$. During step $k$, the composite system evolves by a one-step dynamics of the class introduced in Sec.~\ref{sect:setting}. Assumptions~\ref{asm:local_mech}--\ref{asm:no_unrealized} hold within each step, with a parent structure $\DepX{i}^{(k)}$ that may differ from step to step. The initial distribution of step $k$ is the final distribution of step $k-1$. Composing steps genuinely extends the expressive power of the framework; it is not a bookkeeping convention. A step is the elementary unit within which every mechanism reads only states frozen at the beginning of that step. A protocol in which some mechanism must read a state \emph{after} it has changed, such as an evaluation device reading the output of an earlier operation, cannot be represented as a single step of the class of Sec.~\ref{sect:setting}. The reading must be placed in a later step.

The single-step results compose along such a chain into a bound on its net work balance, the natural measure of the final gain. Superscripts $(k)$ denote quantities evaluated on step $k$; in particular,
$\decIk:=\Sum{i=2}{\Nx}\decIiMu$,
$\accIk:=\Sum{i=2}{\Nx}\accIiMu$, and
$\incIk:=\Sum{i=2}{\Nx}\incIiMu$
are the destroyed correlations \eqref{eq:def_decIij}, their controller-shared parts \eqref{eq:dec_split}, and the created correlations \eqref{eq:def_incIij} of step $k$. Since the labeling of the subsystems is chosen anew for each step (Sec.~\ref{sect:setting}), these step quantities depend on the labeling adopted for that step. The bounds below hold for every admissible choice, and no canonical labeling is singled out. A step is \emph{fully blind} when $\DepX{i}^{(k)}=\emptyset$ for all $i$. At such a step, every destroyed correlation is hidden from its controllers, so $\accIk =0$. Assume additive bare energies,
\begin{IEEEeqnarray}{lCl}
    E(\Xtot) = \Sum{i=1}{\Nx} E_i(\Xs{i}),
    \label{eq:additive_energy}
\end{IEEEeqnarray}
write $\Floc{i} := \langlee E_i\ranglee - \boltz\Temperature\,\shaent{\Xs{i}}$ for the local free energy of subsystem $i$, and let $\Wext$ now denote the net work extracted over the whole chain, $\Delta$ likewise referring to the whole chain.

\begin{proposition}[Chain work bound]
    \label{prop:chain_work}
    Let every step of the chain satisfy Assumptions~\ref{asm:local_mech}--\ref{asm:no_unrealized}, and let the bare energies be additive, \eqref{eq:additive_energy}. Then
    \begin{IEEEeqnarray}{lCl}
        \invTemp\Wext \;\leq\;
        -\invTemp\Sum{i=1}{\Nx}\Delta\Floc{i}
        + \Sum{k=1}{K}\left(\accIk  - \incIk \right).
        \IEEEeqnarraynumspace
        \label{eq:chain_work}
    \end{IEEEeqnarray}
    In particular, since at fully blind steps $\accIk =0$ and $\incIk \geq0$ \footnote{At a fully blind step, $\incIiMu\geq0$ follows from the screening property: splitting the region of $\muI{\Xs{i}\prm}{\Xs{\minusI}\prm}$ by $\Yeng{\Xs{i}}$ and $\Yeng{\Xs{\minusI}}$ gives $\incIiMu = \muI{\Xs{i}\prm}{\Xs{\minusI}\prm\mid\Xs{i}} + \muI{\Xs{i}\prm}{\Xs{\minusI}\prm;\Xs{i}\mid\Xs{\minusI}}$, and the second term vanishes because $\Xs{\minusI}\prm\perp(\Xs{i},\Xs{i}\prm)\mid\Xs{\minusI}$ under blind updates (Lemma~\ref{lem:screening}), leaving a nonnegative conditional mutual information. For general parent structures, every $\incIiMu$ remains nonnegative under Assumption~\ref{asm:no_unrealized} (Lemma~\ref{lem:acc_sum} of Appendix~\ref{ap:multistage}); without that assumption it can be negative.},
    \begin{IEEEeqnarray}{lCl}
        \invTemp\Wext \;\leq\;
        -\invTemp\Sum{i=1}{\Nx}\Delta\Floc{i}
        \quad\text{\emph{(fully blind chains)}},
        \label{eq:blind_work}
    \end{IEEEeqnarray}
    and, since the entropy production of the chain is nonnegative (Lemma~\ref{lem:joint_second_law}; for this, and hence for \eqref{eq:general_work}, Assumptions~\ref{asm:local_mech}--\ref{asm:acyclic} suffice) and $\TotalCorrelation(t_K)\geq0$,
    \begin{IEEEeqnarray}{lCl}
        \invTemp\Wext \;\leq\;
        -\invTemp\Sum{i=1}{\Nx}\Delta\Floc{i}
        + \TotalCorrelation(t_0)
        \quad\text{\emph{(any chain)}}.
        \IEEEeqnarraynumspace
        \label{eq:general_work}
    \end{IEEEeqnarray}
\end{proposition}
\noindent The derivation is elementary. Applied step by step, Theorem~\ref{thm:main} composes into an additive entropy-production ledger for the chain. The first law converts it into \eqref{eq:chain_work} exactly as in the single-step case. It is carried out in Appendix~\ref{ap:multistage}. For $K=1$, Eq.~\eqref{eq:chain_work} reduces to Corollary~\ref{cor:work_bound} under the additivity assumption.

Proposition~\ref{prop:chain_work} delineates what can, and what cannot, be claimed about the role of information for the final proceeds of a chain. First, Eq.~\eqref{eq:blind_work} states that, by either of the two routes available to it, \emph{a fully blind chain can never credit internal correlations to its work balance}. If a blind step destroys a correlation, the destruction raises $-\Delta F$ by exactly the amount that the composed dissipation bound \eqref{eq:composed_ledger} subtracts, and the balance is unchanged. If the correlation is left untouched, it never enters the balance in the first place. The case distinction that plagues informal arguments, ``the correlation is either dissipated or left unused'', is thus absorbed into a single inequality: only the local free energies $\Floc{i}$, whose exploitation requires no knowledge of the joint state, remain as work resources for a fully blind chain.

Second, the pair \eqref{eq:blind_work}--\eqref{eq:general_work} bounds the \emph{value of information} for an entire chain. Fix a \emph{conversion task} in which the final marginal and the final bare energy of every subsystem are prescribed, thereby fixing every $\Delta\Floc{i}$. The length of the chain, the parent structures, and the channels remain at the protocol's disposal. For such a task, the blind bound \eqref{eq:blind_work} is not merely an upper limit but the exact blind optimum. It is approached, in the quasistatic limit, by chains that transform each subsystem separately (Lemma~\ref{lem:blind_attain} of Appendix~\ref{ap:multistage}). Subtracting this optimum from \eqref{eq:general_work} shows that the gap between the optimal informed and the optimal blind work balance is at most $\boltz\Temperature\,\TotalCorrelation(t_0)$, namely $\boltz\Temperature$ times the internal correlation available at the start. This bound holds \emph{no matter how long the chain is and no matter how large the free energy collected at its end is}. Section~\ref{sect:ex_cartridge} computes both optima exactly for one such task and finds their gap equal to $\boltz\Temperature\,\TotalCorrelation(t_0)$: the bound on the value of information is tight.

The saturation of the value of information at $\boltz\Temperature\,\TotalCorrelation(t_0)$ presupposes, however, that the agent may inject external work freely along the way and repay it from later proceeds. If the work that can be injected \emph{before} the final resource is unlocked is limited, as it is for any autonomous device that must start on a small battery, the situation changes qualitatively. A correlation whose own thermodynamic value is only $\boltz\Temperature\,\TotalCorrelation$ then decides whether a final gain of arbitrary size is collected at all. Information thus becomes indispensable for the final proceeds \emph{conditionally on the budget}. Section~\ref{sect:ex_cartridge} exhibits this mechanism with explicit constants. A general theory of budget-constrained chains, in particular the optimal trade-off between measurement, memory, and gate traversal, is left for future work.\section{Examples}
\label{sect:ex}

We apply the results in three examples that proceed from attribution to scale and then to indispensability. The minimal bit model of Sec.~\ref{sect:ex_bit} isolates the mechanism-sensitive verdicts that no distribution-level account can deliver. The disposable container of Sec.~\ref{sect:ex_container} reproduces the informed--blind contrast at macroscopic scale, where a single-bit key decides the fate of an extensive correlation reservoir. The correlation-keyed cartridge of Sec.~\ref{sect:ex_cartridge} places that contrast under a per-run work budget, making a microscopic correlation indispensable for collecting a macroscopic resource unrelated to it.

\subsection{Example 1: Mechanism-Sensitive Attribution in a Minimal Bit Model}
\label{sect:ex_bit}

Our first example uses the smallest setting that can display what the mechanism-sensitive bound \eqref{eq:main_no_new_obs} of Theorem~\ref{thm:main} predicts but distribution-level accounting cannot. The model comprises a family of four one-step processes on three or four bits whose members differ only in their reading pattern, namely, which mechanism reads which initial variable. Three of the four variants share the entire joint distribution $P(\Xtot,\Xtotprm)$ bit for bit. Consequently, every functional of the statistics, including endpoint marginals, local free energies, and the net correlation change $\delitot$, takes identical values across them. The right-hand side of \eqref{eq:main_no_new_obs} does not: it vanishes for one variant and equals $\ln 2$ for the other two. An explicit protocol attains its value in each case, exactly for the keyed bijections and quasistatically for the blind resets. The bound follows the reading pattern rather than the distribution. This is the operational content of defining the informational parents through the kernel rather than through the joint statistics (Sec.~\ref{sect:setting}). The fourth variant exposes a bookkeeping gap in net accounting, whereby correlation created in one place is silently credited against correlation blindly destroyed in another. It turns this gap into a quantitative, attainable separation. Table~\ref{tbl:ex_bit} collects the four variants and their verdicts. A fifth reading pattern on the same stage, an informed reset, is introduced at the end of this subsection; it extends the separation to the quantities generated by accounts that read mechanisms rather than statistics.

\begin{table*}[t]
    \caption{\label{tbl:ex_bit}%
        The four variants of the minimal bit model of Sec.~\ref{sect:ex_bit}. Information quantities are in nats; $\Dissipation$ is the right-hand side of the main inequality \eqref{eq:main_no_new_obs}, and ``min $\DeltaStot$'' is its exact infimum over implementations, attained by the keyed bijections and approached in the quasistatic limit by the blind resets of the protocols described in the text. The last three columns compare the work bound \eqref{eq:work_bound} with the conventional net-accounting bound $\Wext\leq-\Delta F$ and with the attained value. Variants (a), (b), and (b$'$) share the joint distribution of $(\Xtot,\Xtotprm)$ bit for bit, so every distribution-level account assigns them identical verdicts; only the kernels---and hence the parents---differ. In variant (c) the net change of the total correlation vanishes, yet running the step costs $\boltz\Temperature\ln 2$ of work.
    }
    \resizebox{\textwidth}{!}{%
    \begin{tabular}{llcccccc}
        \toprule
        Variant & Reading pattern & $\delitot$ & $\Dissipation$ & min $\DeltaStot$ & Eq.~\eqref{eq:work_bound} & $\Wext\leq-\Delta F$ & attained $\Wext$ \\
        \colrule
        (a) informed destroyer & $\DepX{2}=\{\Xs{3}\}$, others $\emptyset$ & $-\ln 2$ & $0$ & $0$ & $0$ & $0$ & $0$ \\
        (b) bystander knowledge & all $\DepX{i}=\emptyset$ & $-\ln 2$ & $\ln 2$ & $\ln 2$ & $-\boltz\Temperature\ln 2$ & $0$ & $-\boltz\Temperature\ln 2$ \\
        (b$'$) audited destruction & as (b), plus $\DepX{1}=\{\Xs{2},\Xs{3}\}$ & $-\ln 2$ & $\ln 2$ & $\ln 2$ & $-\boltz\Temperature\ln 2$ & $0$ & $-\boltz\Temperature\ln 2$ \\
        (c) creation vs.\ payment & $\DepX{1}=\{\Xs{4}\}$, others $\emptyset$ & $0$ & $\ln 2$ & $\ln 2$ & $-\boltz\Temperature\ln 2$ & $0$ & $-\boltz\Temperature\ln 2$ \\
        \botrule
    \end{tabular}
    }
\end{table*}

\paragraph*{Common stage.}
Variants (a), (b), and (b$'$) act on three bits with vanishing bare energies and are prepared in the maximally correlated state $\Xs{1}=\Xs{2}=\Xs{3}=Z$ with $Z$ uniform. Every pair of subsystems carries mutual information $\ln 2$. In all three variants, subsystem $2$ ends the step in the reset state $\Xs{2}\prm=0$, while subsystems $1$ and $3$ end in their initial values. Subsystem $2$ therefore decouples from the rest, whereas the correlation between subsystems $1$ and $3$ survives. The joint distribution of $(\Xtot,\Xtotprm)$ is the same in all three variants, as is every quantity computed from it. In particular, $\Delta\Floc=\boltz\Temperature\ln 2$ because the marginal entropy of subsystem $2$ drops by $\ln 2$, and $\delitot=-\ln 2$. Hence $\Delta F=\Delta\Floc+\boltz\Temperature\,\delitot=0$, and the conventional net-accounting bound, given by the free-energy decomposition displayed in the Introduction, reads $\Wext\leq-\Delta F=0$. The transition kernel distinguishes the variants. As stipulated in Sec.~\ref{sect:setting}, it is defined on the whole state space rather than on the support of the initial distribution and thereby determines the informational parents.

\paragraph*{(a) Informed destroyer.}
Subsystem $2$ is updated by the keyed bijection $\Xs{2}\prm=\Xs{2}\oplus\Xs{3}$ with $\DepX{2}=\{\Xs{3}\}$. Subsystems $1$ and $3$ are left untouched. The declared parent is minimal because the output varies with $\Xs{3}$ at either value of $\Xs{2}$. On the initial data $\Xs{2}=\Xs{3}$, the exclusive-or gives $\Xs{2}\prm=0$ deterministically. Here the one mechanism that moves reads the one variable carrying what its move destroys. The $\ln 2$ of correlation lost between subsystem $2$ and the preceding block is destroyed by the move of subsystem $2$ alone. Its rescue region $\Yeng{\infPa{2}}=\Yeng{\Xs{3}}$ contains the loss entirely, while the correlation between subsystem $3$ and its preceding block survives the step. Every summand of \eqref{eq:main_no_new_obs} vanishes, so $\Dissipation=0$. The value is attained: the local entropy production \eqref{eq:yuragi_k} vanishes for every subsystem (see the implementation paragraph below), giving $\DeltaStot=0$ and $\Wext=0$. This result saturates the conventional second law and both work bounds simultaneously.

\paragraph*{(b) Bystander knowledge.}
Keep the same initial state and endpoint, but reset subsystem $2$ blindly: $\Xs{2}\prm=0$ with $\DepX{2}=\emptyset$. Every other subsystem is again untouched. Off the support of the initial distribution, the two kernels differ. The exclusive-or of (a) maps $(\Xs{2},\Xs{3})=(0,1)$ to $\Xs{2}\prm=1$, whereas the blind reset maps it to $\Xs{2}\prm=0$. On the initial data, however, they generate identical trajectories: the joint distribution of $(\Xtot,\Xtotprm)$ in (b) coincides, bit for bit, with that of (a). Subsystem $1$ still holds everything that is being destroyed, $\muI{\Xs{1}}{\Xs{2}}=\ln 2$; the mechanism that acts simply does not read it. The verdict of \eqref{eq:main_no_new_obs} flips. The loss is again caused by the move of subsystem $2$, but its rescue region $\Yeng{\infPa{2}}$ is now empty, so $\Dissipation=\ln 2$. Performing the reset as a quasistatic isothermal compression of the marginal of subsystem $2$ attains this value, giving $\DeltaStot=\ln 2$ and $\Wext=-\boltz\Temperature\ln 2$. This is equality in \eqref{eq:work_bound}, while the net bound still reads $\Wext\leq0$ and misses the mandatory expenditure by $\boltz\Temperature\ln 2$. The pair (a)--(b) presents the locality statement below \eqref{eq:def_destroyed_region} in its sharpest form. A single reading line decides between zero dissipation and $\ln 2$, yet nothing in the statistics of the two processes distinguishes them. The number $\ln 2$ in (b), taken by itself, is not new: it is the modularity cost of running mutually blind mechanisms \cite{Boyd_2018,wolpert2019stochastic}. What the mechanism-sensitive bound adds is the attribution around that number. The charge is removed by one reading line in (a), left in place by knowledge that is present but unread in this variant, left in place by knowledge read on the wrong side in (b$'$), and hidden behind a vanishing net change in (c).

\paragraph*{(b$'$) Audited destruction.}
To the blind reset of (b), add an auditor. Subsystem $1$ applies the parity check $\Xs{1}\prm=\Xs{1}\oplus\Xs{2}\oplus\Xs{3}$ with $\DepX{1}=\{\Xs{2},\Xs{3}\}$, reading, at the moment of destruction, both variables whose correlation is being destroyed. The parents are minimal. At either value of $\Xs{1}$, the output varies with $\Xs{2}$ at fixed $\Xs{3}$ and with $\Xs{3}$ at fixed $\Xs{2}$. The reading pattern also respects the parents-later convention \eqref{eq:parents_later}. On the initial data, the parity $\Xs{2}\oplus\Xs{3}$ vanishes identically, so $\Xs{1}\prm=\Xs{1}$ almost surely. The joint distribution of $(\Xtot,\Xtotprm)$, not merely the endpoint, is again that of (b). This is the mirror image of the lesson drawn from the mutual-overwriting process of Appendix~\ref{ap:cex_xor}. There, an output was marginally \emph{independent} of a variable its kernel genuinely reads; here, an update is pathwise \emph{trivial} on the initial data while its kernel genuinely reads two variables. In both directions, the kernel rather than the realized statistics fixes the parents.

The verdict is unchanged, $\Dissipation=\ln 2$, and is attained exactly as in (b). The parity check is a bijection given $(\Xs{2},\Xs{3})$ and therefore runs at zero entropy cost. The reason for the unchanged verdict is worth displaying. The auditor's reading does register on the books. The blind spot of subsystem $1$ shrinks to measure $\shaent{\Xs{1}\mid\Xs{2},\Xs{3}}=0$, so the effective parent region \eqref{eq:103YY} of the block preceding subsystem $2$ grows to the entire block region, as large as it can be. But this region is the rescue region for losses caused by moves \emph{inside the block}. The $\ln 2$ at stake, by contrast, is destroyed by the move of subsystem $2$ itself, whose rescue region is $\Yeng{\infPa{2}}=\emptyset$ as in (b). The auditor watches the destruction without being able to avert it. The parity check is free of charge, and it rescues nothing. Knowledge is credited only where it sits on the causal line of the destruction. The rescue regions of \eqref{eq:main_no_new_obs} are attached per destroying side rather than pooled across the step.

\paragraph*{(c) Creation does not pay for destruction.}
The last variant uses four bits to expose the bookkeeping gap noted above. Net accounting silently credits correlation created in one place against correlation blindly destroyed in another. Prepare the correlated pair $\Xs{2}=\Xs{3}=Z$ as before, an independent uniform source bit $\Xs{4}$, and a register $\Xs{1}=0$, and let one step perform
\begin{IEEEeqnarray}{c}
    \Xs{1}\prm=\Xs{1}\oplus\Xs{4},\qquad
    \Xs{2}\prm=0,\qquad
    \Xs{3}\prm=\Xs{3},\qquad
    \Xs{4}\prm=\Xs{4},
    \label{eq:ex_bit_c}
\end{IEEEeqnarray}
with $\DepX{1}=\{\Xs{4}\}$ and every other parent set empty. The labeling respects the parents-later convention \eqref{eq:parents_later}, and the minimality of $\DepX{1}$ is checked as in (a). Within the step, the blind reset of subsystem $2$ destroys the $\ln 2$ of correlation of the pair $(\Xs{2},\Xs{3})$. Meanwhile, the keyed bijection at subsystem $1$ copies the source into the register and creates $\ln 2$ of fresh correlation in the pair $(\Xs{1},\Xs{4})$. The two entries cancel in every distribution-level account, giving $\delitot=0$. The local entropies balance as well ($\shaent{\Xs{1}}$ rises by $\ln 2$, $\shaent{\Xs{2}}$ falls by $\ln 2$), so $\Delta\Floc=0$ and $\Delta F=0$. The net bound therefore reads $\Wext\leq0$: at the level of distributions, nothing happened. The ledger records the same step differently. In the block decomposition \eqref{eq:decIiMu_decomp}, the loss appears at subsystem $3$. The correlation destroyed between subsystem $3$ and its preceding block is caused by a move inside the block, namely, the blind reset of subsystem $2$. It lies entirely inside that mover's blind spot and hence outside the effective parent region, so $\Dissipation=\ln 2$. The created correlation cannot offset this charge. It enters the work bound separately and with a negative sign, $\Sum{i=2}{\Nx}\accIiMu=0$ and $\Sum{i=2}{\Nx}\incIiMu=\ln 2$, so Corollary~\ref{cor:work_bound} gives $\Wext\leq-\boltz\Temperature\ln 2$. The step cannot be run without a net expenditure of $\boltz\Temperature\ln 2$ of work. This expenditure is realized: the implementation below attains $\DeltaStot=\ln 2$ and $\Wext=-\boltz\Temperature\ln 2$. Creation does not pay for destruction. The cancellation happens in the distribution, not in the ledger. Variant (b) also provides a useful contrast. There, the net account at least registers that correlation has been lost, $\delitot=-\ln 2$, and merely misprices the loss. In (c), the net account is silent, reporting an unchanged correlation stock for a step that irretrievably dissipates $\boltz\Temperature\ln 2$.

The map \eqref{eq:ex_bit_c} should not be confused with the four-bit process of the footnote below Assumption~\ref{asm:no_unrealized}. That process also copies and erases within one step, and it breaks the assumption. The difference lies in the fate of the created correlation. In the footnote's process, the copy is directed at a partner whose own move erases the correlation within the same step, so the correlation is unrealized. In \eqref{eq:ex_bit_c}, the partner $\Xs{4}$ stands still, the copied correlation survives into the final state, and every term of the form \eqref{eq:asm_no_unrealized} vanishes. Variant (c) thus also marks, from inside, where the boundary of Assumption~\ref{asm:no_unrealized} runs. The assumption forbids not simultaneous creation and destruction, but creation aimed at a target that is itself being overwritten.

\paragraph*{Implementations and tightness.}
All announced values are attained within the class of Sec.~\ref{sect:setting}: exactly by the keyed bijections and, in the quasistatic limit, by the blind resets. Every channel above belongs to one of three elementary types. A \emph{static} subsystem is left untouched. A \emph{keyed bijection} (the exclusive-or updates of (a), (b$'$), and (c)) is, given the initial state of its parents, a bijection of the subsystem's state space. As in Appendix~\ref{ap:cex_xor}, it can be implemented isothermally and reversibly, with $\invTemp\Qs{i}=0$, and it satisfies the local detailed balance condition \eqref{eq:local_detail_subsystem} with the inverse map as the reversed channel. A \emph{blind reset} is a quasistatic isothermal compression of the marginal of subsystem $2$ from the uniform bit to the state $0$, with work input $\boltz\Temperature\ln 2$ and released heat $\invTemp\Qs{2}=\ln 2$. For each type, the local entropy production \eqref{eq:yuragi_k} vanishes. For a keyed bijection, the channel is deterministic and invertible given the parents, so $\shaent{\Xs{i}\prm\mid\DepX{i}}=\shaent{\Xs{i}\mid\DepX{i}}$ and $\Qs{i}=0$. For the quasistatic reset, the released heat exactly compensates the entropy drop. Hence Theorem~\ref{thm:ledger} holds with equality in every variant. Since $\Ucost=0$ (Lemma~\ref{lem:U_vanish}) and $\Rreb=0$ (Corollary~\ref{cor:no_rebate}), the total entropy production equals the right-hand side of \eqref{eq:main_no_new_obs} exactly. The assumptions hold by construction. Each kernel factorizes into the listed channels (Assumption~\ref{asm:local_mech}), every reading pattern above is acyclic (Assumption~\ref{asm:acyclic}), and no variant generates unrealized correlation (Assumption~\ref{asm:no_unrealized}). In (a) and (b), only one subsystem moves and everything it reads stands still. In (b$'$), the only reading mover is the auditor, whose output coincides almost surely with its initial value. In (c), the created correlation survives, as discussed above. Within essentially one initial state, the family thus spans the full range of the bound: covered destruction at zero cost, blind destruction at full cost, knowledge wasted at the wrong location, and a mandatory expenditure invisible to net accounting.

\paragraph*{Beyond shared statistics: the record under a fixed declaration.}
The variants above separate the ledger from every account whose input is the realized distribution. Sequential and modular accounts \cite{ito2013information,Wolpert_2020,WolpertMinEP2020,Boyd_2018,wolpert2019stochastic} are not of this kind: they take declared reading patterns as input, and, supplied with the minimal parents, their sequential realizations recover the verdicts of Table~\ref{tbl:ex_bit} for all four variants. The question that matters is therefore one level up. Can the attribution itself (which state change destroyed the correlation, and what the destroying mechanism read) be recovered from the stage balances and fixed-bipartition values those accounts generate once a declaration of the reading patterns is fixed? The following pair shows that it cannot.

Alongside the exclusive-or of (a), consider one more reading pattern on the same stage: subsystem $2$ is reset to the parity of its two static neighbours,
\begin{IEEEeqnarray*}{lCl}
    \Xs{2}\prm \;=\; \Xs{1}\oplus\Xs{3},
    \qquad
    \DepX{2}=\{\Xs{1},\Xs{3}\},
\end{IEEEeqnarray*}
an \emph{informed reset}, denoted (a$''$): it reads the variables carrying the destroyed correlation, but not the bit it overwrites. On the initial data $\Xs{1}=\Xs{2}=\Xs{3}$ its trajectories coincide bit for bit with those of the blind reset (b). Being a conditional reset, it also admits the uniform-sweep implementation of (b), a quasistatic isothermal compression from the flat potential to the target state, here the parity of the two neighbours. The sweep is deliberately mismatched to the deterministic conditional distribution and hence dissipative, releasing $\invTemp\Qs{2}=\ln 2$ on every trajectory. The two processes can therefore be run with identical trajectory-heat tables. A keyed bijection can never do this \footnote{With the reversed channel normalized and supported on the reversals of the forward transitions (Sec.~\ref{sect:setting}), local detailed balance gives $\sum_{x}P(x'\mid x,\mathrm{pa})\,e^{-\invTemp q(x\to x')}=1$ for every output $x'$ reachable under the given parents, where $\invTemp q$ is the trajectory heat. For a bijection given the parents, each output is reached from exactly one input, so the trajectory heat vanishes on every transition. The heat table of variant (a) therefore never matches that of a reset: calorimetry alone separates (a) from (b). The pair compared here is chosen so that even calorimetry agrees.}. Now declare the reading pattern $\mathrm{pa}_2=\{\Xs{1},\Xs{3}\}$ and $\mathrm{pa}_i=\emptyset$ otherwise. The pattern is acyclic; relabeling the subsystems in reverse topological order puts it in the parents-later convention \eqref{eq:parents_later}, and the verdicts below are unchanged under this relabeling. The declaration is valid for both processes, since declared parents may exceed minimal ones. Under it, the two processes share the realized statistics, the trajectory heats, every stage quantity of every parents-later sequential realization, and the value that the fixed-block modularity difference takes (as algebra, irrespective of whether the block is admissible for the modular account) on every bipartition (Lemma~\ref{lem:declaration_measurability} of Appendix~\ref{ap:declaration_measurability}). What does separate them within those accounts is not one of the values compared here but a structural judgement: the blind reset admits three nontrivial modular blocks, the informed reset none. Judging admissibility at the level of kernels is precisely the input whose sound construction is the minimal-parents definition of Sec.~\ref{sect:setting}; read at the level of the shared statistics, the judgement must return the same answer for both. The ledger, by contrast, separates them as a value: the exact infimum of $\DeltaStot$ over implementations of the two kernels is $\ln 2$ for the blind reset and $0$ for the informed one, each approached within the ideal-limit convention of Sec.~\ref{sect:setting} (Table~\ref{tbl:ex_bit_declared}).

That an eraser reading correlated references can run at zero cost is not new: it is the classical saving of information-assisted erasure \cite{Bennett_1982,sagawa2008second,Parrondo_2015}, and the same separation for a pair of perfectly correlated bits appears as the solitary-process erasure example (Example~9) of Ref.~\cite{wolpert2019stochastic}; here the saving equals the mutual information $\muI{\Xs{2}}{\Xs{1},\Xs{3}}=\ln 2$ between the erased bit and its references. The pair shows that the stage balances and fixed-bipartition values of mechanism-level accounts are blind to this saving under a shared valid declaration. Whether the destroyed correlation was read by the mechanism that destroyed it is not a function of what happened (the realized statistics and the heats) but of what the kernels would have done on initial states that did not occur. The distinction carries operational weight. The informed kernel admits an implementation that runs $\ln 2$ cheaper without changing a single realized probability; the blind kernel does not.

These exact coincidences, like the shared statistics of the variants themselves, hold on the maximally correlated ensemble; on an ensemble of full support the statistics determine the kernel pointwise, and no such pair exists. This is not a knife edge. The direction of the degeneracy deserves emphasis. Mix the initial ensemble with the uniform distribution at weight $\epsilon$, so that all eight initial states occur, and run the same two kernels. The per-stage entropy-production balances and information drops of both parents-later schedules remain exactly equal at every $\epsilon$. The fixed-block modularity values drift apart only at order $\epsilon\ln(1/\epsilon)$, with the largest drift at $\epsilon=10^{-5}$ being $7\times10^{-5}$. The minimal costs differ by $\muI{\Xs{2}}{\Xs{1},\Xs{3}}$, which tends to $\ln 2$ and equals $0.6931$ there. Assumptions~\ref{asm:local_mech}--\ref{asm:no_unrealized} hold, and both bounds remain saturated in the convention of Sec.~\ref{sect:setting}, at every $\epsilon$ \footnote{The same degenerate stage admits a companion pair aimed at the jointly destroyed component $\atamaOmegaIMuMu$: a triple blind reset against a reset of subsystem $1$ to the parity of the other two bits, with bounds $2\ln 2$ against $\ln 2$ and matching stage balances and block values. Unlike the pair of Table~\ref{tbl:ex_bit_declared}, that pair does not survive the mixing: off the maximally correlated ensemble its informed member violates Assumption~\ref{asm:no_unrealized}, and only the general ledger of Theorem~\ref{thm:ledger} governs it there.}. The degenerate ensemble is thus the point where the residual drift closes, not a pathology that produces the separation.

\begin{table*}[t]
    \caption{\label{tbl:ex_bit_declared}%
        The second separation of Sec.~\ref{sect:ex_bit}: blind against informed reset under the common valid declaration $\mathrm{pa}_2=\{\Xs{1},\Xs{3}\}$. Both processes start from the maximally correlated three-bit state and, in the uniform-sweep implementations, share the joint distribution of $(\Xtot,\Xtotprm)$ and the heat $\ln 2$ released on every trajectory; by Lemma~\ref{lem:declaration_measurability}, every stage quantity of every parents-later sequential realization and the value of the fixed-block modularity difference on every bipartition then coincide between them, and both matched runs produce $\DeltaStot=\ln 2$. The kernels and their minimal parents do not coincide, and neither do the last two columns: $\Dissipation$ is the ledger bound of Theorem~\ref{thm:main} for this ensemble and each kernel's minimal parents, and min $\DeltaStot$ the corresponding infimum over implementations, approached in the convention of Sec.~\ref{sect:setting}.}
    \begin{ruledtabular}
    \begin{tabular}{llccc}
        Process & Kernel of subsystem 2 & $\DepX{2}$ & $\Dissipation$ & min $\DeltaStot$ \\
        \colrule
        (b) blind reset & $\Xs{2}\prm=0$ & $\emptyset$ & $\ln 2$ & $\ln 2$ \\
        (a$''$) informed reset & $\Xs{2}\prm=\Xs{1}\oplus\Xs{3}$ & $\{\Xs{1},\Xs{3}\}$ & $0$ & $0$ \\
    \end{tabular}
    \end{ruledtabular}
\end{table*}

\subsection{Example 2: Work Extraction from a Disposable Container}
\label{sect:ex_container}

\begin{figure}[ht]
    \centering
    \subfloat[Initial State]{
        \includegraphics[width=0.28\textwidth]{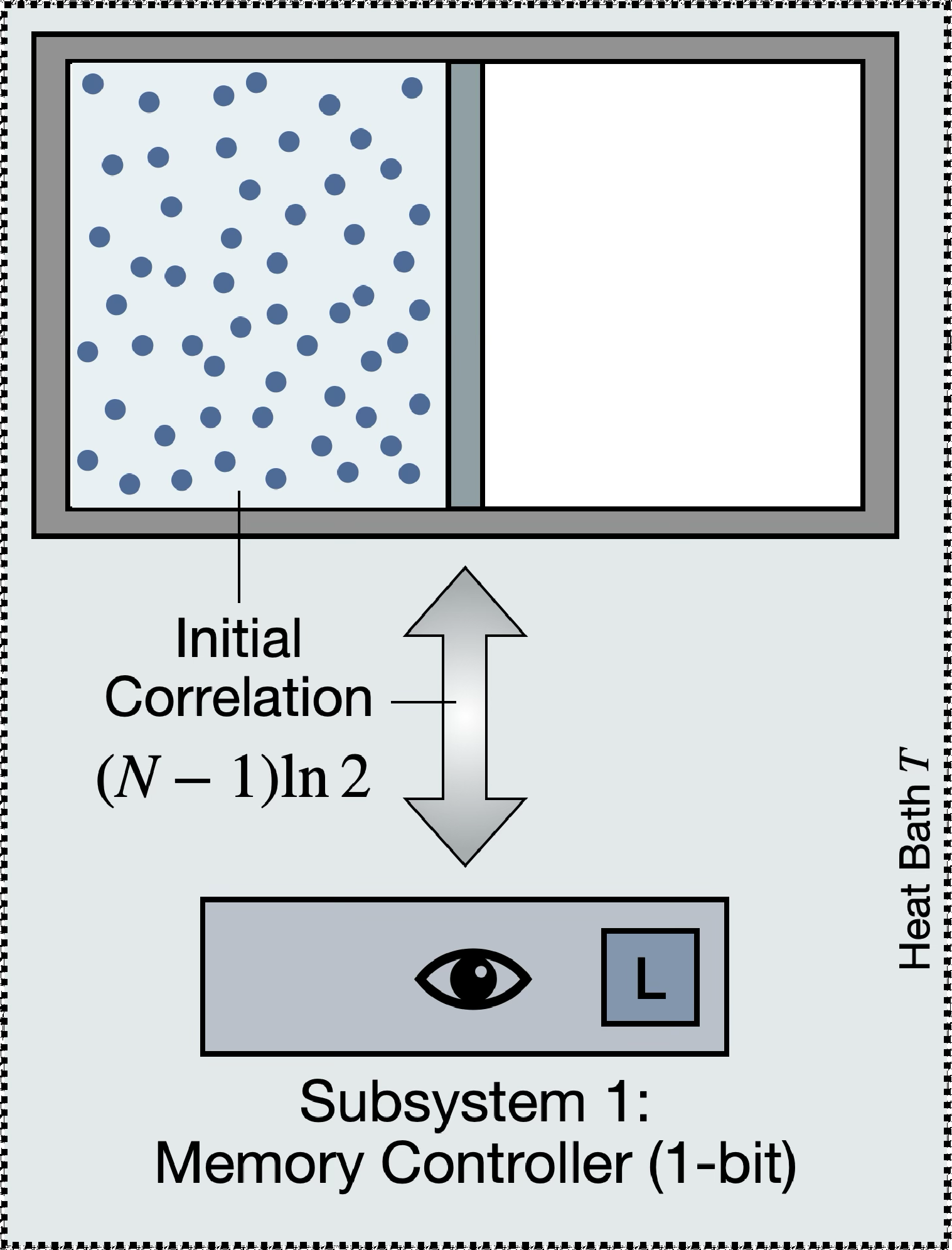}
    }
    \subfloat[Feedback Control Process]{
        \includegraphics[width=0.28\textwidth]{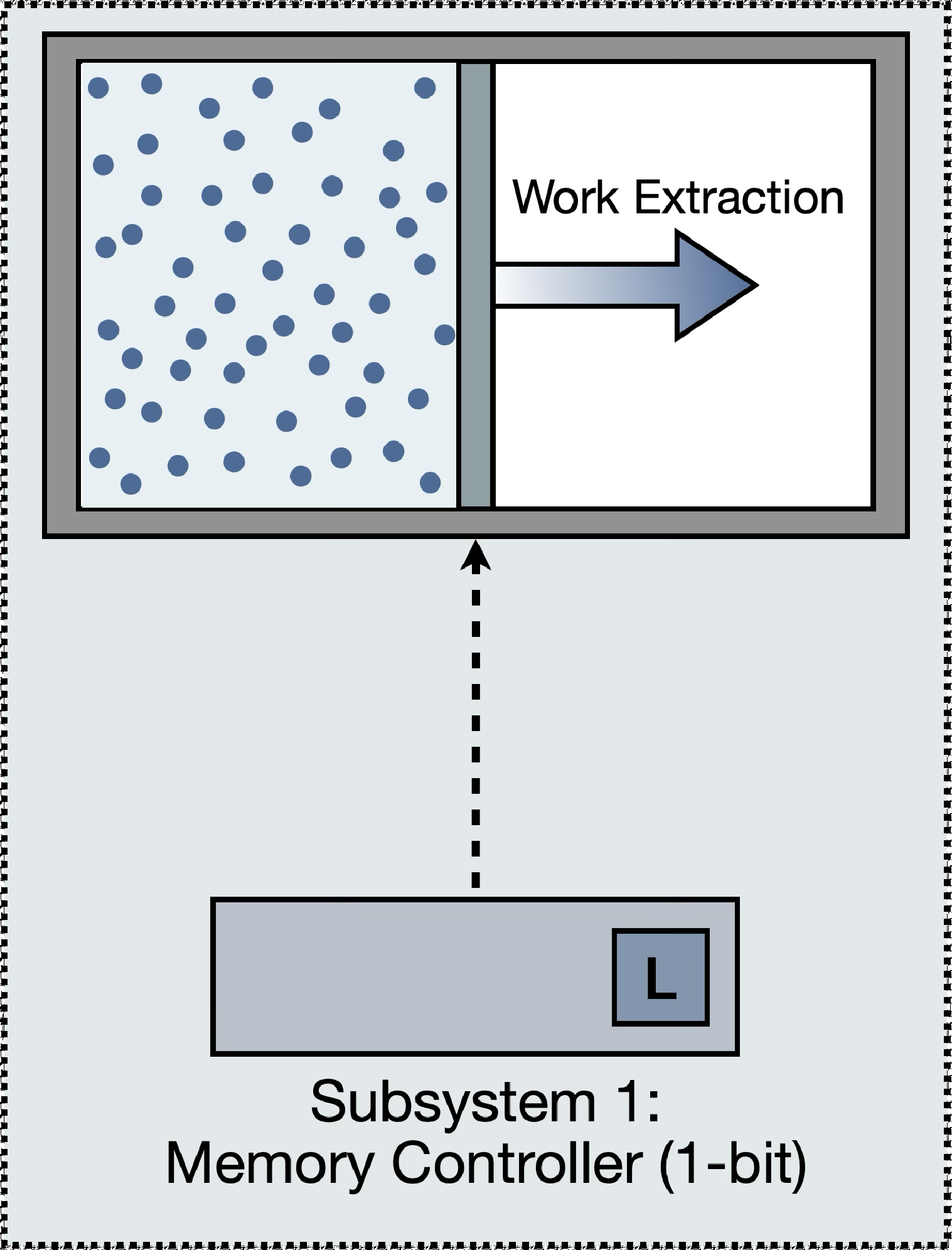}
    }
    \subfloat[Final State]{
        \includegraphics[width=0.28\textwidth]{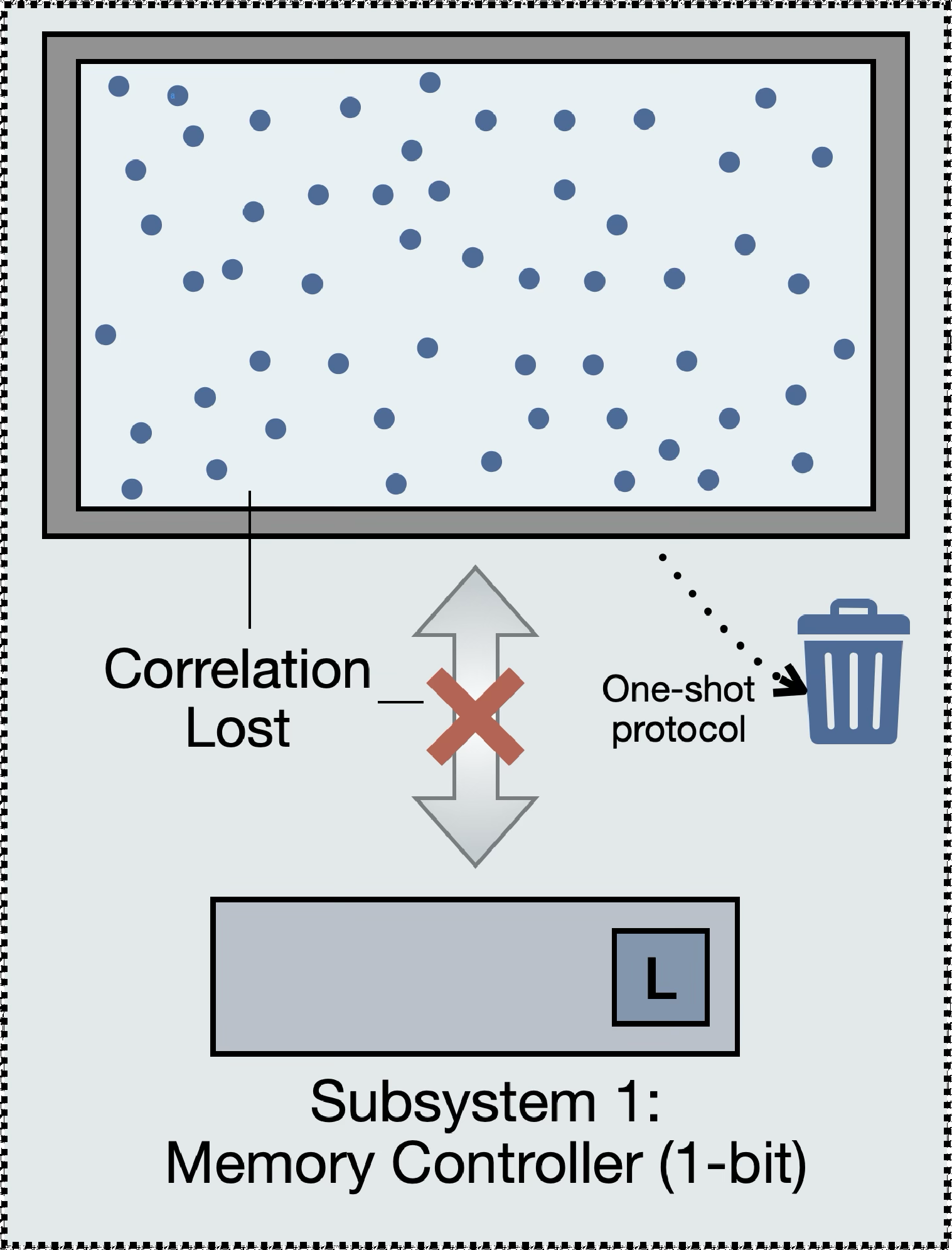}
    }
    \caption{
        Work extraction from internal correlations through feedback control.
        (a) Initial State: A composite system consisting of a 1-bit memory controller (Subsystem 1) and $N-1$ gas molecules. The molecules are initially confined to one of the two partitions with a probability of 0.5, and their collective position has already been observed by the controller. This configuration establishes a macroscopic amount of internal correlation $\TotalCorrelation = (N-1)\ln 2$ within the system.
        (b) Feedback Control Process: The agent performs feedback control by manipulating the barrier position based on the state of the 1-bit memory.
        During this process, the gas undergoes isothermal expansion at temperature $T$, converting heat from the bath into physical work.
        (c) Final State: The pre-existing internal correlations are entirely consumed (``Correlation Lost'') and converted into physical work. Unlike traditional cyclic heat engines, this model treats the container as a disposable resource, focusing on the one-off conversion of informational resources into work without a memory reset operation.
    }
    \label{fig:ex}
\end{figure}

We demonstrate the physical implications of the derived bound regarding informational indispensability by considering a composite system consisting of $N$ subsystems, as illustrated in Fig.~\ref{fig:ex}.
Subsystem 1 is a controller with the 1-bit memory.
The remaining subsystems, designated as $2:N$, comprise $N-1$ ideal gas particles contained within a vessel.
The entire system is in contact with a heat bath at temperature $T$.

Initially, a barrier divides the container into two equal sections. The molecules are localized in one of these sections with a probability of 0.5. This configuration induces mutual information between any two molecules, $I(X_j; X_k) = \ln 2$ for $j, k \ge 2$. Under the accounting convention made explicit at the end of this subsection, the agent's ideal memory charge is only $\ln 2$, whereas the total correlation of the system reaches a macroscopic scale, $\TotalCorrelation = (N-1) \ln 2$. This informational resource is proportional to the system size.
Analyzing how this resource is consumed requires defining the structural interdependencies among the subsystems. We posit that the agent's memory state remains constant during the barrier operation, while the dynamics of the gas particles are influenced solely by the agent's control through barrier movements. Consequently, when the barrier is conditioned on the memory in the feedback scenario below, the dependencies between systems are:
\begin{IEEEeqnarray}{lCr}
\DepX{k}=
\begin{cases}
\emptyset & (k=1)\\
\{\Xs{1}\} & (k=2,3,\cdots,\Nx).
\end{cases}
\label{eq:DXXFB105}
\end{IEEEeqnarray}
Here, $X_1$ represents the controller's memory.
We analyze the process in which the pre-existing correlations are consumed as the molecules expand and decouple throughout the container.
As stipulated in Sec.~\ref{sect:setting}, the kernels are specified on the whole state space. Under feedback, the barrier program conditions on the memory. A molecule caught on the wrong side of the moving barrier is compressed rather than released, so each molecule's channel genuinely differs between the two memory values and the declared parent sets \eqref{eq:DXXFB105} are minimal. Assumption~\ref{asm:no_unrealized} holds as well. The moving molecules build no correlation toward any partner because each update decouples its subsystem, and the memory they read is static. Since no new correlations arise via time evolution, we can analyze this case with inequality \eqref{eq:main_no_new_obs}.

While the setup is inspired by the Szilard engine, our model fundamentally departs from the traditional cyclic framework by treating the container as a disposable resource.
In this one-off operation, the container is presented as a resource to the controller with the gas molecules already confined to one of its two partitions, even though the controller may initially lack knowledge of their specific location.
After completing the work extraction, the agent disposes of the container without performing a reset operation to restore it to its initial state.
This model lets us evaluate the work extracted specifically from the consumption of pre-existing internal correlations.

We evaluate the specific role of information in this setup by contrasting two operational scenarios: feedback control ($\FB$) and non-feedback control (\OpenLoopWithoutMeasure).
In the $\FB$ scenario, the agent can infer the collective initial positions of all molecules by observing the state of just a single particle and can thereby leverage the pre-existing internal correlations.
This implies that the agent's memory is privy to the initial confinement of the molecules, such that $\muI{\Xs{1}}{\Xj}=\ln2$ for all $1<j\leq\Nx$.
By contrast, in the \OpenLoopWithoutMeasure scenario, the initial joint state, including the correlated memory, is the same. The memory is not read by any moving mechanism. The information is present in the system yet unused, and the barrier program runs without conditioning on the molecules' location.

In the $\FB$ scenario, the verdict is that of variant (a) of Sec.~\ref{sect:ex_bit}, scaled from one shared bit to $N-1$ of them. Each mechanism that moves reads, through the memory, exactly what its move destroys. To apply \eqref{eq:main_no_new_obs}, we relabel so that the parents-later convention \eqref{eq:parents_later} holds. This places the frozen memory last, while the display \eqref{eq:DXXFB105} keeps the memory-first numbering for readability. Every destroyed correlation then lies inside its movers' rescue regions. The memory holds a copy of each molecule's initial state throughout the step and is read by every mechanism that moves. Thus, no destroying move is blind to any part of the correlation it destroys: neither a molecule's $\ln 2$ with the rest of the system, nor the correlation the moving molecules destroy with the memory. Every summand of \eqref{eq:main_no_new_obs} vanishes, and entropy production can reach zero in the quasistatic limit ($\EntropyProduction = 0$).
Under these conditions, the agent extracts a gross work of $\boltz \Temperature(N-1) \ln 2$ from the expansion of the molecules.
This saturates the work bound \eqref{eq:work_bound}: the marginal distribution of each molecule is unchanged, so $\Delta\Floc = 0$, while the destroyed correlation is fully shared with the controller, $\sum_{i=2}^{\Nx}\accIiMu = (N-1)\ln 2$.
The gross extraction receives the correlated memory as part of the initial state. As an accounting convention, we additionally charge the preparation of the 1-bit memory at the Landauer cost $\boltz\Temperature\ln2$ (a measurement-and-record step from a blank memory, performed before the step analyzed here). The process then yields a net gain of $\boltz\Temperature(\Nx-2) \ln 2$, which scales with the system size $\Nx$ while the agent's operational cost remains that of a single bit.

In the \OpenLoopWithoutMeasure scenario, the barrier program does not condition on the memory, so $\infPaI = \emptyset$ for all $i$, and the verdict flips to that of variant (b) of Sec.~\ref{sect:ex_bit} at scale. The correlations among the particles change exactly as in the $\FB$ scenario. Every destroyed correlation, however, is now hidden from the mechanisms that destroy it. The right-hand side of \eqref{eq:main_no_new_obs} equals the full destroyed correlation, the entropy production remains at $\EntropyProduction = (N-1) \ln 2$ even in the quasistatic limit, and the $\boltz \Temperature(N-1) \ln 2$ of free energy released by the vanishing internal correlations is entirely dissipated. Equivalently, $\accIiMu = 0$ in the work bound \eqref{eq:work_bound}. The bound then reduces to $\Wext \leq -\Delta\Floc = 0$. Although the uninformed agent bears no operational cost at all, its ceiling is zero; the macroscopic reservoir is worthless to an agent that cannot read it.

These results realize the scaling announced in the Introduction. The container carries a macroscopic reservoir of information-based free energy, $\boltz\Temperature(N-1)\ln 2$, yet it is unexploitable unless the agent can leverage the correlations through observation. The observation needed is a single bit. Two remarks delimit what the example does and does not show. First, its two scenarios are the calibration poles of the bound. The $\FB$ pattern \eqref{eq:DXXFB105} is fixed-memory feedback of the standard information-engine kind \cite{sagawa2008second,Parrondo_2015}. The memory is prepared first and only read thereafter. This pattern realizes the covering regime discussed below \eqref{eq:work_bound}, so its verdict is equally delivered by the conventional bound $\Wext\leq-\Delta F$. The \OpenLoopWithoutMeasure verdict, that free energy released without extraction is dissipated, is elementary. The role of this example is calibration and scale. The verdicts that separate the mechanism-sensitive ledger from every distribution-level account are those of Sec.~\ref{sect:ex_bit}. Second, the stake here is still the correlation itself. The extracted work is fed, nat for nat, by the correlation destroyed. The final example removes this feature. It makes a correlation the indispensable key to a resource that is not informational at all.\subsection{Example 3: Correlation-Keyed Cartridge}
\label{sect:ex_cartridge}

We construct a minimal, self-contained example in which the final resource is manifestly unrelated to internal correlations: a locked free-energy cartridge. The example attains the chain work bounds of Proposition~\ref{prop:chain_work} with equality: the general bound \eqref{eq:chain_work} on its blind side and the bound \eqref{eq:general_work} on its informed side. It also exhibits the conditional indispensability announced in Sec.~\ref{sect:indispensability}: under a budget constraint, a correlation worth $n\boltz\Temperature\ln 2$ decides the fate of an arbitrarily large cartridge. The claim concerns scaling. The correlation required grows only logarithmically with the size of the locked resource; it does not provide a route to macroscopic energy harvesting. The physically meaningful regime is thermal, as the implementation sketched at the end of this section makes concrete.

\paragraph*{Setup.}
The composite system consists of $2n+2$ subsystems in contact with a single bath at temperature $\Temperature$:
(i) a \emph{reference register} $A=(A_1,\dots,A_n)$, frozen throughout (identity mechanisms);
(ii) a \emph{working medium} $B=(B_1,\dots,B_n)$, each $B_j$ a particle in a symmetric double well with states $\{\mathrm{L},\mathrm{R}\}$ of equal energy;
(iii) an \emph{access bit} $K$ with two states $\{0,1\}$ of equal energy, initialized at $K=0$;
(iv) a \emph{cartridge} $C$ with two states $\{\mathrm{charged},\mathrm{spent}\}$ of bare energies $E_C(\mathrm{charged})=\Fcart$ and $E_C(\mathrm{spent})=0$, initialized charged. Its stored free energy is released only through the gate mechanism of step 3 below, which couples it to $K$.
Initially, each pair is perfectly correlated, $B_j=A_j\in\{\mathrm{L},\mathrm{R}\}$ with probability $1/2$ each, and distinct pairs are independent, so that
\begin{IEEEeqnarray}{lCl}
    \muI{A}{B} = \TotalCorrelation(t_0) = n\ln 2,
    \label{eq:ex2_resource}
\end{IEEEeqnarray}
while each $A_j$ and $B_j$ marginal, uniform over two states of equal energy, is at equilibrium. The initial state therefore contains three nonequilibrium resources: the internal correlation \eqref{eq:ex2_resource}, the charged cartridge, and the sharp access bit, whose pure state $K=0$ carries a local free energy of $\boltz\Temperature\ln2$ above the uniform equilibrium of its two equal-energy states. The access bit is part of the device specification, and the gate is supplied initialized. Its $\boltz\Temperature\ln2$ is never harvested by the protocols below. In the chain accounting, the local free energy released as the marginal of $K$ softens is consumed, nat for nat, in building the correlation of $K$ with $B$; the two contributions cancel (see the derivation of \eqref{eq:ex2_blind}). A device re-used for another run would have to repay up to the same $\boltz\Temperature\ln2$ to re-initialize $K$, a per-device constant that leaves the $n$-scaling claims of this section untouched.

The chain consists of three steps, each of the class of Sec.~\ref{sect:setting}:
\begin{enumerate}
    \item (\emph{correlation use}) Each $B_j$ is updated by a local channel $P(B_j\prm\mid B_j,\DepX{B_j})$ with parents $\DepX{B_j}\in\bigl\{\emptyset,\{A_j\}\bigr\}$ chosen by the protocol. This per-bit restriction is essential to the budget analysis below; see the remark following \eqref{eq:ex2_threshold}.
    \item (\emph{evaluation}) The fixed mechanism with $\DepX{K}=\{B\}$ applies the controlled-NOT $K\prm=K\oplus\mathbf{1}\{B=\mathrm{LL\cdots L}\}$, defined on the whole state space. At each value of the parent $B$, the update is a bijection between the two equal-energy states of $K$ and is therefore logically reversible. It is implemented at zero work and zero heat, with the same controlled-NOT as the reversed channel (as for the keyed bijections of Sec.~\ref{sect:ex_bit}). On the support, where $K=0$ initially, it flips $K\to1$ iff $B=\mathrm{LL\cdots L}$. The parent set is minimal because the output varies with $B$ at either value of $K$.
    \item (\emph{release}) The fixed mechanism with $\DepX{C}=\{K\}$ applies, on the whole state space, the controlled involution that exchanges charged and spent iff $K=1$ and leaves $C$ frozen for $K=0$. At each value of the parent, the map is a bijection. The exchange is run quasistatically against a work reservoir, so the energy difference $\Fcart$ is delivered as work, at zero heat, on the trajectories with $K=1$. On the support, where $C$ starts charged, the cartridge is spent, and $\Fcart$ is collected, exactly on the pass trajectories. The parent set $\{K\}$ is again minimal.
\end{enumerate}
The mechanisms of steps 2 and 3 are part of the device specification and are not modifiable by the protocol (a \emph{tamper-proof gate}); the protocol chooses only the channels acting on $B$ in step 1. Step 2 reads the \emph{final} state of step 1, which, as remarked in Sec.~\ref{sect:indispensability}, forces the multistage description. Every protocol considered below satisfies Assumptions~\ref{asm:local_mech}--\ref{asm:no_unrealized} at every step. Assumptions~\ref{asm:local_mech} and \ref{asm:acyclic} hold by construction. Each step's kernel is a product of the listed full-state channels, and the parent graphs are acyclic: they are edgeless in step 1, contain the single edge $B\to K$ in step 2, and contain $K\to C$ in step 3. Local detailed balance (Assumption~\ref{asm:ldb}) holds channelwise. The gates of steps 2 and 3 are keyed bijections admitting their inverses as reversed channels, as in Sec.~\ref{sect:ex_bit}; the step-1 channels are elementary one-bit operations whose implementations are specified where they are used. For Assumption~\ref{asm:no_unrealized}, whenever at least one member of a pair $(j,k)$ is static at a step, the conditioning in \eqref{eq:asm_no_unrealized} fixes that member at both times, and the term vanishes, as in the standard feedback setting of Sec.~\ref{sect:setting}. The only subsystems updated within the same step are the $B_j$ in step 1. There, the triples $(A_j,B_j,B_j\prm)$ are independent across $j$ because the channels act pairwise on independent pairs, so those terms also vanish. Hence Proposition~\ref{prop:chain_work} and the composed ledger \eqref{eq:composed_ledger} of Appendix~\ref{ap:multistage} apply.

\paragraph*{Informed protocol.}
Choose $\DepX{B_j}=\{A_j\}$ and let the channel be the conditional swap: the two wells of $B_j$ are exchanged iff $A_j=\mathrm{R}$, so that $B_j\prm=\mathrm{L}$ deterministically. At each fixed parent value, the local map is a bijection between states of equal energy and is implemented reversibly at zero work and zero heat. The step completely destroys the pair correlations, $\decIk[1]=n\ln2$, but every destroyed atom is shared with the controller that destroyed it, $\accIk[1]=n\ln2$. Step 1 therefore contributes nothing to the composed ledger \eqref{eq:composed_ledger}; indeed, $\DeltaStot^{(k)}=0$ at every step. The gate opens with certainty, and the cartridge is harvested:
\begin{IEEEeqnarray}{lCl}
    \Wext^{\mathrm{informed}} = \Fcart .
    \label{eq:ex2_informed}
\end{IEEEeqnarray}
This result saturates the general bound \eqref{eq:general_work}. The marginal of each $B_j$ sharpens from uniform to deterministic, so $-\invTemp\sum_{i}\Delta\Floc{i} = \invTemp\Fcart - n\ln2$, and adding $\TotalCorrelation(t_0)=n\ln2$ gives exactly $\invTemp\Fcart$. The role of the correlation here is specific: it is \emph{not} converted into work; step 1 extracts none. It is converted into the \emph{purification} of $B$. The nonequilibrium free energy of $A\cup B$ is unchanged while its correlation part is transformed into sharp marginals. The sharp state opens the gate, which releases the cartridge. The harvested free energy $\Fcart$ preexists in the cartridge and is unrelated to $A$ and $B$.

\paragraph*{Blind protocols.}
Let $\DepX{B_j}=\emptyset$ for all $j$. Because the channels act independently on independent pairs, the final bits $B_1\prm,\dots,B_n\prm$ are mutually independent. Writing $q_j:=P(B_j\prm=\mathrm{L})$, the gate opens with probability $P=\prod_j q_j$. The chain is not fully blind because the gates of steps 2 and 3 keep their parents. The applicable bound is therefore the general chain bound \eqref{eq:chain_work}, evaluated term by term. The local terms give
$-\invTemp\sum_{i}\Delta\Floc{i}
=\invTemp P\Fcart-\sum_{j}\bigl[\ln2-H_2(q_j)\bigr]+2H_2(P)$,
where $H_2(q):=-q\ln q-(1-q)\ln(1-q)$. The steered media contribute the sum, while $K$ and $C$, whose final marginals are mixtures of weight $P$, contribute $H_2(P)$ each. The correction terms are fixed by the mechanisms. Step 1 is blind and destroys the pair correlations while sharing none of the loss, $\accIk[1]=0$. The gates of steps 2 and 3 destroy nothing, $\accIk[2]=\accIk[3]=0$, but each \emph{creates} fresh correlation: first between $K\prm$ and $B\prm$, then between $C\prm$ and both, with exactly $\incIk[2]=\incIk[3]=H_2(P)$. The created correlations enter \eqref{eq:chain_work} with a negative sign and cancel the $+2H_2(P)$ of the local terms:
\begin{IEEEeqnarray}{lCl}
    \invTemp\,\mathbb{E}[\Wext]
    \;\leq\; \invTemp P\,\Fcart
    - \sum_{j=1}^{n}\bigl[\ln2 - H_2(q_j)\bigr].
    \IEEEeqnarraynumspace
    \label{eq:ex2_blind}
\end{IEEEeqnarray}
The bound is tight throughout. Quasistatic blind steering of each marginal to $q_j$, at injected work $\boltz\Temperature[\ln2-H_2(q_j)]$ per bit, followed by the two reversible gates, attains \eqref{eq:ex2_blind} with equality in the quasistatic limit. It saturates the composed ledger \eqref{eq:composed_ledger} with $\DeltaStot^{(1)}=\decIk[1]=n\ln2$ and $\DeltaStot^{(2)}=\DeltaStot^{(3)}=0$.

Two optimization problems must be distinguished. For the \emph{fixed conversion task} of Sec.~\ref{sect:indispensability}, in which the final marginals are prescribed to be the pass state, $q_j=1$, the bound \eqref{eq:ex2_blind} specializes to the exact blind optimum,
\begin{IEEEeqnarray}{lCl}
    \sup_{\mathrm{blind}}\;\mathbb{E}[\Wext] = \Fcart - n\boltz\Temperature\ln2
    \quad\text{\emph{(fixed conversion task)}},
    \label{eq:ex2_blind_opt}
\end{IEEEeqnarray}
This optimum is approached in the quasistatic limit by the outright erasure of each $B_j$. The informed protocol completes the same task at zero injected work and collects \eqref{eq:ex2_informed}. Thus, the informed--blind gap for the task equals $\boltz\Temperature\,n\ln2=\boltz\Temperature\,\muI{A}{B}$ exactly, independently of $\Fcart$. The cartridge term, common to both sides, cancels from the difference. The value of information saturates its general bound of Sec.~\ref{sect:indispensability} and never exceeds the correlation actually held, however large the cartridge.

If instead the \emph{expected yield} is maximized with the final marginals left free, the deterministic reset is optimal only asymptotically. At finite $\invTemp\Fcart$, the right-hand side of \eqref{eq:ex2_blind} is maximized by a \emph{soft} steering to an interior point. For $n=1$, the maximizer is $q^{*}=\bigl(1+e^{-\invTemp\Fcart}\bigr)^{-1}$, with optimal value $\ln\bigl(1+e^{\invTemp\Fcart}\bigr)-\ln2$. This exceeds the value $\invTemp\Fcart-\ln2$ of the reset $q=1$ by $\ln\bigl(1+e^{-\invTemp\Fcart}\bigr)$, a difference that is exponentially small for $\invTemp\Fcart$ large but positive at every finite $\Fcart$. The steering protocol above attains it in the quasistatic limit. For general $n$, the first-order condition is $\ln[q_j/(1-q_j)]=\invTemp\Fcart\prod_{k\neq j}q_k$. Because its left-hand side diverges as $q_j\to1$, every maximizer lies at an interior symmetric point, $q_j^{*}=1-\Theta\bigl(e^{-\invTemp\Fcart}\bigr)$ at fixed $n$ (verified numerically for $n\leq3$). The informed side moves in lockstep: informed and blind steering to the same marginals differ in injected work by $\boltz\Temperature\ln2$ per bit, identically in the marginals, so the conditional swap of \eqref{eq:ex2_informed} is likewise beaten by the same exponentially small margin. The corrections cancel in the difference. The informed--blind gap thus equals $n\boltz\Temperature\ln2$ exactly at every $\Fcart$, for the fixed task and for the expected yield alike.

\paragraph*{Budget constraint: conditional indispensability.}
Suppose that the external work injectable \emph{before} the cartridge is unlocked is bounded by $b$ \emph{in every single run}. Along each realization of the protocol, including any randomness used in choosing the channels, the injected work never exceeds $b$ (a ``starter battery'' of capacity $b$; the work released by $C$ becomes available only after step 3). The per-run reading is the natural one for a battery, and it is essential. A budget imposed merely on the \emph{average} injection is defeated by coupling the erasure to a nearly free coin: with probability $p$, erase all $n$ bits outright at injected work $n\boltz\Temperature\ln2$; otherwise, do nothing. For $p=\invTemp b/(n\ln2)$, the mean injection equals $b$, yet the gate opens with probability at least $p$. This dependence is linear in $b$, in place of the exponentially small bounds derived below. The per-run budget closes this loophole. The bounds below then hold conditionally on each realization of the protocol's randomness, with the same constant $b$, and hence also on average. The informed protocol is unaffected by the budget because it draws no external work at all. For a blind protocol, applying \eqref{eq:blind_work} to step 1 alone shows that steering the marginals to $(q_1,\dots,q_n)$ requires an average injected work of at least $\boltz\Temperature\sum_j[\ln2-H_2(q_j)]$. The budget therefore enforces $\sum_j[\ln2-H_2(q_j)]\leq\invTemp b$. Pinsker's inequality $\ln2-H_2(q)\geq2\left(q-\tfrac12\right)^2$ then bounds the pass probability \footnote{Indeed, $P=\prod_jq_j\leq e^{-\sum_j(1-q_j)}$, while $\sum_j(1-q_j)=\tfrac{n}{2}-\sum_j(q_j-\tfrac12)\geq\tfrac{n}{2}-\sqrt{n\sum_j(q_j-\frac12)^2}\geq\tfrac{n}{2}-\sqrt{n\invTemp b/2}$ by the Cauchy--Schwarz and Pinsker inequalities.}:
\begin{IEEEeqnarray}{lCl}
    P \;\leq\; \exp\left(-\frac{n}{2}+\sqrt{\frac{n\invTemp b}{2}}\right),
    \label{eq:ex2_pinsker}
\end{IEEEeqnarray}
and hence, by \eqref{eq:ex2_blind},
\begin{IEEEeqnarray}{lCl}
    \mathbb{E}[\Wext]
    \;\leq\;
    e^{-n/2+\sqrt{n\invTemp b/2}}\,\Fcart .
    \label{eq:ex2_budget}
\end{IEEEeqnarray}
In particular, by \eqref{eq:ex2_budget}, whenever
\begin{IEEEeqnarray}{lCl}
    n \;\geq\; 2\invTemp b + 4\ln(\invTemp\Fcart),
    \label{eq:ex2_threshold}
\end{IEEEeqnarray}
the blind expectation is thermodynamically negligible, $\mathbb{E}[\Wext]\leq\boltz\Temperature$ \footnote{For $\invTemp\Fcart\geq1$, so that $L:=\ln(\invTemp\Fcart)\geq0$: with $n=2\invTemp b+4L$ one has $-n/2+\sqrt{n\invTemp b/2}=-\invTemp b-2L+\sqrt{\invTemp b(\invTemp b+2L)}\leq-L$, since $\sqrt{x(x+2L)}\leq x+L$. For $\invTemp\Fcart<1$ the claim is trivial: $\mathbb{E}[\Wext]\leq P\,\Fcart<\boltz\Temperature$.}, while the informed protocol still collects the full $\Fcart$.

The derivation of \eqref{eq:ex2_pinsker} relies on the restriction, built into step 1, that each blind channel acts on its own $B_j$ alone. This restriction renders the final bits independent, so the pass probability factorizes as $P=\prod_jq_j$ and is controlled by the marginal-by-marginal budget accounting. The restriction is load-bearing, not a cosmetic simplification. The phenomenon itself is expected to survive its removal. For auxiliary-free blind step-1 protocols under the same per-run budget (channels acting jointly on all of $B$, with flat energy levels), the detailed fluctuation theorem \cite{Crooks_1999,seifert2012stochastic} yields the tilted bound $P\leq2^{-n}e^{\invTemp b}$\footnote{For flat energy levels and the uniform initial distribution of $B$, summing the detailed fluctuation theorem over the trajectories that end in the pass state gives $\sum_{\mathrm{path}\to\mathrm{LL\cdots L}}P[\mathrm{path}]\,e^{-\invTemp W[\mathrm{path}]}\leq2^{-n}$, where $W[\mathrm{path}]$ is the work injected along the trajectory. The per-run budget caps $W[\mathrm{path}]\leq b$ on every trajectory, so the left-hand side is at least $e^{-\invTemp b}P$. The one-shot reading of the auxiliary charge is essential for the same reason the budget is per-run: an auxiliary fully charged with probability $\epsilon$ and thermal otherwise carries an average free energy of at most $\epsilon\,n\boltz\Temperature\ln2$, yet a zero-work swap unlocks the gate with probability $\approx\epsilon$; charging the one-shot free energy $\boltz\Temperature\ln(\epsilon2^{n}+1-\epsilon)$ prices this attack tightly.}. This bound is tighter than \eqref{eq:ex2_pinsker} throughout the nontrivial regime $\invTemp b\leq n\ln2$ and sharpens the threshold \eqref{eq:ex2_threshold} to $n\ln2\geq\invTemp b+\ln(\invTemp\Fcart)$. For protocols assisted by auxiliary systems, we \emph{conjecture} the same bound, provided any initial nonequilibrium free energy of the auxiliary is charged to $b$ in the one-shot (max-relative-entropy) sense \cite{del2011thermodynamic,faist2015the}. The footnote prices the natural mixing attack tightly under this convention, but a formal treatment of general auxiliaries is left to future work. We retain the per-bit setting in the text for its explicit protocols and self-contained accounting.

Three features of the budget-constrained regime deserve emphasis.
(i) \emph{Information is amplified along the chain.} A correlation worth only $n\boltz\Temperature\ln2$, together with the mechanisms that read it, decides whether an arbitrarily large $\Fcart$ is collected. By \eqref{eq:ex2_threshold}, a sufficient correlation grows only logarithmically with the size of the gain (and linearly with the budget). This is a sufficiency threshold; no matching lower bound on the necessary correlation is claimed. This does not contradict the saturation of the value of information at $\boltz\Temperature\,\muI{A}{B}$: that saturation presupposes unlimited borrowing, which is exactly what the budget removes.
(ii) \emph{The final resource is manifestly unrelated to the correlation.} The cartridge free energy exists prior to, and independently of, $A$ and $B$; no correlation is ``converted into'' the harvested work. What the correlation provides, at zero work cost but only to an informed device, is the purification of $B$ that opens the gate. The gain is mediated by a chain of enablements, correlation $\to$ definite output state $\to$ access $\to$ free energy, rather than by a direct correlation-to-work conversion. This is the precise sense in which information can be indispensable for proceeds that are not themselves informational.
(iii) \emph{The structural assumptions are irreducible.} If the protocol could act on $K$ directly, or if $B$ started concentrated on the pass state, blind protocols would succeed at no cost. The tamper-proofness of the gate and the exponentially small prior weight of the pass state cannot be derived from thermodynamics; they are properties of the device that must be posited. The example shows that they can be realized by explicit, physically legitimate mechanisms. Once they hold, the indispensability of information for the final gain is quantitative and robust.

\paragraph*{Implementation sketch.}
The ingredients of this example have related precedents in single-electron experiments: correlated charge states prepared by measurement-based feedback \cite{Koski2014experimental}, gated series arrays of islands \cite{Mills_2019}, per-run metering of injected work \cite{Koski_2014,Koski2014experimental}, and erasure at close to minimal dissipation \cite{Scandi_2022}. These precedents identify single-electron platforms as a candidate testbed, although integration, leakage budgeting, and work accounting remain nontrivial \footnote{The hardware leakage of a series gate is an imperfection $\epsilon_{\mathrm{leak}}$ distinct from the statistical factor $2^{-n}$: the latter is the prior weight of the pass state, the former a bypass probability, to be kept small enough that $\epsilon_{\mathrm{leak}}\invTemp\Fcart$ is negligible on the thermodynamic scale under test. Likewise, the per-run work budget would have to be reconstructed from calibrated gate trajectories and charge transitions within a declared accounting boundary, rather than identified with a cooling-power constraint.}. For small pre-unlock budgets, $\invTemp b=O(1)$, a device of $n\approx10$--$13$ pairs with $\invTemp\Fcart\sim10^{2}$--$10^{3}$ would reach the threshold regime of the sharpened fluctuation-theorem bound noted above (established there for auxiliary-free protocols), $n\ln2\geq\invTemp b+\ln(\invTemp\Fcart)$. The per-bit threshold \eqref{eq:ex2_threshold} is more demanding, $n\approx20$--$30$ in the same regime, and larger budgets require $n$ to grow linearly with $\invTemp b$ under either bound. The distinctive observable is not the value of feedback information \cite{sagawa2008second,Toyabe_2010}, nor the blind rare-event bound alone, but the informed--blind unlocking contrast under a fixed pre-existing correlation and a declared per-run budget. The informed controller reaches the pass state with unit probability at zero injected work, with the explicit construction saturating the equality of Theorem~\ref{thm:main}. The blind unlocking probability obeys, in the stated protocol classes, the fluctuation-theorem-like bound $P\leq2^{-n}e^{\invTemp b}$. To our knowledge, this contrast has not been isolated in existing single-electron information-thermodynamics experiments, which have measured $\boltz\Temperature\ln2$-scale work extraction \cite{Koski_2014}, autonomous demon refrigeration and power generation \cite{Koski_2015,Chida_2017}, and near-minimal erasure \cite{Scandi_2022}. It is also distinct from work extraction with feedback control using limited resources \cite{Hartle_2026}, where the constrained resources are the measurement quality, the memory, and the allowed protocol repertoire, whereas here the constrained resource is a pre-existing correlation whose budgeted unlocking law is the target. The applied value of such a device would be metrological rather than motive: a testbed for $\boltz\Temperature$-scale accounting of correlated registers, and a benchmark for single-electron work metrology and for thermodynamic-scale accounting in ultra-low-dissipation logic platforms. We leave its specification to future work.\section{Conclusion}
\label{sect:conclusion}

We have derived a second law for classical multipartite systems in contact with a single heat bath, resolved by the state change causing each correlation loss and by the variables read by that side's mechanisms. For a single synchronous step, every component is updated by its own local mechanism, which reads a snapshot of a fixed subset of the others. The total entropy production obeys an exact ledger inequality that is valid for arbitrary reading patterns, including reciprocal ones. Explicit processes saturate it with equality (Theorem~\ref{thm:ledger}). Under acyclic reading, and for protocols that generate no unrealized correlation, the ledger collapses to the transparent statement of Theorem~\ref{thm:main}. The entropy produced is bounded below, in aggregate, by the destroyed internal correlation minus its shared part, namely, the part visible to the controllers of the side whose move caused the destruction. Information elsewhere in the system does not by itself generate the aggregate reading-pattern credit; what matters is whether it enters the parent structure of the acting mechanisms. The bound interpolates between two limiting regimes. Under fully blind dynamics, every destroyed correlation is dissipated in full. When the parents cover all destroyed correlations, the conventional second law is recovered. Converted into a work bound, it states that the work extractable beyond the local free energies is fed, in aggregate, exclusively by the shared part of the destroyed correlation. The attribution cannot be read off the statistics. In the minimal bit model of Sec.~\ref{sect:ex_bit}, three one-step processes share the joint distribution of $(\Xtot,\Xtotprm)$ bit for bit. Every distribution-level account therefore assigns them identical verdicts, yet a single reading line decides between zero dissipation and $\ln 2$ (Table~\ref{tbl:ex_bit}). The bound follows the reading pattern, not the distribution. The acyclicity cannot simply be dropped. Under reciprocal reading, the collapse can fail because an explicit mutual-refresh process makes the rebate unavoidable (Appendix~\ref{ap:cex_refresh}); the full ledger nonetheless remains exact.

Composed along conversion chains, these results sharpen the exchange view of information: a fully blind chain can never credit internal correlations to its work balance. For a fixed conversion task, the value of information is capped by $\boltz\Temperature$ times the correlation initially present, and this cap is attained with equality. The budgeted regime then departs qualitatively from the exchange view. When the work that can be advanced before a resource is unlocked is bounded, as it is for any autonomous device starting on a small battery, a finite correlation becomes, within a stated device model, the precondition for collecting an unrelated free-energy resource. The model has a tamper-proof gate with an exponentially unlikely pass state and blind operations acting bit by bit; the general blind case remains open. The size of the resource enters the sufficient correlation only logarithmically, whereas the budget enters linearly (Sec.~\ref{sect:ex_cartridge}). The ingredients of the correlation-keyed cartridge have related precedents in single-electron experiments. The device sketched in Sec.~\ref{sect:ex_cartridge} would isolate an observable that, to our knowledge, has not been measured: the informed--blind unlocking contrast under a declared per-run budget. The applied value of such a device is metrological rather than motive: a testbed for $\boltz\Temperature$-scale accounting of correlated registers, and a benchmark for single-electron work metrology and ultra-low-dissipation logic.

The structure isolated by the example is also commonplace in engineered systems: frozen correlation registers whose loss carries a real price. Examples include alignment marks patterned in the same exposure as the layer they reference and state records whose loss must be repaid by recalibration \cite{den_Boef_2016,Orji_2018}. The ledger derived here identifies the thermodynamic floor of that structure, the part of the price of lost correlation that no engineering refinement can remove. The costs that currently dominate such systems lie many orders of magnitude above this floor, and we make no claim that the floor is where their present prices originate.

Several problems remain open. The general theory of budget-constrained chains awaits development, including the optimal trade-off between measurement, memory, and gate traversal already flagged in Sec.~\ref{sect:indispensability}. The conjectured nonnegativity \eqref{eq:Ucost_sum} of the aggregate unrealized-correlation cost, verified for $\Nx\leq5$ by exact linear-programming certificates, lacks a structure-independent proof. The fluctuation-theorem bound for blind protocols assisted by arbitrary auxiliaries calls for a formal treatment. The experimental specification of the single-electron test, including leakage budgeting and per-run work reconstruction within a declared accounting boundary, remains to be worked out. All of these are internal to the framework. The framework itself is confined to a single synchronous step of a classical system, the very step structure that separates it from continuous-time multipartite jump processes. Extending the mechanism-sensitive attribution remains open for continuous time, where mechanisms read continuously rather than through a declared snapshot, and for quantum systems.

\section*{Data availability}
No experimental data were created or analyzed in this study. Code for the computer-assisted verifications reported here---the exact linear-programming certificates for \eqref{eq:Ucost_sum} and the numerical consistency checks of the examples---is available from the authors upon request.

\bibliographystyle{apsrev4-2}
\bibliography{akihito/corpus-all}%
\appendix

\section{Decomposition of Total Correlation}
\label{ap:delitotTotalIC}

We establish the following decomposition of the total correlation:
\begin{IEEEeqnarray}{lCl}
\TotalCorrelation
=
\sum_{i=2}^{N} \muI{X_i}{X_{\minusI}}
\nonumber
\end{IEEEeqnarray}
From the definition of the total correlation,
\begin{IEEEeqnarray}{lCl}
    \TotalCorrelation &=& D_{KL}[P(\Xtot) || P(X_1)P(X_2)\cdots P(X_\Nx)]
    \\
    &=& \Sum{i=1}{\Nx} \shaent{X_i} - \shaent{\Xtot}
\end{IEEEeqnarray}

Substituting the entropy decomposition \eqref{eq:EntropySplit}, with $Z=\emptyset$, for $\shaent{\Xtot}$ cancels the individual entropies and leaves
\begin{IEEEeqnarray}{lCl}
    \TotalCorrelation
    &=& \sum_{i=2}^{N} \muI{X_i}{X_{1:i-1}},
\end{IEEEeqnarray}
as claimed.

\section{Entropy bound for a subsystem}
\label{ap:fluctuation_subsystem}

We derive Eq.~\eqref{eq:yuragi_k} from the premise formulated in Eq.~(\ref{eq:local_detail_subsystem}) and Assumption~\ref{asm:local_mech}, whose screening property identifies the conditional law $P(\Xs{j}\prm\mid\Xs{j},\DepXJ)$ with the channel entering \eqref{eq:local_detail_subsystem}.
Let $P\cj$  denote a conjugate probability distribution of $P$. Assume first that $P\cj$ assigns no mass outside the support of $P$; the general case, quantified below Eq.~\eqref{eq:AAPPPone}, replaces the exact identities by their $1-\lambda$ forms. We determine $P\cj$ by requiring the following relations:
\begin{IEEEeqnarray}{rCl}
P\cj(\DepXJ) &=& P(\DepXJ),
\label{eq:PdPXX0}
\\
P\cj(\Xj'\mid\DepXJ)&=&P(\Xj'\mid\DepXJ),
\label{eq:PdPXX1}
\\
P\cj(X_j\mid X_j\dashh,\DepX{j})&=&P\B(X_j\mid X_j\dashh,\DepX{j}).
\label{eq:PdPXX2}
\end{IEEEeqnarray}
Define $\Xtotmin:=\Xtot\setminus\Xsub\setminus\Zd$ and $\Xtotmin\prm:=\Xtotprm\setminus\Xsub\prm\setminus\Zd$.
The chain rule gives
\begin{IEEEeqnarray}{lCr}
\frac{P\cj(\Xsub\prm,\Xsub,\Zd)}{P(\Xsub,\Xsub\prm,\Zd)}
=
\frac{P\cj(\Xsub\prm,\Zd) P\cj(\Xsub\mid\Xsub\prm,\Zd)}
{P(\Xsub,\Zd) P(\Xsub\prm\mid\Xsub,\Zd)}
\IEEEeqnarraynumspace
\label{eq:PfracP_B1}
\end{IEEEeqnarray}
and
\begin{IEEEeqnarray}{lCl}
P(\Xtot,\Xtotprm)
= P(\Xtotmin,\Xtotmin\prm,\Zd\prm\mid\Xsub,\Xsub\prm,\Zd)
P(\Xsub,\Xsub\prm,\Zd).
\IEEEeqnarraynumspace
\label{eq:PXXPXXPXX_B2}
\end{IEEEeqnarray}
It follows that
\begin{widetext}
\begin{IEEEeqnarray}{rCl}
\IEEEeqnarraymulticol{3}{l}{
\langlee\frac{P\cj(\Xsub\prm,\Xsub,\Zd)}{P(\Xsub,\Xsub\prm,\Zd)}\ranglee
=
\int P(\Xtot,\Xtotprm) \frac{P\cj(\Xsub\prm,\Xsub,\Zd)}{P(\Xsub,\Xsub\prm,\Zd)} d\Xtot d\Xtotprm
}
\\
&\EqualText{(\ref{eq:PXXPXXPXX_B2})}&
\underbrace{
\int P(\Xtotmin,\Xtotmin\prm,\Zd\prm\mid\Xsub,\Xsub\prm,\Zd)
d\Xtotmin d\Xtotmin\prm d\Zd\prm}_{=1}
\NonumberNewline
&&
\times
\underbrace{
\int \cancel{P(\Xsub,\Xsub\prm,\Zd)}\frac{P\cj(\Xsub\prm,\Xsub,\Zd)}{\cancel{P(\Xsub,\Xsub\prm,\Zd)}} d\Xsub d\Xsub\prm d\Zd }_{=1}
\nonumber
\\
&=& 1.
\label{eq:sahen1_A1}
\end{IEEEeqnarray}
The second factor equals unity when the conjugate distribution places no mass outside the support of $P$, over which the expectation runs; the general case is stated below Eq.~\eqref{eq:AAPPPone}.
We also obtain
\begin{IEEEeqnarray}{rCl}
\langlee\frac{P\cj(\Xsub\prm,\Xsub,\Zd)}{P(\Xsub,\Xsub\prm,\Zd)}\ranglee
&\EqualText{(\ref{eq:PfracP_B1})}& \langlee\frac{P\cj(\Xsub\prm,\Zd) P\cj(\Xsub\mid\Xsub\prm,\Zd)}
{P(\Xsub,\Zd) P(\Xsub\prm\mid\Xsub,\Zd)}\ranglee
\\
&=& \langlee \frac{P\cj(\Xsub\prm\mid\Zd)\cancel{P\cj(\Zd)}P\cj(\Xsub\mid\Xsub\prm,\Zd)}{P(\Xsub\mid\Zd)\cancel{P(\Zd)}P(\Xsub\prm\mid\Xsub,\Zd)}\ranglee
\\
&=& \langlee \exp\left\{-\left[
s\cj(\Xsub\prm\mid\Zd) - s(\Xsub\mid\Zd)
\right.
\right.
\right.
\NonumberNewline
&&
\left.
\left.
\left.
+ \ln\frac{P(\Xsub\dashh\mid \Xsub,\Zd)}{P\cj(\Xsub\mid \Xsub\dashh,\Zd)}
\right]\right\} \ranglee,
\label{eq:eqplnPPP9}
\end{IEEEeqnarray}
where $s(a\mid b):=-\ln P(a\mid b)$ and $s\cj(a\mid b):=-\ln P\cj(a\mid b)$ denote the surprisals of $P$ and $P\cj$, as in Appendix~\ref{ap:fluctuation_internal}.
Combining Eqs.~(\ref{eq:PdPXX1}), (\ref{eq:PdPXX2}), (\ref{eq:sahen1_A1}), and (\ref{eq:eqplnPPP9}) gives
\begin{IEEEeqnarray}{lCr}
\langlee \exp\left\{-\left[
s(\Xsub\prm\mid\Zd) - s(\Xsub\mid\Zd)
\right.\right.\right.
\NonumberNewlineQQ
\left.\left.\left.
+ \ln\frac{P(\Xsub\dashh\mid \Xsub,\DepXJ)}{P\B(\Xsub\mid \Xsub\dashh,\DepXJ)}
\right]\right\} \ranglee
= 1.
\label{eq:AAPPPone}
\IEEEeqnarraynumspace
\end{IEEEeqnarray}
\end{widetext}
Two clarifications delimit Eq.~\eqref{eq:AAPPPone}. The equality holds whenever the conjugate distribution assigns no mass outside the support of $P$---in particular whenever the initial conditional law $P(\Xsub\mid\DepXJ)$ has full support. In general the cancellation in \eqref{eq:sahen1_A1} runs over the support of $P$ only, and the average equals $1-\lambda$, where $\lambda:=P\cj[\,P=0\,]\in[0,1)$ is the conjugate mass on forward-null events; the informed uniform-sweep implementation of Sec.~\ref{sect:ex_bit}, run on the maximally correlated ensemble, realizes $\lambda=\tfrac12$. Since $1-\lambda\leq1$, applying Jensen's inequality to the general form, under the premises expressed in Eq.~(\ref{eq:local_detail_subsystem}), still yields inequality~(\ref{eq:yuragi_k}); every averaged statement below is unaffected by $\lambda$.

The derivation requires only that the channel of subsystem $j$ depend on nothing beyond $(\Xs{j},\DepXJ)$ and that $\Zd$ remain fixed during the update. Accordingly, the construction \eqref{eq:PdPXX0}--\eqref{eq:PdPXX2}, the identity \eqref{eq:AAPPPone} in its general $1-\lambda$ form, and the averaged inequality remain valid when $\Zd$ is enlarged from $\DepXJ$ to any collection $Z\supseteq\DepXJ$ of variables frozen while $j$ is updated. This rereading uses two conventions. First, the backward kernel in the third defining condition \eqref{eq:PdPXX2} retains its conditioning $\DepXJ$; only \eqref{eq:PdPXX0} and \eqref{eq:PdPXX1} involve the enlarged $Z$. Second, $Z$ may include final-state variables. Here, frozen means only that they do not change while $j$ is updated. The proof of Lemma~\ref{lem:joint_second_law} below uses both conventions. This yields the strengthened per-subsystem inequality
\begin{IEEEeqnarray}{lCl}
\shaent{\Xs{j}\prm\mid Z} - \shaent{\Xs{j}\mid Z} + \invTemp\Qs{j} \;\geq\; 0,
\qquad Z \supseteq \DepXJ.
\label{eq:yuragi_k_strong}
\end{IEEEeqnarray}
The following consequence is used in the derivation of Eq.~\eqref{eq:general_work} in the main text and in Appendix~\ref{ap:multistage}.

\begin{lemma}[Joint second law]
\label{lem:joint_second_law}
Under Assumptions~\ref{asm:local_mech}--\ref{asm:acyclic}, $\DeltaStot \geq 0$ for a single step, and $\DeltaStot^{(1:K)}\geq0$ for every multistage chain of Sec.~\ref{sect:indispensability}.
\end{lemma}
\begin{proof}
Under the parents-later convention \eqref{eq:parents_later}, updating the subsystems in the order $j=1,\dots,\Nx$ realizes the dynamics \eqref{eq:asm_local_mech} exactly (Sec.~\ref{sect:setting}). Apply \eqref{eq:yuragi_k_strong} to the $j$-th substep with $Z_j = (\Xs{1:j-1}\prm, \Xs{\plusX{j}})$, the variables frozen at that substep; $Z_j\supseteq\DepXJ$ holds by \eqref{eq:parents_later}. Writing $S_j := \shaent{(\Xs{1:j}\prm,\Xs{\plusX{j}})}$ for the joint entropy of the intermediate state, the conditional-entropy difference in \eqref{eq:yuragi_k_strong} equals $S_j - S_{j-1}$, so summing over $j$ telescopes to $\Delta\Htot + \invTemp\Qtot = \DeltaStot \geq 0$. Additivity over steps \eqref{eq:step_additivity} extends the statement to chains.
\end{proof}

\noindent We emphasize the division of labor with Appendix~\ref{ap:fluctuation_internal}: the exact integral fluctuation theorem there requires the initial distribution to factorize along the parent structure and to have full support, whereas the average statement above holds for arbitrary initial distributions.\section{Fluctuation Theorem for Correlated Subsystems}
\label{ap:fluctuation_internal}

Under additional assumptions on the initial data, the inequality \eqref{eq:2ndlaw1} follows, via Jensen's inequality, from an exact integral fluctuation theorem for the composite system. The first additional assumption is that the initial distribution factorizes along the parent structure,
\begin{IEEEeqnarray}{lCl}
    P(\Xtot) = \prod_{j=1}^{\Nx} P(\Xs{j}\mid\DepXJ),
    \label{eq:markov_initial}
\end{IEEEeqnarray}
That is, the initial state is a Bayesian network over the informational parents, with the product taken along the parents-later order of Assumption~\ref{asm:acyclic}. Under the parents-later convention \eqref{eq:parents_later}, this states that, at the initial time, $\DepXJ$ screens $\Xs{j}$ from all remaining later variables. We emphasize that \eqref{eq:markov_initial} is used only in this appendix: the derivation of \eqref{eq:2ndlaw1} in Sec.~\ref{sect:generalized-ineq} rests on the per-subsystem inequality \eqref{eq:yuragi_k} of Appendix~\ref{ap:fluctuation_subsystem} and requires no condition on the initial distribution. The condition cannot be dropped from the fluctuation theorem itself. For two blindly relaxing subsystems ($\DepX{1}=\DepX{2}=\emptyset$) with correlated initial data, the average in \eqref{eq:detail_gen_secondlaw} genuinely departs from unity. The second additional assumption is that the initial distribution have full support. It is equally indispensable: a point-mass initial state factorizes trivially, yet for two parentless subsystems whose channels randomize completely (with $q_j=0$ and full-support channels), every forward trajectory carries $\Delta s_{tot}=\ln4$ while $q_{tot}=\DelIstocha=0$, so the average in \eqref{eq:detail_gen_secondlaw} equals $\tfrac14$. In general the average equals $1-\Lambda$, where $\Lambda$ is the conjugate mass on events of zero forward probability, exactly as for the per-subsystem identity of Appendix~\ref{ap:fluctuation_subsystem}; since $1-\Lambda\leq1$, the averaged inequality \eqref{eq:2ndlaw1} follows from the general form by Jensen's inequality for arbitrary initial data satisfying \eqref{eq:markov_initial}.

We denote stochastic quantities by lowercase letters,
\begin{IEEEeqnarray}{lCl}
    s(\Xs{j}\mid\DepXJ) := -\ln P(\Xs{j}\mid\DepXJ),
    \\
    i(\Xs{j};\DepXJ) := \ln\frac{P(\Xs{j},\DepXJ)}{P(\Xs{j})P(\DepXJ)},
\end{IEEEeqnarray}
We similarly write $s_{tot} := -\ln P(\Xtot)$ and $i(\Xk;\Xs{1:k-1})$. Primed arguments are evaluated in the final distribution, and the prefix $\Delta$ takes final minus initial values along a trajectory. The local detailed balance ratio in Assumption~\ref{asm:ldb} defines the stochastic dissipated heat of subsystem $j$,
\begin{IEEEeqnarray}{lCl}
    \invTemp q_j := \ln\frac{P(\Xs{j}\prm\mid\Xs{j},\DepXJ)}{P\B(\Xs{j}\mid\Xs{j}\prm,\DepXJ)},
    \qquad
    q_{tot} := \sum_{j=1}^{\Nx} q_j,
    \nonumber
\end{IEEEeqnarray}
so that $\langlee \invTemp q_j\ranglee = \invTemp\Qs{j}$ by Eq.~\eqref{eq:local_detail_subsystem}. We define
\begin{IEEEeqnarray}{lCl}
    \DelIstocha :=
    \sum_{j=1}^{\Nx}\left[ i(\Xs{j}\prm;\DepXJ) - i(\Xs{j};\DepXJ) \right]
    - \Delta i_{tot},
    \nonumber
\end{IEEEeqnarray}
where $\Delta i_{tot} := \sum_k \Delta i(\Xk;\Xs{1:k-1})$ is the stochastic counterpart of \eqref{eq:def_delitot}. By construction, $\langlee \DelIstocha \ranglee=\UsableNewAll - \delitot$ is the right-hand side of Eq.~\eqref{eq:2ndlaw1}.

\begin{proposition}
    Under Assumptions~\ref{asm:local_mech}--\ref{asm:acyclic}, the initial condition \eqref{eq:markov_initial}, and full support of the initial distribution,
    \begin{IEEEeqnarray}{lCl}
    \langlee \exp\left[-\left(\Delta s_{tot} + \invTemp q_{tot} - \DelIstocha\right) \right] \ranglee = 1.
    \label{eq:detail_gen_secondlaw}
    \IEEEeqnarraynumspace
    \end{IEEEeqnarray}
\end{proposition}

\begin{proof}
For each $j$, let $P\cj$ be the conjugate distribution on $(\Xs{j}\prm,\Xs{j},\DepXJ)$ specified by Eqs.~\eqref{eq:PdPXX0}--\eqref{eq:PdPXX2} of Appendix~\ref{ap:fluctuation_subsystem}. By the chain rule and the screening property [which identifies the conditional law $P(\Xs{j}\prm\mid\Xs{j},\DepXJ)$ with the channel of Assumption~\ref{asm:local_mech}], each conjugate ratio is pointwise
\begin{IEEEeqnarray}{lCl}
    \frac{P\cj(\Xs{j}\prm,\Xs{j},\DepXJ)}{P(\Xs{j},\Xs{j}\prm,\DepXJ)}
    = \exp\left\{-\left[\Delta s(\Xs{j}\M\mid\DepXJ) + \invTemp q_j\right]\right\},
    \nonumber
\end{IEEEeqnarray}
where $\Delta s(\Xs{j}\M\mid\DepXJ) := s(\Xs{j}\prm\mid\DepXJ) - s(\Xs{j}\mid\DepXJ)$, only $\Xs{j}$ evolving. Summing the exponents over $j$, using $s(\Xs{j}\M\mid\DepXJ)=s(\Xs{j}\M)-i(\Xs{j}\M;\DepXJ)$ and the pointwise entropy decomposition $\sum_j \Delta s(\Xs{j}) = \Delta s_{tot} + \Delta i_{tot}$ [the trajectory-wise form of \eqref{eq:F18YY}], we obtain
\begin{IEEEeqnarray}{lCl}
    \sum_{j=1}^{\Nx}\left[\Delta s(\Xs{j}\M\mid\DepXJ) + \invTemp q_j\right]
    = \Delta s_{tot} + \invTemp q_{tot} - \DelIstocha.
    \nonumber
\end{IEEEeqnarray}
It therefore suffices to show that $\langlee \prod_j P\cj/P \ranglee = 1$. Taking the average with the joint law $P(\Xtot,\Xtotprm)=P(\Xtot)\prod_j P(\Xs{j}\prm\mid\Xs{j},\DepXJ)$ of Assumption~\ref{asm:local_mech}, the channels cancel factor by factor against the denominators $P(\Xs{j},\Xs{j}\prm,\DepXJ) = P(\Xs{j},\DepXJ)\,P(\Xs{j}\prm\mid\Xs{j},\DepXJ)$, and the summation over the final states factorizes, because $\Xs{j}\prm$ appears in the $j$-th ratio only:
\begin{IEEEeqnarray}{lCl}
    \langlee \prod_{j=1}^{\Nx}\frac{P\cj(\Xs{j}\prm,\Xs{j},\DepXJ)}{P(\Xs{j},\Xs{j}\prm,\DepXJ)}\ranglee
    = \sum_{\Xtot} P(\Xtot) \prod_{j=1}^{\Nx}\frac{P\cj(\Xs{j}\mid\DepXJ)}{P(\Xs{j}\mid\DepXJ)},
    \IEEEeqnarraynumspace
    \label{eq:appC_channels_cancel}
\end{IEEEeqnarray}
where $P\cj(\Xs{j}\mid\DepXJ) = \sum_{\Xs{j}\prm} P(\Xs{j}\prm\mid\DepXJ)\,P\B(\Xs{j}\mid\Xs{j}\prm,\DepXJ)$. Inserting the factorization \eqref{eq:markov_initial} for $P(\Xtot)$ cancels the denominators, leaving $\sum_{\Xtot}\prod_j P\cj(\Xs{j}\mid\DepXJ)$. Under the parents-later convention, $\DepXJ$ involves only variables with indices larger than $j$, so summing over $\Xs{1},\Xs{2},\dots,\Xs{\Nx}$ in this order eliminates the factors one at a time, each summation contributing unity. Hence the right-hand side of \eqref{eq:appC_channels_cancel} equals $1$, which proves \eqref{eq:detail_gen_secondlaw}.
\end{proof}

Applying Jensen's inequality to \eqref{eq:detail_gen_secondlaw} under full support, or to its $1-\Lambda$ form otherwise, recovers Eq.~\eqref{eq:2ndlaw1} whenever \eqref{eq:markov_initial} holds. The proof shows where that condition enters. Without \eqref{eq:markov_initial}, the mismatch between $P(\Xtot)$ and $\prod_j P(\Xs{j}\mid\DepXJ)$ leaves a nontrivial weight in \eqref{eq:appC_channels_cancel}, and the average genuinely departs from unity. The per-subsystem fluctuation relation of Appendix~\ref{ap:fluctuation_subsystem} holds without any such condition, and so does the inequality \eqref{eq:2ndlaw1}.

\section{Formulas of Shannon Information Measures}
\label{ap:shannon_properties}

We collect the basic formulas for Shannon information used in our analysis.
The identities introduced in this section may be used in the main text and the appendices without further notice.
Consider random variables $\Ra, \Rb, \Rc, \Rz$ and a sequence of random variables $\Ra_{1:n}=\set{A_1, A_{2},\ldots,A_{n}}$. The following identities are well established in the literature \cite{cover1999elements}:
\begin{IEEEeqnarray}{rCl}
    \shaent{\Ra\mid\Rz} &=& \shaent{\Ra\mid\Rb,\Rz} + \MutualInfo{\Ra}{\Rb\mid\Rz}{},
    \label{eq:SRARBSHARB_D1}
    \IEEEeqnarraynumspace
    \\
    \MutualInfo{\Ra}{(\Rb,\Rc)\mid\Rz}{}
    &=&
    \MutualInfo{\Ra}{\Rb\mid\Rz}{} + \MutualInfo{\Ra}{\Rc\mid\Rb,\Rz}{}.
    \label{eq:chain_MI6}
\end{IEEEeqnarray}

The interaction information, also referred to as the multivariate mutual information, is defined recursively by extending Eq.~(\ref{eq:SRARBSHARB_D1}) \cite{mcgill1954multivariate,fano1961transmission,srinivasa2005review}:
\begin{IEEEeqnarray}{lCl}
    \IEEEeqnarraymulticol{3}{l}{
        \MutualInfo{\Ra_1}{\Ra_2;\dots;\Ra_{n}\mid\Rz}{}
    }
    \nonumber\\* \quad\quad
    &:=& \MutualInfo{\Ra_1}{\Ra_2;\dots;\Ra_{n-1}\mid\Rz}{}
    - \MutualInfo{\Ra_1}{\Ra_2;\dots;\Ra_{n-1}\mid\Ra_{n},\Rz}{}.
    \IEEEeqnarraynumspace
    \label{eq:def_interaction_cond}
\end{IEEEeqnarray}
For $n=3, \Ra_1=\Ra, \Ra_2=\Rb$ and $\Ra_3=\Rc$, the recursion gives:
\begin{IEEEeqnarray}{lCr}
    \MutualInfo{A}{B\mid\Rz}{}
    =
    \MutualInfo{A}{B;C\mid\Rz}{}
    +\MutualInfo{A}{B\mid C,\Rz}{}.
    \IEEEeqnarraynumspace
    \label{eq:MABCMA134}
\end{IEEEeqnarray}

The following proposition decomposes the joint entropy into the entropies of the individual variables and the correlations among them:
\begin{proposition}
    \begin{IEEEeqnarray}{rCl}
            \shaent{A_{1:\Ngen}\mid Z}
        =
        \sum_{j=1}^\Ngen \shaent{A_j\mid Z}
        - \sum_{j=2}^\Ngen I(A_j;A_{1:j-1}\mid Z).
        \label{eq:EntropySplit}
    \end{IEEEeqnarray}
\end{proposition}
\begin{proof}
    Telescoping the joint entropy by the chain rule gives
    \begin{IEEEeqnarray}{rCl}
        \shaent{A_{1:\Ngen}\mid Z}
        &=& \sum_{j=1}^{\Ngen}\shaent{A_j\mid A_{1:j-1},Z}.
        \nonumber
    \end{IEEEeqnarray}
    Each summand obeys
    \begin{IEEEeqnarray}{rCl}
        \shaent{A_j\mid A_{1:j-1},Z}
        &=& \shaent{A_j\mid Z}
        - \muI{A_j}{A_{1:j-1}\mid Z}
        \nonumber
    \end{IEEEeqnarray}
    by Eq.~(\ref{eq:SRARBSHARB_D1}), which yields Eq.~(\ref{eq:EntropySplit}).
\end{proof}
Applying Eq.~(\ref{eq:EntropySplit}) with $Z=\emptyset$ at the final and initial times and subtracting according to the difference convention of Eq.~(\ref{eq:conv_mui}) gives
\begin{IEEEeqnarray}{lCl}
        \Delta\shaent{A_{1:\Ngen}}
    =
    \sum_{j=1}^\Ngen \Delta\shaent{A_j}
    - \sum_{j=2}^\Ngen \Delta I(A_j;A_{1:j-1}).
    \label{eq:F18YY}
\end{IEEEeqnarray}

An analogous partition holds for the mutual information between the sequence and a further variable $\Rb$. Replacing the conditioning $\Rz$ by $(\Rb,\Rz)$ in Eq.~(\ref{eq:EntropySplit}) and subtracting the result from Eq.~(\ref{eq:EntropySplit}) itself turns each entropy difference into a mutual information by Eq.~(\ref{eq:SRARBSHARB_D1}), $\shaent{\Ra\mid\Rz}-\shaent{\Ra\mid\Rb,\Rz}=\muI{\Ra}{\Rb\mid\Rz}$. The recursion~(\ref{eq:def_interaction_cond}) similarly turns each difference of correlation terms into an interaction information, $\muI{A_j}{A_{1:j-1}\mid Z}-\muI{A_j}{A_{1:j-1}\mid B,Z}=\muI{A_j}{A_{1:j-1};B\mid Z}$. Thus,
\begin{IEEEeqnarray}{rCl}
        \muI{A_{1:\Ngen}}{\Rb\mid Z}
    =
    \sum_{j=1}^\Ngen \muI{A_j}{B\mid Z}
    - \sum_{j=2}^\Ngen I(A_j;A_{1:j-1};B\mid Z).
    \label{eq:XXHHI71}
\end{IEEEeqnarray}

\section{Set-Theoretic Shorthand for Shannon Information Quantities}
\label{ap:shorthand}

The derivations in Appendix~\ref{ap:derivation} involve long chains of identities between multivariate information quantities. We use the following shorthand to keep them manageable in the appendices. In the main text, it appears only in the restriction notation of Eq.~\eqref{eq:def_Rreb}.

First, subsystem indices stand for the corresponding random variables, and Shannon information quantities are abbreviated as
\begin{IEEEeqnarray}{lCl}
i\mid j := \shaent{\Xs{i}\mid \Xs{j}},
\\
\muInfo{i}{j\mid k}:=\muI{\Xs{i}}{\Xs{j}\mid\Xs{k}}.
\end{IEEEeqnarray}
The same applies to the index ranges of Sec.~\ref{sect:setting}: a bare token $\minusI$ or $\plusI$ stands for the joint variable $\Xs{\minusI}$ or $\Xs{\plusI}$, so that, e.g., $\muInfo{i}{\minusI} = \muI{\Xs{i}}{\Xs{\minusI}}$. We stress that $\prec$ and $\succ$ are part of these tokens and not relational operators; in particular, a conditioning bar followed by such a token, as in $\muInfo{i}{j\mid \plusI}$, simply denotes conditioning on $\Xs{\plusI}$.

For a compact notation of changes during the time evolution, a superscript $\sstar$ on a variable signifies that it has evolved to its final state. The prefix $\Delta$ takes the difference with respect to the quantity in which the starred variables are evaluated at the initial time:
\begin{IEEEeqnarray}{rCl}
\Delta \muI{A\M}{B\M\mid C\M} &:=& \muI{A\prm}{B\prm\mid C\prm} - \muI{A}{B\mid C},
\\
\Delta \muI{A\M}{B\M\mid C} &:=& \muI{A\prm}{B\prm\mid C} - \muI{A}{B\mid C},
\\
\Delta \muI{A\M}{B\mid C} &:=& \muI{A\prm}{B\mid C} - \muI{A}{B\mid C},
\\
\Delta \muI{A}{B\mid C\M} &:=& \muI{A}{B\mid C'} - \muI{A}{B\mid C}.
\end{IEEEeqnarray}
In the index shorthand, the $\Delta$ symbol is omitted; for example, $\muInfo{i\M}{j\M} := \Delta\muI{\Xs{i}\M}{\Xs{j}\M}$.

This shorthand rests on the fact that Shannon information quantities are equivalent to a signed measure on a set space \cite{yeung1991new,yeung2012first}. Random variables correspond to sets (regions of an information diagram), and the standard operators have set-theoretic analogues. The semicolon (`;') corresponds to set intersection ($\cap$), the vertical bar (`|') to set difference ($\setminus$), and the comma (`,') to set union ($\cup$). Manipulations of information quantities thereby reduce to elementary set operations. For instance, the distributive law holds:
\begin{IEEEeqnarray}{lCl}
\muInfo{\left(\muInfo{i}{j}+\muInfo{k}{l}\right)}{m} = \muInfo{i}{j}[m]+\muInfo{k}{l}[m],
\\
\left(\muInfo{i}{j}+\muInfo{k}{l}\right)\mid{m} = \muInfo{i}{j\mid m}+\muInfo{k}{l\mid m}.
\IEEEeqnarraynumspace
\end{IEEEeqnarray}
Recall from Sec.~\ref{sect:components} that $\Yeng{Z}$ denotes the set (region) corresponding to a random variable $Z$. If $\Yeng{A}\cap\Yeng{B}=\emptyset$, we refer to $A$ and $B$ as \textit{exclusive} components. The decomposition of an information quantity into a sum of such exclusive components is referred to as \textit{partitioning}.

Every expression in this shorthand is the value of the same signed measure evaluated on a set. An identity between two such expressions then holds as soon as each \emph{atom} of the information diagram (i.e., each minimal cell generated by the regions of the individual variables) receives the same coefficient on both sides. Several proofs in Appendix~\ref{ap:derivation} verify identities directly in this atom-wise manner. Such identities are insensitive to the values of the underlying measure and may therefore be intersected with arbitrary regions or applied to signed combinations of regions without further justification.

As a worked example, we verify the decomposition \eqref{eq:decIiMu_decomp} of the destroyed correlation used in Sec.~\ref{sect:indispensability}. In shorthand, the destroyed and newly created correlations of Eqs.~\eqref{eq:def_decIij} and \eqref{eq:def_incIij} read
\begin{IEEEeqnarray}{lCl}
    \decIiMu = -i\M\scl \minusI\M \scl i \scl \minusI,
    \qquad
    \incIiMu = i\M\scl \minusI\M \mid (i \scl \minusI).
    \nonumber
\end{IEEEeqnarray}
Using the identity $i'\scl \minusI' =(i',\minusI') - \minusI'\mid i' - i'\mid \minusI'$, we obtain
\begin{IEEEeqnarray}{lCl}
\decIiMu
&=& i\scl \minusI - i'\scl \minusI' \scl i\scl \minusI
\\
&=& i\scl \minusI - i\scl \minusI \scl(i',\minusI') + i\scl \minusI \scl \minusI'\mid i' + i\scl \minusI\scl i'\mid \minusI'
\\
&=&
i\scl \minusI \mid i', \minusI' + i\scl \minusI\scl \minusI' \mid i' + i\scl \minusI \scl i' \mid  \minusI'
\\
&=& \atamaOmegaIMuMu + \centerSeibunIMu + \bottomSeibunIMuMu,
\end{IEEEeqnarray}
which is Eq.~\eqref{eq:decIiMu_decomp}.\section{Derivation of the Main Inequality}
\label{ap:derivation}

We derive the identity \eqref{eq:199DD}, which underlies the ledger inequality \eqref{eq:158FF} of Sec.~\ref{sect:mainresult}. We use the shorthand of Appendix~\ref{ap:shorthand} throughout. The derivation has three stages. First, we record the screening property in the mutual-information form through which the causal structure enters the derivation (Proposition~\ref{prop:screening_mi}) and establish two elementary counting lemmas for regions of the information diagram (Propositions~\ref{prop:first_occ} and \ref{prop:redistribution}). Second, we use these lemmas to recast the right-hand side of Eq.~\eqref{eq:2ndlaw1} into the structural form \eqref{eq:154CC}. Third, we convert the structural form, term by term, into the information components of Sec.~\ref{sect:components}. At this stage, the effective collective parent region $\effPa$ of Eq.~\eqref{eq:103YY} enters through Proposition~\ref{prop:k_mid_k}.

\subsection{Counting lemmas}
\label{ap:counting_lemmas}

The causal structure of the dynamics enters the derivation through the following consequence of the screening property. Recall that $\Ra$ is independent of $\Rb$ conditioned on $\Rz$, written $\Ra\perp\Rb\mid\Rz$, when $P(\Ra,\Rb\mid\Rz)=P(\Ra\mid\Rz)\,P(\Rb\mid\Rz)$ \cite{yeung2012first}.

\begin{proposition}
    \label{prop:screening_mi}
    Let $\Ra$, $\Rb$, and $\Rz$ be collections of variables of the process. If the evolved state of $\Ra$ is screened from $\Rb$ by $(\Ra,\Rz)$, i.e., $\Ra\prm\perp\Rb\mid(\Ra,\Rz)$---as holds under Assumption~\ref{asm:local_mech} for $\Ra=\Xs{i}$, $\Rz=\DepX{i}$, and any collection $\Rb$ of the remaining variables, initial or final, by Lemma~\ref{lem:screening}(ii)---then
    \begin{IEEEeqnarray}{cCc}
        \MutualInfo{\Ra\prm}{\Rb\mid\Ra,\Rz}{}
        =0.
        \label{eq:IPCPDCE_D78}
    \end{IEEEeqnarray}
\end{proposition}
\begin{proof}
    The premise is the factorization $P(\Ra\prm,\Rb\mid\Ra,\Rz) = P(\Ra\prm\mid\Ra,\Rz)\,P(\Rb\mid\Ra,\Rz)$, under which the logarithm defining the conditional mutual information, $\langlee\ln\frac{P(\Ra\prm,\Rb\mid\Ra,\Rz)}{P(\Ra\prm\mid\Ra,\Rz)P(\Rb\mid\Ra,\Rz)}\ranglee$, vanishes identically.
\end{proof}

\begin{proposition}[first-occurrence partition]
    \label{prop:first_occ}
    For any finite list $D_1,\dots,D_r$ of regions of the information diagram,
    \begin{IEEEeqnarray}{lCl}
        \Sum{j=1}{r} D_j
        &=&
        \bigcup_{j=1}^{r} D_j
        + \Sum{j=1}{r} D_j\scl\biggl(\bigcup_{l<j}D_l\biggr),
        \label{eq:first_occ_sum}
        \\
        \bigcup_{j=1}^{r} D_j
        &=&
        \Sum{j=1}{r} D_j\mid\bigcup_{l<j}D_l,
        \label{eq:first_occ_disjoint}
    \end{IEEEeqnarray}
    where a sum of regions denotes the sum of their measures.
\end{proposition}
\begin{proof}
    Fix an atom $t$ contained in at least one of the regions and let $j_0(t)$ be the smallest index with $t\in D_{j_0(t)}$. In \eqref{eq:first_occ_disjoint}, $t$ lies in the piece $j=j_0(t)$ and in no other, and every atom of the union arises in this way. In \eqref{eq:first_occ_sum}, $t$ contributes once through the union and once through the correction term for every index $j>j_0(t)$ with $t\in D_j$, so its total coefficient on the right-hand side equals the number of indices $j$ with $t\in D_j$, which is its coefficient on the left-hand side. Both identities thus hold atom by atom.
\end{proof}

\begin{proposition}[redistribution]
    \label{prop:redistribution}
    For any region $A$ and any index $i$,
    \begin{IEEEeqnarray}{lCl}
        A\mid\plusI + \sum_{k=i+1}^{\Nx}k\scl A\mid\plusX{k}= A.
        \label{eq:166DD}
    \end{IEEEeqnarray}
    Consequently, for any regions $A_2,\dots,A_{\Nx}$,
    \begin{IEEEeqnarray}{lCl}
        \SumIN\left(
            A_i
            + \sum_{k=2}^{i-1}
            i\scl A_k
        \middle)\relright{|}\plusI
        = \SumIN A_i.
        \label{eq:174DD}
    \end{IEEEeqnarray}
\end{proposition}
\begin{proof}
    Splitting $A$ by the region of $\Xs{\plusI}$ gives $A = A\mid\plusI + A\scl\plusI$. Applying \eqref{eq:first_occ_disjoint} to the family $D_m=\Yeng{\Xs{m}}$, $m=\Nx,\Nx-1,\dots,i+1$, ordered by decreasing index, and intersecting every term with $A$,
    \begin{IEEEeqnarray}{lCl}
        A\scl\plusI = \Sum{k=i+1}{\Nx} k\scl A\mid\plusX{k},
        \label{eq:nu_partition}
    \end{IEEEeqnarray}
    which proves \eqref{eq:166DD}.
    For \eqref{eq:174DD}, fix $j\in\{2,\dots,\Nx\}$ and collect the terms containing $A_j$ on the left-hand side: they are $A_j\mid\plusX{j}$ (from $i=j$) and $k\scl A_j\mid\plusX{k}$ for $k=j+1,\dots,\Nx$ (from $i=k$), which sum to $A_j$ by \eqref{eq:166DD}.
\end{proof}

Both propositions hold atom by atom and depend linearly on each region involved. They therefore remain valid when the regions are replaced by finite signed combinations of regions, such as the $\Delta$-quantities of Appendix~\ref{ap:shorthand}; we use this extension below without further comment.

\subsection{Structural form of the right-hand side of Eq.~\eqref{eq:2ndlaw1}}
\label{ap:structural_form}

We recast the two constituents of the right-hand side of Eq.~\eqref{eq:2ndlaw1} in parallel structural forms.

\begin{proposition}
    The following identity holds:
    \begin{IEEEeqnarray}{lCl}
        \UsableNew{}
        &=& \Sum{i=2}{\Nx}
        \left(
            i\M\scl\minusI\scl\infPaI
        \right.
        \NonumberNewline
        &&\quad\left.
            + \sum_{k\in\minusI}i\scl k\M\scl\infPaK
        \middle)\relright{|}\plusI
        \IEEEeqnarraynumspace
        \label{eq:92YY}
    \end{IEEEeqnarray}
\end{proposition}
\begin{proof}
    Since $\Yeng{\infPaI}\subseteq\Yeng{\Xs{1:\Nx\setmin i}}$, the definition \eqref{eq:def_eta} takes the form
    \begin{IEEEeqnarray}{lCl}
        \UsableNew{i}
        &=& i\M\scl\left(1:\Nx\setmin i\right)\scl\infPaI.
        \nonumber
    \end{IEEEeqnarray}
    Splitting the region of $\Xs{1:\Nx\setmin i}$ into the part outside $\Yeng{\Xs{\plusI}}$ and the rest, and applying \eqref{eq:nu_partition} to the latter,
    \begin{IEEEeqnarray}{lCl}
        \UsableNew{i}
        &=& i\M\scl\minusI\scl\infPaI\mid\plusI
        + i\M\scl\plusI\scl\infPaI
        \\
        &=& i\M\scl\minusI\scl\infPaI\mid\plusI
        + \Sum{m=i+1}{\Nx} i\M\scl m\scl\infPaI\mid\plusX{m}.
    \end{IEEEeqnarray}
    We now sum over $i=1,\dots,\Nx$. In the double sum, each pair of subsystems contributes exactly once, with the star carried by the lower-index member; relabeling the pair $(i,m)$ as $(k,i)$,
    \begin{IEEEeqnarray}{lCl}
        \Sum{i=1}{\Nx}\Sum{m=i+1}{\Nx} i\M\scl m\scl\infPaI\mid\plusX{m}
        = \Sum{i=2}{\Nx}\sum_{k\in\minusI} i\scl k\M\scl\infPaK\mid\plusI.
    \end{IEEEeqnarray}
    Since the $i=1$ term of the diagonal sum vanishes ($\minusX{1}=\emptyset$), this proves \eqref{eq:92YY}.
\end{proof}

To reformulate $\delitot$, we define the following quantity:
\begin{IEEEeqnarray}{lCl}
    \turiBfull{A}{B} := \muInfo{A\M}{B}+ \muInfo{A}{B\M}-\muInfo{A\M}{B\M},
    \\
    \turiB{i}:= \turiBfull{i}{\minusI}.
    \label{eq:def_turiB}
\end{IEEEeqnarray}
This quantity expresses $\delitot$ in a form parallel to that of $\UsableNew{}$.
\begin{proposition}
    The following identity holds:
    \begin{IEEEeqnarray}{lCl}
        \delitot
        =
        \Sum{i=2}{\Nx}
        \left(
            i\M\scl\minusI
            +\sum_{k\in\minusI}i\scl k\M
            -\turiB{i}
        \middle)\relright{|}\plusI
        \label{eq:77YZ}
    \end{IEEEeqnarray}
\end{proposition}
\begin{proof}
    By definition, $\delitot=\Sum{i=2}{\Nx}\delIP{i}$ with $\delIP{i}=i\M\scl\minusI\M$.
    Applying the partition \eqref{eq:XXHHI71}---with $Z=\emptyset$ and $B=\Xs{i}$, the block entering as the first argument via the symmetry of the mutual information---to $\muI{\Xs{i}}{\Xs{\minusI}\prm}$ and to $\muI{\Xs{i}}{\Xs{\minusI}}$, and subtracting,
    \begin{IEEEeqnarray}{lCl}
        i\scl\minusI\M
        &=&
        \sum_{k\in\minusI}\left( i\scl k\M - i\scl\delIP{k} \right),
        \label{eq:chain_mu_star}
    \end{IEEEeqnarray}
    the interaction terms pairing into $i\scl k\prm\scl\minusK\prm - i\scl k\scl\minusK = i\scl\delIP{k}$ (the interaction sum in \eqref{eq:XXHHI71} starts at the second entry; the $k=1$ term added here vanishes, $\delIP{1}=0$).
    On the other hand, the definition \eqref{eq:def_turiB} reads $\delIP{i} = i\M\scl\minusI + i\scl\minusI\M - \turiB{i}$. Substituting \eqref{eq:chain_mu_star} into it and moving the correction terms to the left-hand side,
    \begin{IEEEeqnarray}{lCl}
        \delIP{i} + \sum_{k\in\minusI} i\scl\delIP{k}
        =
        i\M\scl\minusI
        + \sum_{k\in\minusI} i\scl k\M
        - \turiB{i}.
        \label{eq:per_i_structural}
    \end{IEEEeqnarray}
    Finally, applying the redistribution identity \eqref{eq:174DD} with $A_i = \delIP{i}$ (note that the $k=1$ term in \eqref{eq:per_i_structural} vanishes because $\delIP{1}=0$) and inserting \eqref{eq:per_i_structural},
    \begin{IEEEeqnarray}{lCl}
        \delitot
        = \SumIN\left(\delIP{i} + \sum_{k=2}^{i-1} i\scl\delIP{k}\middle)\relright{|}\plusI
        = \SumIN\left(
            i\M\scl\minusI
            +\sum_{k\in\minusI}i\scl k\M
            -\turiB{i}
        \middle)\relright{|}\plusI.
        \nonumber
    \end{IEEEeqnarray}
\end{proof}

The preceding results give the following expression for the difference $\UsableNew{} - \delitot$:
\begin{proposition}
    \begin{IEEEeqnarray}{lCl}
        \UsableNew{} - \delitot
        &=&
        -\Sum{i=2}{\Nx}\left({\muInfo{i\M}{\minusI\mid\infPaI}+\sum_{k\in\minusI}\muInfo{i}{k\M\mid\infPa{k}}}-\turiB{i}\middle)\relright|\plusI
        \label{eq:154CC}
    \end{IEEEeqnarray}
\end{proposition}
\begin{proof}
    Subtracting \eqref{eq:77YZ} from \eqref{eq:92YY},
    \begin{IEEEeqnarray}{lCl}
        \UsableNew{} - \delitot
        =
        -\Sum{i=2}{\Nx}
        \left(
            i\M\scl\minusI
            - i\M\scl\minusI\scl\infPaI
            -\turiB{i}
            \right.
            \NonumberNewlineQQ\qquad
            \left.
            +\sum_{k\in\minusI}\left(
            i\scl k\M
            - i\scl k\M\scl\infPaK
            \right)
        \middle)\relright{|}\plusI
        \label{eq:156CC}
        \IEEEeqnarraynumspace
    \end{IEEEeqnarray}
    Since $ i\M\scl\minusI - i\M\scl\minusI\scl\infPaI = i\M\scl\minusI\mid\infPaI$ and $ i\scl k\M - i\scl k\M\scl\infPaK = i\scl k\M\mid\infPaK$, Eq.~\eqref{eq:156CC} is equivalent to Eq.~\eqref{eq:154CC}.
\end{proof}

\subsection{Conversion into information components}
\label{ap:conversion}

For general arguments $u$, $v$, and $w$, the information components defined in Sec.~\ref{sect:components} take the following form in the present shorthand:
\begin{IEEEeqnarray}{lCl}
\atamaOmega{u}{v}{w}
= \muInfo{u}{v\mid(u,w)\prm},
\label{eq:atamaOmega_gen}
\\
\centerSeibun{u,v} = u\scl v\scl v' \mid u',
\label{eq:centerSeibun_gen}
\\
\bottomSeibun{u,v,w}
= u\scl v \scl u'\mid w',
\qquad
\bottomSeibun{u,v} := \bottomSeibun{u,v,v},
\label{eq:bottomSeibun_gen}
\\
\botObsSeibunUVW = \bottomSeibun{u,v}\scl w'\mid w
= u\scl v\scl u'\scl w'\mid v',w,
\label{eq:botObs_gen}
\\
\memotouSeibunL{u,v} = \muInfo{u}{v'\mid u',v},
\\
\memotouSeibunR{u,v}
= u'\scl v\mid u,v',
\\
\memotouSeibun{u,v} = \memotouSeibunL{u,v} + \memotouSeibunR{u,v},
\\
\memotouSeibunLL{u,v,w} = u\scl v\scl w'\mid (u',v', w),
\label{eq:memotouLL_gen}
\end{IEEEeqnarray}
with the abbreviations $\atamaOmegaIMuMu=\atamaOmega{i}{\minusI}{\minusI}$, $\centerSeibunIMu=\centerSeibun{i,\minusI}$, $\bottomSeibunIMuMu=\bottomSeibun{i,\minusI}$, and $\memotouSeibunIMu=\memotouSeibun{i,\minusI}$.

The causal structure of the dynamics has two further consequences.

\begin{proposition}
    The following equalities hold:
    \begin{IEEEeqnarray}{lCl}
        \muInfo{i\M}{k\mid\infPaI}
        =
        -\left(
        \atamaOmegaIKK + \centerSeibunIK
        \middle)\relright|\infPaI.
        \label{eq:g1313xx}
        \\
        \muInfo{i}{k\M\mid\infPaK} =
        -(\atamaOmega{i}{k}{k} + \bottomSeibun{i,k})\mid\infPaK.
        \label{eq:propH252CC}
    \end{IEEEeqnarray}
\end{proposition}

\begin{proof}
    First, we prove Eq.~\eqref{eq:g1313xx}.
    From the definitions \eqref{eq:atamaOmega_gen} and \eqref{eq:centerSeibun_gen}, we have
    \begin{IEEEeqnarray}{lCl}
        \text{RHS} &=& \muInfo{i\M}{i}[k\mid\infPaI]
        \\
        &=& \left(\muInfo{i\prm}{i}[k] - \muInfo{i}{k}\middle)\relright|\infPaI.
    \end{IEEEeqnarray}
    On the other hand,
    \begin{IEEEeqnarray}{lCl}
        \text{LHS} &=& \muInfo{i\prm}{k\mid\infPaI}-\muInfo{i}{k\mid\infPaI}
        \\
        &=&
        \muInfo{i\prm}{i}[k\mid\infPaI]
        + \muInfo{i\prm}{k\mid i, \infPaI}
        -\muInfo{i}{k\mid\infPaI}.
    \end{IEEEeqnarray}
    Given that it is conditioned on $\infPaI$, we can apply Eq.~\eqref{eq:IPCPDCE_D78}, yielding
    \begin{IEEEeqnarray}{lCl}
        \muInfo{i\prm}{k\mid i, \infPaI} = 0
    \end{IEEEeqnarray}
    This demonstrates that the left-hand side and the right-hand side are equal.
    Next, we prove Eq.~\eqref{eq:propH252CC}.
    From the definitions \eqref{eq:atamaOmega_gen}, \eqref{eq:centerSeibun_gen}, and \eqref{eq:bottomSeibun_gen}, we have:
    \begin{IEEEeqnarray}{lCl}
        \atamaOmega{u}{v}{v}=\atamaOmega{v}{u}{u},
        \label{eq:atamaSymmetry}
        \\
        \centerSeibun{u,v} = \bottomSeibun{v,u}.
        \label{eq:centerBottomEqual}
    \end{IEEEeqnarray}
    By swapping the indices $i$ and $k$ in Eq.~\eqref{eq:g1313xx}, we obtain:
    \begin{IEEEeqnarray}{lCl}
        \muInfo{i}{k\M\mid\infPaK}
        &=&
        -(\atamaOmega{k}{i}{i} + \centerSeibun{k,i})\mid\infPaK.
    \end{IEEEeqnarray}
    Substituting Eqs.~\eqref{eq:atamaSymmetry} and \eqref{eq:centerBottomEqual} into this expression yields Eq.~\eqref{eq:propH252CC}.
\end{proof}

Two extensions of Eqs.~\eqref{eq:g1313xx} and \eqref{eq:propH252CC} follow directly and will be used repeatedly. An arbitrary collection $W$ of variables of the process may be appended to every conditioning, and the single subsystem $k$ may be replaced by any collection of subsystems disjoint from $i$ (in particular, by the block $\Xs{\minusI}$). The derivations combine region identities with vanishings supplied by Eq.~\eqref{eq:IPCPDCE_D78}. The region identities hold atom by atom and hence survive both changes. The vanishings extend as well. The collection $\Rb$ in Proposition~\ref{prop:screening_mi} absorbs both the enlarged second argument and the appended variables, while $\MutualInfo{\Xs{i}\prm}{(\Rb,W)\mid\Xs{i},\DepX{i}}{}=0$ splits by the chain rule \eqref{eq:chain_MI6} into two nonnegative conditional mutual informations, each of which therefore vanishes. In particular, $\MutualInfo{\Xs{i}\prm}{\Rb\mid\Xs{i},\DepX{i},W}{}=0$. The same reasoning applies to the identities derived from Eqs.~\eqref{eq:g1313xx} and \eqref{eq:propH252CC} below, in particular Eqs.~\eqref{eq:172DD} and \eqref{eq:162DD}; we invoke these extended forms without further comment.

For the following manipulations, we use the complement of $\effPa$ within the block region, namely the union of the per-subsystem blind spots:
\begin{IEEEeqnarray}{lCl}
    \blind := \Yeng{\Xs{\minusI}}\setminus\effPa
    = \bigcup_{n\in\minusI} D_n,
    \qquad
    D_n := \Yeng{\Xs{n}}\setminus\Yeng{\infPa{n}}.
    \label{eq:def_blind}
\end{IEEEeqnarray}
For any quantity $K$ whose region is contained in $\Yeng{\Xs{\minusI}}$ (as is the case for every $\effPa$-conditioned component appearing below), the two regions are complementary within the region of $K$. Thus,
\begin{IEEEeqnarray}{lCl}
    K\mid\effPa = K\scl\blind;
    \label{eq:blind_translation}
\end{IEEEeqnarray}
We therefore switch freely between the two forms. The region $\effPa$ enters the derivation through the following instance of the first-occurrence partition.

\begin{proposition}
    \label{prop:k_mid_k}
    For every $i$,
    \begin{IEEEeqnarray}{lCl}
        \sum_{k\in\minusI}k\mid\infPaK
        &=&
        \minusI\mid\effPa
        + \sum_{k\in\minusI}k\scl\minusK\mid(\infPaK, \effPa[k]),
        \label{eq:101YY}
    \end{IEEEeqnarray}
    and, for any region $R$,
    \begin{IEEEeqnarray}{lCl}
        \sum_{k\in\minusI}R\scl k\mid\infPaK
        &=&
        R\scl\minusI\mid\effPa
        + \sum_{k\in\minusI}R\scl k\scl\minusK\mid(\infPaK, \effPa[k]).
        \label{eq:101YYmult}
    \end{IEEEeqnarray}
\end{proposition}
\begin{proof}
    The $k$-th summand on the left-hand side of \eqref{eq:101YY} is the measure of $D_k$ while, by \eqref{eq:def_blind} and \eqref{eq:blind_translation}, the two terms on the right-hand side are the measures of $\blind=\bigcup_{k\in\minusI}D_k$ and of $D_k\scl\blind[k]$ with $\blind[k]=\bigcup_{l<k}D_l$. The identity is therefore the first-occurrence partition \eqref{eq:first_occ_sum} applied to the family $D_1,\dots,D_{i-1}$. Since \eqref{eq:first_occ_sum} holds atom by atom, every term may be intersected with $R$, which yields \eqref{eq:101YYmult}.
\end{proof}

\begin{proposition}
    The following identity holds:
    \begin{IEEEeqnarray}{lCl}
        \sum_{k\in\minusI}\atamaOmegaIKMu\mid\infPaK
        &=&
        \atamaOmegaIMuMu\mid\effPa
        + \sum_{k\in\minusI}
        \atamaOmega{i}{k\scl\minusK}{\minusI}\mid(\infPaK, \effPa[k]).
        \label{eq:219CC}
    \end{IEEEeqnarray}
\end{proposition}
\begin{proof}
    Apply \eqref{eq:101YYmult} with $R = i\mid (i\prm,\minusI\prm)$. The left-hand side becomes $\sum_{k\in\minusI} i\scl k\mid(i\prm,\minusI\prm,\infPaK) = \sum_{k\in\minusI}\atamaOmegaIKMu\mid\infPaK$; the first term on the right-hand side becomes $i\scl\minusI\mid(i\prm,\minusI\prm,\effPa) = \atamaOmegaIMuMu\mid\effPa$; and the $k$-th correction term becomes $i\scl k\scl\minusK\mid(i\prm,\minusI\prm,\infPaK,\effPa[k]) = \atamaOmega{i}{k\scl\minusK}{\minusI}\mid(\infPaK,\effPa[k])$.
\end{proof}

\begin{proposition}
    \begin{IEEEeqnarray}{lCl}
        - \sum_{k\in\minusI}
        i\scl k\M\mid\infPaK
        &=&
        - \sum_{k\in\minusI}
        \MuPrimeSeibunI
        + \atamaOmegaIMuMu\mid\effPa
        \NonumberNewline
        &&+ \sum_{k\in\minusI}
        \left[
            \atamaOmega{i}{k\scl\minusK}{\minusI}\mid(\infPaK,\effPa[k])
            + \bottomSeibunIKMu\mid\infPaK
        \right]
        \label{eq:prop222CCXX}
    \end{IEEEeqnarray}
\end{proposition}
\begin{proof}
    Appending the variables $\Xs{\minusI}\prm$ to the conditioning of Eq.~\eqref{eq:propH252CC}, as licensed below that equation, gives
    \begin{IEEEeqnarray}{lCl}
        -i\scl k\M\mid\infPaK,\minusI'
        = \atamaOmegaIKMu\mid\infPaK + \bottomSeibunIKMu\mid\infPaK,
        \nonumber
    \end{IEEEeqnarray}
    since the added primes absorb $\Xs{k}\prm$ and enlarge the conditioning of both components to $\Xs{\minusI}\prm$.
    Summing over $k$ and inserting \eqref{eq:219CC},
    \begin{IEEEeqnarray}{lCl}
        -\sum_{k\in\minusI}i\scl k\M\mid\infPaK,\minusI'
        &=& \atamaOmegaIMuMu\mid\effPa
        \NonumberNewline
        &&+ \sum_{k\in\minusI}
        \left[
        \atamaOmega{i}{k\scl\minusK}{\minusI}\mid(\infPaK,\effPa[k])
        + \bottomSeibunIKMu\mid\infPaK
        \right]
        \label{eq:153DD}
    \end{IEEEeqnarray}
    On the other hand, decomposing $i\scl k\M\mid\infPaK$ with respect to $\Xs{\minusI}\prm$ yields
    \begin{IEEEeqnarray}{lCl}
        \sum_{k\in\minusI}i\scl k\M\mid\infPaK = \sum_{k\in\minusI}\left(\MuPrimeSeibunI + i\scl k\M\mid\infPaK,\minusI'\right).
        \nonumber
    \end{IEEEeqnarray}
    Substituting Eq.~\eqref{eq:153DD} into the right-hand side of this identity proves Eq.~\eqref{eq:prop222CCXX}.
\end{proof}

\begin{proposition}
    \begin{IEEEeqnarray}{lCl}
        \sum_{k\in\minusI}\bottomSeibunIKMu\mid\infPaK
        &=&
        \bottomSeibunIMuMu\mid\effPa
        \NonumberNewline
        &&+ \sum_{k\in\minusI}
        i\scl i\prm
        \scl
        \atamaOmega{k}{\minusK}{\minusK}\mid\left(\left(k+1:i-1\right)\prm, \infPaK, \effPa[k]\right)
        \IEEEeqnarraynumspace
        \label{eq:154DD}
    \end{IEEEeqnarray}
\end{proposition}
\begin{proof}
    Since $\bottomSeibunIKMu\mid\infPaK = (i\scl i\prm\mid\minusI\prm)\scl k\mid\infPaK$, applying \eqref{eq:101YYmult} with $R=i\scl i\prm\mid\minusI\prm$ gives
    \begin{IEEEeqnarray}{lCl}
        \sum_{k\in\minusI}\bottomSeibunIKMu\mid\infPaK
        &=& i\scl i\prm\scl\minusI\mid(\minusI\prm,\effPa)
        \NonumberNewline
        &&+ \sum_{k\in\minusI} i\scl i\prm\scl k\scl\minusK\mid(\minusI\prm,\infPaK,\effPa[k]).
        \nonumber
    \end{IEEEeqnarray}
    The first term equals $\bottomSeibunIMuMu\mid\effPa$. For the $k$-th correction term, write
    \begin{IEEEeqnarray}{rCl}
        \minusI\prm
        &=& \left(k\prm,\minusK\prm,(k+1:i-1)\prm\right),
        \NonumberNewline
        k\scl\minusK\mid(k\prm,\minusK\prm)
        &=& \atamaOmega{k}{\minusK}{\minusK}.
        \nonumber
    \end{IEEEeqnarray}
    The correction term therefore becomes
    \begin{IEEEeqnarray}{lCl}
        i\scl i\prm\scl k\scl\minusK\mid(\minusI\prm,\infPaK,\effPa[k])
        \NonumberNewline
        = i\scl i\prm\scl\atamaOmega{k}{\minusK}{\minusK}\mid\left((k+1:i-1)\prm,\infPaK,\effPa[k]\right).
        \nonumber
    \end{IEEEeqnarray}
\end{proof}

\begin{proposition}
    For $1<k<i$,
    \begin{IEEEeqnarray}{lCl}
        \atamaOmega{i}{k\scl\minusK}{\minusI}\mid(\infPaK,\effPa[k])
        = (i\mid i\prm)
        \scl
        \atamaOmega{k}{\minusK}{\minusK}\mid\left(\left(k+1:i-1\right)\prm,\infPaK,\effPa[k]\right).
        \label{eq:155DD}
    \end{IEEEeqnarray}
\end{proposition}
\begin{proof}
    By definition, the left-hand side equals $i\scl k\scl\minusK\mid\left(i\prm,\minusI\prm,\infPaK,\effPa[k]\right)$. Expanding $\minusI\prm=\left(k\prm,\minusK\prm,(k+1:i-1)\prm\right)$ as in the previous proof yields the right-hand side.
\end{proof}

The last of Eqs.~\eqref{eq:prop222CCXX}, \eqref{eq:154DD}, and \eqref{eq:155DD} carries the restriction $1<k<i$, but this is immaterial because both sides vanish for $k=1$ ($\minusX{1}=\emptyset$). Splitting the factor $i$ with respect to $i\prm$ in the combined correction terms then gives
\begin{IEEEeqnarray}{lCl}
    - \sum_{k\in\minusI}
    i\scl k\M\mid\infPaK
    &=&
    - \sum_{k\in\minusI}
    \MuPrimeSeibunI
    + \atamaOmegaIMuMu\mid\effPa
    + \bottomSeibunIMuMu\mid\effPa
    \NonumberNewlineQQ
    &&
    + \SumKinMu i \scl
    \atamaOmega{k}{\minusK}{\minusK}\mid\left(\left(k+1:i-1\right)\prm, \infPaK, \effPa[k]\right)
    \IEEEeqnarraynumspace
    \label{eq:157DD}
\end{IEEEeqnarray}

\begin{proposition}
    \begin{IEEEeqnarray}{lCl}
        i\scl k\scl \minusK\prm\mid (k\prm,\minusK,\infPaK)
        + i\scl k\prm\scl \minusK\mid (k,\minusK\prm,\effPa[k])
        \NonumberNewlineQQ
        = \botObsSeibunIKMuk\mid\infPaK
        + \botObsSeibunIMukK\mid\effPa[k]
        + \memotouSeibunLL{i,k,\minusK}\mid\infPaK
        + \memotouSeibunLL{i,\minusK,k}\mid\effPa[k]
    \label{eq:168DD}
    \end{IEEEeqnarray}
\end{proposition}
\begin{proof}
    Decomposing each term on the left-hand side with respect to $i\prm$ gives:
    \begin{IEEEeqnarray}{lCl}
        i\scl k\scl \minusK\prm\mid (k\prm,\minusK,\infPaK)
        + i\scl k\prm\scl \minusK\mid (k,\minusK\prm,\effPa[k])
        \NonumberNewlineQQ
        =
        i\scl i\prm\scl k\scl \minusK\prm\mid (k\prm,\minusK,\infPaK)
        + i\scl i\prm\scl k\prm\scl \minusK\mid (k,\minusK\prm,\effPa[k])
        \NonumberNewlineQQ\quad
        + i\scl k\scl \minusK\prm\mid (i\prm, k\prm,\minusK,\infPaK)
        + i\scl k\prm\scl \minusK\mid (i\prm, k,\minusK\prm,\effPa[k])
        \nonumber
    \end{IEEEeqnarray}
    From the definitions in Eqs.~\eqref{eq:botObs_gen} and \eqref{eq:memotouLL_gen}, the four terms on the right-hand side equal, in order, $\botObsSeibunIKMuk\mid\infPaK$, $\botObsSeibunIMukK\mid\effPa[k]$, $\memotouSeibunLL{i,k,\minusK}\mid\infPaK$, and $\memotouSeibunLL{i,\minusK,k}\mid\effPa[k]$.
\end{proof}

\begin{proposition}
    \begin{IEEEeqnarray}{lCl}
        -\SumKinMu
        \MuPrimeSeibunI
        &=&
        \SumKinMu
        \Bigg[
        i\scl\left\{
            \bottomSeibunKMuMu\mid\effPa[k]
            + \centerSeibunKMu\mid\infPaK
            \right.
            \NonumberNewline
            &&\qquad\quad
            \left.
            + \atamaOmega{k}{\minusK}{\minusK}\scl\left(k+1:i-1\right)\prm\mid(\infPaK,\effPa[k])
        \right\}
        \NonumberNewline
        &&+ \botObsSeibunIKMuk\mid\infPaK
        + \botObsSeibunIMukK\mid\effPa[k]
        \NonumberNewline
        &&+ \memotouSeibunLL{i,k,\minusK}\mid\infPaK
        + \memotouSeibunLL{i,\minusK,k}\mid\effPa[k]
        \Bigg]
        \IEEEeqnarraynumspace
        \label{eq:159DD}
    \end{IEEEeqnarray}
\end{proposition}
\begin{proof}
    We first transform a single summand:
    \begin{IEEEeqnarray}{lCl}
        -\MuPrimeSeibunI
        &=& i\scl k\scl\minusI\prm\mid\infPaK - i\scl k\prm\scl\minusI\prm\mid\infPaK
        \\
        &=&
        \left(
            i\scl k\scl k\prm\mid\infPaK + i\scl k\scl(\minusI\setmin k)\prm\mid k\prm,\infPaK
        \right)
        \NonumberNewline
        &&- \left(
            i\scl k\scl k\prm\mid\infPaK + i\scl k\prm\mid k,\infPaK
        \right)
        \IEEEeqnarraynumspace
        \\
        &=& i\scl k\scl(\minusI\setmin k)\prm\mid k\prm,\infPaK.
        \nonumber
    \end{IEEEeqnarray}
    Here, the second equality splits both terms by $\Yeng{\Xs{k}\prm}\subseteq\Yeng{\Xs{\minusI}\prm}$, and the third uses $i\scl k\prm\mid (k,\infPaK)=0$, which follows from Eq.~\eqref{eq:IPCPDCE_D78}, the final state $\Xs{k}\prm$ entering as the first argument via the symmetry of the mutual information.
    Decomposing the region of $\Xs{\minusI\setmin k}\prm$ into $\Yeng{\Xs{\minusK}\prm}$ and the rest,
    \begin{IEEEeqnarray}{lCl}
        -\SumKinMu\MuPrimeSeibunI
        &=&
        \SumKinMu \left(
            i\scl k\scl \minusK\prm\mid (k\prm,\infPaK)
            \right.
            \NonumberNewline
            &&\qquad\quad
            \left.
            + i\scl k\scl (k+1:i-1)\prm\mid (k\prm,\minusK\prm,\infPaK)
        \right).
        \label{eq:172DD}
        \IEEEeqnarraynumspace
    \end{IEEEeqnarray}
    The first summand is split by $\Yeng{\Xs{\minusK}}$:
    \begin{IEEEeqnarray}{lCl}
        i\scl k\scl \minusK\prm\mid (k\prm,\infPaK)
        =
        i\scl \centerSeibun{k,\minusK}\mid\infPaK
        +
        i\scl k\scl \minusK\prm\mid (k\prm,\minusK, \infPaK).
        \label{eq:173DD}
    \end{IEEEeqnarray}
    For the second summand of \eqref{eq:172DD}, note that $(k\prm,\minusK\prm)=(1:k)\prm$ and apply the first-occurrence partition \eqref{eq:first_occ_disjoint} to the primed regions $D_j=\Yeng{\Xs{j}\prm}$, $j=k+1,\dots,i-1$, in increasing order; then reorganize the double sum and apply \eqref{eq:101YYmult} with $R=i\scl j\prm\mid\minusJ\prm$:
    \begin{IEEEeqnarray}{lCl}
        \SumKinMu
            i\scl k\scl (k+1:i-1)\prm\mid ((1:k)\prm,\infPaK)
        \NonumberNewlineQQ
        =
        \SumKinMu\Sum{j=k+1}{i-1}
        i\scl k\scl j\prm\mid (\minusJ\prm, \infPaK)
        \\
        \qquad
        =
        \Sum{j=2}{i-1}\sum_{k\in\minusJ}
        \left(i\scl j\prm\mid\minusJ\prm\right)\scl k\mid\infPaK
        \\
        \qquad
        \EqualText{\eqref{eq:101YYmult}}
        \Sum{j=2}{i-1}
        i\scl j\prm\scl\minusJ\mid(\minusJ\prm,\effPa[j])
        \NonumberNewlineQQ\qquad\qquad
        +\Sum{j=2}{i-1}\sum_{k\in\minusJ}
        i\scl j\prm\scl k\scl\minusK\mid(\minusJ\prm,\infPaK,\effPa[k]).
        \IEEEeqnarraynumspace
        \label{eq:181DD}
    \end{IEEEeqnarray}
    The first sum of \eqref{eq:181DD} is split by $\Yeng{\Xs{j}}$:
    \begin{IEEEeqnarray}{lCl}
        i\scl j\prm\scl\minusJ\mid(\minusJ\prm,\effPa[j])
        =
        i\scl\bottomSeibun{j,\minusJ,\minusJ}\mid\effPa[j]
        + i\scl j\prm\scl \minusJ\mid (j,\minusJ\prm,\effPa[j]).
        \label{eq:181DDfirst}
    \end{IEEEeqnarray}
    In the second sum of \eqref{eq:181DD}, we revert the order of summation and reassemble the primed block by \eqref{eq:first_occ_disjoint}, now applied within $i\scl k\scl\minusK\mid(\infPaK,\effPa[k])$:
    \begin{IEEEeqnarray}{lCl}
        \Sum{j=2}{i-1}\sum_{k\in\minusJ}
        i\scl j\prm\scl k\scl\minusK\mid(\minusJ\prm,\infPaK,\effPa[k])
        \NonumberNewlineQQ
        =
        \SumKinMu\Sum{j=k+1}{i-1}
        i\scl k\scl\minusK\scl j\prm\mid(\minusJ\prm,\infPaK,\effPa[k])
        \\
        \qquad
        =
        \SumKinMu
        i\scl k\scl\minusK\scl (k+1:i-1)\prm\mid((1:k)\prm,\infPaK,\effPa[k])
        \IEEEeqnarraynumspace
        \\
        \qquad
        =
        \SumKinMu
        i\scl \atamaOmega{k}{\minusK}{\minusK}\scl (k+1:i-1)\prm\mid(\infPaK,\effPa[k]),
        \label{eq:181DDsecond}
    \end{IEEEeqnarray}
    where the last equality uses $(1:k)\prm=(k\prm,\minusK\prm)$ and $k\scl\minusK\mid(k\prm,\minusK\prm)=\atamaOmega{k}{\minusK}{\minusK}$.
    Substituting \eqref{eq:181DDfirst} and \eqref{eq:181DDsecond} into \eqref{eq:181DD}, and then \eqref{eq:181DD} and \eqref{eq:173DD} into \eqref{eq:172DD} (relabeling $j\to k$ in the diagonal terms), we obtain
    \begin{IEEEeqnarray}{lCl}
        -\SumKinMu\MuPrimeSeibunI
        &=&
        \SumKinMu
        i\scl\left\{
            \centerSeibun{k,\minusK}\mid\infPaK
            + \bottomSeibun{k,\minusK,\minusK}\mid\effPa[k]
            \right.
            \NonumberNewline
            &&\qquad\quad
            \left.
            + \atamaOmega{k}{\minusK}{\minusK}\scl(k+1:i-1)\prm\mid(\infPaK,\effPa[k])
        \right\}
        \NonumberNewlineQQ
        &&+\SumKinMu
        \left(
            i\scl k\scl \minusK\prm\mid (k\prm,\minusK, \infPaK)
            \right.
            \NonumberNewline
            &&\qquad\quad
            \left.
            + i\scl k\prm\scl \minusK\mid (k,\minusK\prm,\effPa[k])
        \right).
        \IEEEeqnarraynumspace
        \nonumber
    \end{IEEEeqnarray}
    Substituting Eq.~\eqref{eq:168DD} into the second line yields Eq.~\eqref{eq:159DD}.
\end{proof}

\begin{proposition}
    \begin{IEEEeqnarray}{lCl}
        - \sum_{k\in\minusI}
        i\scl k\M\mid\infPaK
        &=&
        \atamaOmegaIMuMu\mid\effPa
        + \bottomSeibunIMuMu\mid\effPa
        \NonumberNewline
        &&+ \sum_{k=2}^{i-1}\left(
            \botObsSeibunIKMuk\mid\infPaK
            + \botObsSeibunIMukK\mid\effPa[k]
            \right.
            \NonumberNewline
            &&\qquad\quad
            \left.
            + \memotouSeibunLL{i,k,\minusK}\mid\infPaK
            + \memotouSeibunLL{i,\minusK,k}\mid\effPa[k]
        \right)
        \NonumberNewlineQQ
        &&
        + \sum_{k=2}^{i-1}
        i\scl\left\{
            \atamaOmega{k}{\minusK}{\minusK}\mid(\infPaK,\effPa[k])
            + \centerSeibunKMu\mid\infPaK
            \right.
            \NonumberNewline
            &&\qquad\quad
            \left.
            + \bottomSeibunKMuMu\mid\effPa[k]
        \right\}
        \label{eq:162DD}
    \end{IEEEeqnarray}
\end{proposition}
\begin{proof}
    Every term on the right-hand side of \eqref{eq:159DD} vanishes for $k=1$, so the sums there may be restricted to $k\geq2$. Substituting \eqref{eq:159DD} into \eqref{eq:157DD} and combining
    \begin{IEEEeqnarray}{lCl}
        \atamaOmega{k}{\minusK}{\minusK}\scl(k+1:i-1)\prm\mid(\infPaK,\effPa[k])
        \NonumberNewlineQQ
        + \atamaOmega{k}{\minusK}{\minusK}\mid\left((k+1:i-1)\prm,\infPaK,\effPa[k]\right)
        = \atamaOmega{k}{\minusK}{\minusK}\mid(\infPaK,\effPa[k]),
        \nonumber
    \end{IEEEeqnarray}
    an instance of $X\scl Y + X\mid Y = X$, yields \eqref{eq:162DD}.
\end{proof}

\subsection{Proof of the main identity}
\label{ap:final_assembly}

The last ingredient expresses $\turiB{i}$ through the components of Sec.~\ref{sect:components}.

\begin{proposition}
    \begin{IEEEeqnarray}{lCl}
        \turiB{i} = \memotouSeibunIMu - \atamaOmegaIMuMu - i'\scl\minusI'\mid i,\minusI.
        \label{eq:141FF}
    \end{IEEEeqnarray}
\end{proposition}
\begin{proof}
    By definition, we have
    \begin{IEEEeqnarray}{lCl}
        \turiB{i} = i\M\scl\minusI + i\scl\minusI\M - i\M\scl\minusI\M.
        \label{eq:142FF}
    \end{IEEEeqnarray}
    Note that
    \begin{IEEEeqnarray}{lCl}
        i\M\scl\minusI &=& i'\scl\minusI - i\scl\minusI
        \\
        &=& i\scl\minusI\scl i' + i'\scl\minusI\mid i
        - i\scl\minusI\mid i' - i\scl\minusI\scl i'
        \\
        &=& i'\scl\minusI\mid i - i\scl\minusI\mid i'.
    \end{IEEEeqnarray}
    Furthermore,
    \begin{IEEEeqnarray}{lCl}
        i\scl\minusI\mid i' &=& i\scl\minusI\mid i', \minusI' + i\scl\minusI\scl\minusI'\mid i'
        \\
        &=& \atamaOmegaIMuMu + \centerSeibunIMu
    \end{IEEEeqnarray}
    and
    \begin{IEEEeqnarray}{lCl}
        i'\scl\minusI\mid i &=& i'\scl\minusI\mid i, \minusI' + i'\scl\minusI\scl\minusI'\mid i
        \\
        &=&
        \memotouSeibunRIMu + i'\scl\minusI\scl\minusI'\mid i.
    \end{IEEEeqnarray}
    It follows that
    \begin{IEEEeqnarray}{lCl}
        i\M\scl\minusI = -\atamaOmegaIMuMu - \centerSeibunIMu + \memotouSeibunRIMu + i'\scl\minusI\scl\minusI'\mid i.
        \label{eq:150FF}
    \end{IEEEeqnarray}
    Similarly, we find
    \begin{IEEEeqnarray}{lCl}
        i\scl\minusI\M &=& -\atamaOmegaIMuMu - \bottomSeibunIMuMu + \memotouSeibunLIMu + i\scl i' \scl\minusI'\mid\minusI,
        \label{eq:151FF}
        \\
        i\M\scl\minusI\M &=& - \atamaOmegaIMuMu - \centerSeibunIMu - \bottomSeibunIMuMu + i'\scl\minusI\scl\minusI'\mid i
        \NonumberNewline
        &&+ i\scl i' \scl\minusI'\mid\minusI + i'\scl\minusI'\mid i,\minusI.
        \label{eq:152FF}
    \end{IEEEeqnarray}
    Substituting \eqref{eq:150FF}, \eqref{eq:151FF}, and \eqref{eq:152FF} into \eqref{eq:142FF} yields \eqref{eq:141FF}.
\end{proof}

\begin{proposition}
    \label{prop:rigorous_uhen}
    \begin{IEEEeqnarray}{lCl}
        -\Sum{i=2}{\Nx}
        \left({\muInfo{i\M}{\minusI\mid\infPaI}+\sum_{k\in\minusI}\muInfo{i}{k\M\mid\infPa{k}}}-\turiB{i}\middle)\relright|\plusI
        \NonumberNewlineQQ
        =
        \Sum{i=2}{\Nx}\left\{
        \atamaOmegaIMuMu\mid\infPaI,\effPa
        + \centerSeibunIMu\mid\infPaI
        + \bottomSeibunIMuMu\mid\effPa
        \right.
        \NonumberNewlineQQ\quad
        \left.
        + \left[
        \memotouSeibunIMu
        - \atamaOmegaIMuMu\scl\infPaI\scl\effPa
        - i\prm\scl\minusI\prm\mid i,\minusI
        \right.\right.
        \NonumberNewlineQQ\quad\quad
        \left.\left.
        + \sum_{k=2}^{i-1}\left(
            \botObsSeibunIKMuk\mid\infPaK
            + \botObsSeibunIMukK\mid\effPa[k]
            \right.\right.\right.
            \NonumberNewlineQQ\quad\quad\quad
            \left.\left.\left.
            + \memotouSeibunLL{i,k,\minusK}\mid\infPaK
            + \memotouSeibunLL{i,\minusK,k}\mid\effPa[k]
            \right)
            \middle]\relright|\plusI
        \right\}
        \label{eq:199DD}
        \IEEEeqnarraynumspace
    \end{IEEEeqnarray}
\end{proposition}
\begin{proof}
    Abbreviate the diagonal block
    \begin{IEEEeqnarray}{lCl}
        A_k := \atamaOmega{k}{\minusK}{\minusK}\mid(\infPaK,\effPa[k])
        + \centerSeibunKMu\mid\infPaK
        + \bottomSeibunKMuMu\mid\effPa[k]
        \label{eq:def_Ak}
    \end{IEEEeqnarray}
    and the unrealized-correlation block
    \begin{IEEEeqnarray}{lCl}
        G_i := \sum_{k=2}^{i-1}\left(
            \botObsSeibunIKMuk\mid\infPaK
            + \botObsSeibunIMukK\mid\effPa[k]
            + \memotouSeibunLL{i,k,\minusK}\mid\infPaK
            + \memotouSeibunLL{i,\minusK,k}\mid\effPa[k]
        \right),
        \label{eq:def_Gi}
    \end{IEEEeqnarray}
    so that \eqref{eq:162DD} reads
    $-\sum_{k\in\minusI}i\scl k\M\mid\infPaK = \atamaOmegaIMuMu\mid\effPa + \bottomSeibunIMuMu\mid\effPa + G_i + \sum_{k=2}^{i-1} i\scl A_k$,
    and note that $A_i = \atamaOmegaIMuMu\mid\infPaI,\effPa + \centerSeibunIMu\mid\infPaI + \bottomSeibunIMuMu\mid\effPa$.
    By Eqs.~\eqref{eq:g1313xx}, \eqref{eq:141FF}, and \eqref{eq:162DD}---Eqs.~\eqref{eq:g1313xx} and \eqref{eq:162DD} being invoked in their extended forms, with the block $\Xs{\minusI}$ in place of $\Xs{k}$ in the former and with $\Xs{\plusI}$ appended to every conditioning in both, as licensed below Eq.~\eqref{eq:propH252CC}---the $i$-th summand of the left-hand side of \eqref{eq:199DD} equals
    \begin{IEEEeqnarray}{lCl}
        \left[
        \underbrace{\atamaOmegaIMuMu\mid\infPaI + \atamaOmegaIMuMu\mid\effPa - \atamaOmegaIMuMu}_{(\ast)}
        + \centerSeibunIMu\mid\infPaI
        + \bottomSeibunIMuMu\mid\effPa
        \right.
        \NonumberNewlineQQ\quad
        \left.
        + \memotouSeibunIMu
        - i\prm\scl\minusI\prm\mid i,\minusI
        + G_i
        + \sum_{k=2}^{i-1} i\scl A_k
        \middle]\relright{|}\plusI.
        \label{eq:assembly_i}
        \IEEEeqnarraynumspace
    \end{IEEEeqnarray}
    For the marked combination $(\ast)$, an atom-by-atom check gives
    \begin{IEEEeqnarray}{lCl}
        (\ast) = \atamaOmegaIMuMu\mid\infPaI,\effPa - \atamaOmegaIMuMu\scl\infPaI\scl\effPa:
        \label{eq:ast_combination}
    \end{IEEEeqnarray}
    an atom $t$ of $\Yeng{\atamaOmegaIMuMu}$ carries the coefficient $[t\notin\Yeng{\infPaI}]+[t\notin\effPa]-1$ on the left-hand side and $[t\notin\Yeng{\infPaI}][t\notin\effPa]-[t\in\Yeng{\infPaI}][t\in\effPa]$ on the right-hand side, and the two agree in all four cases.
    Substituting \eqref{eq:ast_combination} into \eqref{eq:assembly_i}, the first three terms combine to $A_i$, so the left-hand side of \eqref{eq:199DD} equals
    \begin{IEEEeqnarray}{lCl}
        \Sum{i=2}{\Nx}\left(A_i + \sum_{k=2}^{i-1} i\scl A_k\middle)\relright{|}\plusI
        \NonumberNewlineQQ
        + \Sum{i=2}{\Nx}\left[
        \memotouSeibunIMu
        - \atamaOmegaIMuMu\scl\infPaI\scl\effPa
        - i\prm\scl\minusI\prm\mid i,\minusI
        + G_i
        \middle]\relright{|}\plusI.
        \nonumber
    \end{IEEEeqnarray}
    Applying the redistribution identity \eqref{eq:174DD} to the first sum replaces it by $\Sum{i=2}{\Nx}A_i$, which is precisely the unconditioned part of the right-hand side of \eqref{eq:199DD}. This completes the proof.
\end{proof}

\subsection{Proof of Lemma~\ref{lem:U_vanish}}
\label{ap:proof_U_vanish}

Fix $i$. With the outer conditioning $\Xs{\plusI}$ of Eq.~\eqref{eq:def_Ucost} attached, $\Ucost$ has the following constituents: the two halves of $\memotouSeibunIMu$, Eq.~\eqref{eq:memotouSeibunSum}, and the four interaction components for each $k$. In the shorthand of Appendix~\ref{ap:shorthand}, they read
\begin{IEEEeqnarray}{lCl}
    \memotouSeibunL{i,\minusI}\mid\plusI &=& i\scl\minusI'\mid(i',\minusI,\plusI),
    \nonumber\\
    \memotouSeibunR{i,\minusI}\mid\plusI &=& i'\scl\minusI\mid(i,\minusI',\plusI),
    \nonumber\\
    \memotouSeibunLL{i,k,\minusK}\mid(\infPaK,\plusI) &=& i\scl k\scl\minusK'\mid(i',k',\minusK,\infPaK,\plusI),
    \nonumber\\
    \botObsSeibunIKMuk\mid(\infPaK,\plusI) &=& i\scl k\scl i'\scl\minusK'\mid(k',\minusK,\infPaK,\plusI),
    \nonumber\\
    \memotouSeibunLL{i,\minusK,k}\mid(\effPa[k],\plusI) &=& i\scl\minusK\scl k'\mid(i',\minusK',k,\effPa[k],\plusI),
    \nonumber\\
    \botObsSeibunIMukK\mid(\effPa[k],\plusI) &=& i\scl\minusK\scl i'\scl k'\mid(\minusK',k,\effPa[k],\plusI),
    \IEEEeqnarraynumspace
    \label{eq:Ucost_constituents}
\end{IEEEeqnarray}
with $k$ ranging over $\{2,\dots,i-1\}$. We show that each of the six is a finite signed combination of terms of the form \eqref{eq:asm_no_unrealized}, and hence vanishes under Assumption~\ref{asm:no_unrealized}. The probabilistic input enters only through Proposition~\ref{prop:asm4_block} below. Every other step is an atom-wise identity of the signed measure in the sense of Appendix~\ref{ap:shorthand}.

The following proposition extends Assumption~\ref{asm:no_unrealized} from pairs of subsystems to pairs of disjoint blocks.

\begin{proposition}[block form of Assumption~\ref{asm:no_unrealized}]
    \label{prop:asm4_block}
    Let $J$ and $K$ be disjoint sets of subsystem indices, and write $\Xs{J}:=(\Xs{j})_{j\in J}$. Under Assumption~\ref{asm:no_unrealized}, for every collection $Z$ of initial and final subsystem states,
    \begin{IEEEeqnarray}{lCl}
        \muI{\Xs{J}\prm}{\Xs{K}\mid \Xs{J},\Xs{K}\prm, Z} \;=\; 0.
        \label{eq:asm4_block}
    \end{IEEEeqnarray}
\end{proposition}
\begin{proof}
    Expanding $\Xs{J}\prm$ and then $\Xs{K}$ entry by entry with the chain rule \eqref{eq:chain_MI6}---applied in the first argument via the symmetry of the mutual information---,
    \begin{IEEEeqnarray}{lCl}
        \muI{\Xs{J}\prm}{\Xs{K}\mid \Xs{J},\Xs{K}\prm, Z}
        \;=\;
        \sum_{j\in J}\sum_{k\in K}
        \muI{\Xs{j}\prm}{\Xs{k}\mid \Xs{j},\Xs{k}\prm, Z_{jk}},
        \nonumber
    \end{IEEEeqnarray}
    where $Z_{jk}$ collects the remaining conditioning entries: $Z$, the states $\Xs{J\setminus\{j\}}$ and $\Xs{K\setminus\{k\}}\prm$, and the already-expanded entries $\Xs{J_{<j}}\prm$ and $\Xs{K_{<k}}$, with $J_{<j}:=\{l\in J: l<j\}$ and $K_{<k}:=\{l\in K: l<k\}$. Since $j\neq k$ and each $Z_{jk}$ is a collection of initial and final subsystem states, every summand is of the form \eqref{eq:asm_no_unrealized} and vanishes.
\end{proof}

The second device removes an interaction entry. Let $\mathcal{K}$ be any component, $W$ any variable, and $C$ any conditioning list:
\begin{IEEEeqnarray}{lCl}
    \mathcal{K}\scl\Yeng{W}\mid C
    \;=\;
    \mathcal{K}\mid C \;-\; \mathcal{K}\mid(W,C),
    \IEEEeqnarraynumspace
    \label{eq:slot_removal}
\end{IEEEeqnarray}
Here, $\mathcal{K}\scl\Yeng{W}$ denotes the component obtained from $\mathcal{K}$ by appending $W$ as an additional interaction entry. Any component carrying $W$ among its semicolon-separated entries can be written in this form, with $\mathcal{K}$ denoting the component with that entry deleted. Identity \eqref{eq:slot_removal} is the recursion \eqref{eq:def_interaction_cond} with the ordering of the entries removed by the signed-measure representation: the semicolon is an intersection, and intersections commute. Both sides evaluate the atoms of the region of $\mathcal{K}\mid C$ that lie inside $\Yeng{W}$. Iterating \eqref{eq:slot_removal} over a set $\mathcal{W}$ of entries expands a component into $2^{|\mathcal{W}|}$ signed terms. Each subset $T\subseteq\mathcal{W}$ contributes one term, with the entries in $T$ moved to the conditioning side with sign $(-1)^{|T|}$ and the remaining entries of $\mathcal{W}$ discarded.

The third device eliminates the region conditionings.

\begin{proposition}[blind-spot expansion]
    \label{prop:blind_expansion}
    Fix $k$ and let $D_n=\Yeng{\Xs{n}}\setminus\Yeng{\infPa{n}}$ be the blind spots of \eqref{eq:def_blind}, so that, recalling the definition below \eqref{eq:103YY}, $\effPa[k]=\Yeng{\Xs{\minusK}}\setminus\blind[k]$ with $\blind[k]=\bigcup_{n\in\minusK}D_n$. Let $\mathcal{K}$ be a component whose region is contained in $\Yeng{\Xs{\minusK}}$, and let $V$ be a collection of variables. Then
    \begin{IEEEeqnarray}{lCl}
        \mathcal{K}\mid(\effPa[k],V)
        \;=\;
        \sum_{\emptyset\neq S\subseteq\minusK}(-1)^{|S|+1}
        \Bigl(\mathcal{K}\scl\bigcap_{n\in S}\Yeng{\Xs{n}}\Bigr)\Bigm|(\infPa{S},V),
        \IEEEeqnarraynumspace
        \label{eq:blind_expansion}
    \end{IEEEeqnarray}
    where $\mathcal{K}\scl\bigcap_{n\in S}\Yeng{\Xs{n}}$ denotes $\mathcal{K}$ with every initial state $\Xs{n}$, $n\in S$, appended as a separate interaction entry, and $\infPa{S}:=\bigcup_{n\in S}\infPa{n}$. In particular, the right-hand side conditions only on collections of variables.
\end{proposition}
\begin{proof}
    By the conditioning convention of Sec.~\ref{sect:components}, the left-hand side is the measure of the region of $\mathcal{K}\mid V$ with $\effPa[k]$ removed. That region is contained in $\Yeng{\Xs{\minusK}}$, within which $\effPa[k]$ and $\blind[k]$ are complementary, so, as in \eqref{eq:blind_translation} (applied to the block $\minusK$),
    \begin{IEEEeqnarray}{lCl}
        \mathcal{K}\mid(\effPa[k],V) \;=\; (\mathcal{K}\mid V)\scl\blind[k].
        \nonumber
    \end{IEEEeqnarray}
    Inclusion--exclusion for the union $\blind[k]=\bigcup_{n\in\minusK}D_n$ holds atom by atom: an atom lying in exactly $m\geq1$ of the $D_n$ receives the coefficient $\sum_{r=1}^{m}(-1)^{r+1}\binom{m}{r}=1$ on the right-hand side of
    \begin{IEEEeqnarray}{lCl}
        (\mathcal{K}\mid V)\scl\blind[k]
        \;=\;
        \sum_{\emptyset\neq S\subseteq\minusK}(-1)^{|S|+1}
        (\mathcal{K}\mid V)\scl\bigcap_{n\in S}D_n.
        \nonumber
    \end{IEEEeqnarray}
    Finally, by the definition of the blind spots,
    \begin{IEEEeqnarray}{lCl}
        \bigcap_{n\in S}D_n
        \;=\;
        \bigcap_{n\in S}\bigl(\Yeng{\Xs{n}}\setminus\Yeng{\infPa{n}}\bigr)
        \;=\;
        \Bigl(\bigcap_{n\in S}\Yeng{\Xs{n}}\Bigr)\Bigm\backslash\Yeng{\infPa{S}},
        \nonumber
    \end{IEEEeqnarray}
    so the $S$-th term intersects the region of $\mathcal{K}\mid V$ with the regions of the $\Xs{n}$, $n\in S$, and removes $\Yeng{\infPa{S}}$: it is the component $\mathcal{K}\scl\bigcap_{n\in S}\Yeng{\Xs{n}}$ conditioned on $(\infPa{S},V)$, which proves \eqref{eq:blind_expansion}.
\end{proof}

\begin{proof}[Proof of Lemma~\ref{lem:U_vanish}]
    \emph{(i) Block components.} The two components in the first two lines of \eqref{eq:Ucost_constituents} are the instances $(J,K)=(\{i\},\minusI)$ and $(J,K)=(\minusI,\{i\})$ of \eqref{eq:asm4_block} with $Z=\Xs{\plusI}$: in standard notation,
    \begin{IEEEeqnarray}{lCl}
        \memotouSeibunR{i,\minusI}\mid\plusI
        &=& \muI{\Xs{i}\prm}{\Xs{\minusI}\mid \Xs{i},\Xs{\minusI}\prm,\Xs{\plusI}},
        \nonumber
        \\
        \memotouSeibunL{i,\minusI}\mid\plusI
        &=& \muI{\Xs{\minusI}\prm}{\Xs{i}\mid \Xs{\minusI},\Xs{i}\prm,\Xs{\plusI}},
        \nonumber
    \end{IEEEeqnarray}
    where the second line uses the symmetry of the mutual information in its two arguments to read $\memotouSeibunL{i,\minusI}$ as the unrealized correlation from the block to $i$. Both vanish by Proposition~\ref{prop:asm4_block}.

    \emph{(ii) Components conditioned on $\infPaK$.} Abbreviate the conditioning list as
    \begin{IEEEeqnarray}{rCl}
        C &:=& (k',\minusK,\infPaK,\plusI).
        \nonumber
    \end{IEEEeqnarray}
    Removing the entry $i$ from the third line of \eqref{eq:Ucost_constituents} by \eqref{eq:slot_removal} leaves two terms,
    \begin{IEEEeqnarray}{lCl}
        \memotouSeibunLL{i,k,\minusK}\mid(\infPaK,\plusI)
        &=&
        k\scl\minusK'\mid(i',C)
        - k\scl\minusK'\mid(i,i',C),
        \IEEEeqnarraynumspace
        \label{eq:LLa_two_terms}
    \end{IEEEeqnarray}
    and removing $i'$ and then $i$ from the fourth line leaves four,
    \begin{IEEEeqnarray}{lCl}
        \botObsSeibunIKMuk\mid(\infPaK,\plusI)
        &=&
        k\scl\minusK'\mid C
        - k\scl\minusK'\mid(i,C)
        \NonumberNewline
        &&-\; k\scl\minusK'\mid(i',C)
        + k\scl\minusK'\mid(i,i',C).
        \IEEEeqnarraynumspace
        \label{eq:botObsa_four_terms}
    \end{IEEEeqnarray}
    Every term on the right-hand sides of Eqs.~\eqref{eq:LLa_two_terms} and \eqref{eq:botObsa_four_terms} is of the form
    \begin{IEEEeqnarray}{lCl}
        k\scl\minusK'\mid(\minusK,k',Z)
        \;=\;
        \muI{\Xs{\minusK}\prm}{\Xs{k}\mid \Xs{\minusK},\Xs{k}\prm,Z},
        \nonumber
    \end{IEEEeqnarray}
    the instance $(J,K)=(\minusK,\{k\})$ of \eqref{eq:asm4_block}: the required entries $\minusK$ and $k'$ sit in $C$, and the rest of the conditioning---$\infPaK$ (a collection of initial states by the definition of the parents, Assumption~\ref{asm:local_mech}), $\plusI$, and the removed entries $i$, $i'$ where present---is a collection of initial and final subsystem states. All six terms vanish.

    \emph{(iii) Components conditioned on $\effPa[k]$.} The regions of $\memotouSeibunLL{i,\minusK,k}$ and of $\botObsSeibunIMukK$ are contained in $\Yeng{\Xs{\minusK}}$ (both carry the entry $\minusK$), so Proposition~\ref{prop:blind_expansion} applies with $V=\plusI$:
    \begin{IEEEeqnarray}{rCl}
        \memotouSeibunLL{i,\minusK,k}\mid(\effPa[k],\plusI)
        \NonumberNewline
        &=&
        \sum_{\emptyset\neq S\subseteq\minusK}(-1)^{|S|+1}
        \NonumberNewline
        &&\quad\times
        \Bigl(i\scl\minusK\scl k'\scl\bigcap_{n\in S}\Yeng{\Xs{n}}\Bigr)
        \NonumberNewline
        &&\quad
        \Bigm|(i',\minusK',k,\infPa{S},\plusI),
        \IEEEeqnarraynumspace
        \label{eq:LLb_expanded}
        \\
        \botObsSeibunIMukK\mid(\effPa[k],\plusI)
        \NonumberNewline
        &=&
        \sum_{\emptyset\neq S\subseteq\minusK}(-1)^{|S|+1}
        \NonumberNewline
        &&\quad\times
        \Bigl(i\scl\minusK\scl i'\scl k'\scl\bigcap_{n\in S}\Yeng{\Xs{n}}\Bigr)
        \NonumberNewline
        &&\quad
        \Bigm|(\minusK',k,\infPa{S},\plusI),
        \IEEEeqnarraynumspace
        \label{eq:botObsb_expanded}
    \end{IEEEeqnarray}
    where the $\Xs{n}$, $n\in S$, enter as separate interaction entries. In each summand of \eqref{eq:LLb_expanded}, iterating \eqref{eq:slot_removal} over the entries $\mathcal{W}=\{\Xs{i}\}\cup\{\Xs{n}\}_{n\in S}$ gives
    \begin{IEEEeqnarray}{lCl}
        \sum_{T\subseteq\mathcal{W}}(-1)^{|T|}\;
        \minusK\scl k'\mid(i',\minusK',k,\infPa{S},\plusI,T),
        \IEEEeqnarraynumspace
        \label{eq:LLb_terminal}
    \end{IEEEeqnarray}
    and in each summand of \eqref{eq:botObsb_expanded}, iterating over $\mathcal{W}'=\{\Xs{i},\Xs{i}\prm\}\cup\{\Xs{n}\}_{n\in S}$ gives
    \begin{IEEEeqnarray}{lCl}
        \sum_{T\subseteq\mathcal{W}'}(-1)^{|T|}\;
        \minusK\scl k'\mid(\minusK',k,\infPa{S},\plusI,T).
        \IEEEeqnarraynumspace
        \label{eq:botObsb_terminal}
    \end{IEEEeqnarray}
    Every term in \eqref{eq:LLb_terminal} and \eqref{eq:botObsb_terminal} is of the form
    \begin{IEEEeqnarray}{lCl}
        \minusK\scl k'\mid(k,\minusK',Z)
        \;=\;
        \muI{\Xs{k}\prm}{\Xs{\minusK}\mid \Xs{k},\Xs{\minusK}\prm,Z},
        \nonumber
    \end{IEEEeqnarray}
    the instance $(J,K)=(\{k\},\minusK)$ of \eqref{eq:asm4_block}: the entries $k$ and $\minusK'$ sit in the conditioning lists of \eqref{eq:LLb_terminal} and \eqref{eq:botObsb_terminal} from the outset, and the rest---$\infPa{S}$ (initial states by Assumption~\ref{asm:local_mech}), $\plusI$, $T$, and $i'$ where present---is a collection of initial and final subsystem states. All terms vanish.

    Every constituent of $\Ucost$ is therefore a finite signed combination of instances of \eqref{eq:asm4_block}, each of which vanishes under Assumption~\ref{asm:no_unrealized}. Hence $\Ucost=0$ for every $i$.
\end{proof}\section{Necessity of the Assumptions: Two Counterexamples}
\label{ap:counterexamples}

Explicit two-subsystem processes show that the strengthenings of the ledger derived in Sec.~\ref{sect:mainresult} fail once reciprocal reading within a single step is admitted. Assumption~\ref{asm:acyclic} cannot be dropped from Corollary~\ref{cor:no_rebate} or from Theorem~\ref{thm:main}, and the mechanism-based definition of the informational parents (Sec.~\ref{sect:setting}) is necessary for the identity \eqref{eq:199DD}. Both processes satisfy the local detailed balance condition \eqref{eq:local_detail_subsystem} and the factorization \eqref{eq:asm_local_mech} with the reciprocal parent structure
\begin{IEEEeqnarray}{lCl}
    \DepX{1}=\{\Xs{2}\},\qquad \DepX{2}=\{\Xs{1}\},
    \label{eq:cex_parents}
\end{IEEEeqnarray}
and both saturate the ledger \eqref{eq:158FF} with equality, thereby also demonstrating its tightness.

For $\Nx=2$ the aggregates of Sec.~\ref{sect:mainresult} take a simple form. With the parents \eqref{eq:cex_parents}, the effective parent region is $\effPa[2]=\Yeng{\Xs{1}}\cap\Yeng{\Xs{2}}$, and a short computation gives
\begin{IEEEeqnarray}{lCl}
    \Dmiss[2] = 0,
    \qquad
    \Ucost[2] = \memotouSeibunL{2,1} + \memotouSeibunR{2,1},
    \NonumberNewline
    \Rreb[2] = \muI{\Xs{1}}{\Xs{2}\mid \Xs{1}\prm,\Xs{2}\prm}
    + \muI{\Xs{1}\prm}{\Xs{2}\prm\mid \Xs{1},\Xs{2}},
    \label{eq:cex_aggregates}
\end{IEEEeqnarray}
where the second rebate term vanishes for both examples by the factorization \eqref{eq:asm_local_mech}.

\subsection{Mutual refresh: the rebate is unavoidable under reciprocal reading}
\label{ap:cex_refresh}

Let $\Xs{1},\Xs{2}$ be binary spins with uniform marginals and symmetric initial correlation $P(X_1=X_2)=a$, $a\in(1/2,1)$. During the step, each subsystem thermalizes completely in a potential set by the \emph{initial} state of its partner:
\begin{IEEEeqnarray}{lCl}
    P\big(\Xs{1}\prm = x \mid \Xs{1},\Xs{2}\big) = P\big(\Xs{1}=x\mid \Xs{2}\big),
    \NonumberNewline
    P\big(\Xs{2}\prm = x \mid \Xs{2},\Xs{1}\big) = P\big(\Xs{2}=x\mid \Xs{1}\big),
    \label{eq:cex_refresh_channel}
\end{IEEEeqnarray}
with the two updates performed independently (a parallel Gibbs-sampler step). Each local channel is a complete relaxation in the potential $E(x;y)=-\invTemp^{-1}\ln P(X_1=x\mid X_2=y)$ and therefore satisfies detailed balance. Since the initial conditional distribution coincides with the Gibbs distribution of that potential, each local process is a relaxation from equilibrium to equilibrium in a fixed potential. Hence the local entropy production \eqref{eq:yuragi_k} vanishes for both subsystems, and by Theorem~\ref{thm:ledger} the ledger \eqref{eq:158FF} holds with equality.

Evaluating \eqref{eq:cex_aggregates} for this process gives $\Rreb[2]=\muI{\Xs{1}}{\Xs{2}\mid\Xs{1}\prm,\Xs{2}\prm}>0$, so that
\begin{IEEEeqnarray}{lCl}
    \DeltaStot
    = \Ucost[2] - \Rreb[2]
    \;<\; \Dmiss[2] + \Ucost[2]:
    \label{eq:cex_refresh_violation}
\end{IEEEeqnarray}
the right-hand side of \eqref{eq:204EE} strictly exceeds the total entropy production. Numerically, at $a=0.8$ one finds $\DeltaStot = 0.169$, $\Ucost[2] = 0.253$, and $\Rreb[2] = 0.084$ in units of nats. The strengthening \eqref{eq:204EE} is therefore false for reciprocal architectures. The example also fixes the physical role of the rebate: the correlation destroyed jointly from both sides, being visible to both controllers, pays back part of the unrealized-correlation cost $\Ucost[2]$ that the same protocol incurs.

\subsection{Mutual overwriting: hidden records and the breakdown of the bookkeeping}
\label{ap:cex_xor}

Let $\Xs{1},\Xs{2}$ be binary with uniform marginals and $P(X_1=X_2)=a\in[1/2,1)$, and let both subsystems be overwritten by the exclusive-or of the initial states:
\begin{IEEEeqnarray}{lCl}
    \Xs{1}\prm = \Xs{2}\prm = \Xs{1}\oplus\Xs{2}.
    \label{eq:cex_xor_map}
\end{IEEEeqnarray}
Given the initial state of its partner, each local map is a bijection of the subsystem's state space. It can therefore be implemented isothermally and reversibly, with $\invTemp\Qs{i}=0$, and satisfies the local detailed balance condition \eqref{eq:local_detail_subsystem} with the inverse map as the reversed channel. The local entropy production \eqref{eq:yuragi_k} vanishes for both subsystems. The kernel factorizes as in \eqref{eq:asm_local_mech}, and every conditional mutual information of the form \eqref{eq:asm_no_unrealized} vanishes because $\Xs{k}=\Xs{j}\oplus\Xs{k}\prm$ is determined by the variables in the conditioning slot. Assumption~\ref{asm:no_unrealized} holds as well. Only Assumption~\ref{asm:acyclic} is violated.

Yet the joint map \eqref{eq:cex_xor_map} is two-to-one: one bit of the composite state is erased at zero heat. Indeed, $\shaent{\Xtotprm}=H_2(a)$ while $\shaent{\Xtot}=\ln 2 + H_2(a)$, where $H_2(a):=-a\ln a-(1-a)\ln(1-a)$, so that
\begin{IEEEeqnarray}{lCl}
    \DeltaStot = \Delta\shaent{\Xtot} + \invTemp\Qtot = -\ln 2 \;<\; 0,
    \label{eq:cex_xor_negative}
\end{IEEEeqnarray}
contradicting the nonnegativity of the total entropy production and, a fortiori, the bound \eqref{eq:main_no_new_obs}, whose right-hand side vanishes for this process. Theorem~\ref{thm:ledger} itself is \emph{not} contradicted: one finds $\Dmiss[2]=\Ucost[2]=0$ and $\Rreb[2]=\muI{\Xs{1}}{\Xs{2}\mid\Xs{1}\prm,\Xs{2}\prm}=\ln 2$, so the ledger \eqref{eq:158FF} holds with equality; the rebate absorbs the entire negative balance.

The mechanism of the breakdown is instructive. Implementing both channels of \eqref{eq:cex_xor_map} within one step requires each mechanism to read the initial value of a variable that is simultaneously being overwritten. This is possible only by keeping the initial values available past their overwriting, in practice by first copying them into external records. The bit erased by the two-to-one compression is then paid for, by Landauer's principle, in the record-erasing machinery, outside the bookkeeping of $\DeltaStot$, which accounts for the two subsystems alone. Assumption~\ref{asm:acyclic} excludes precisely this loophole. Under the parents-later convention the dynamics is realized by updating the subsystems in order, no record is ever needed, and the composite process admits a globally defined time reversal obtained by reversing the schedule. The joint second law then guarantees the nonnegativity of $\DeltaStot$ (Lemma~\ref{lem:joint_second_law}).

Finally, this example shows why the informational parents must be defined through the mechanism factorization of Assumption~\ref{asm:local_mech} rather than through marginal statistical dependence. For the process \eqref{eq:cex_xor_map}, the final state $\Xs{1}\prm$ is marginally independent of both $\Xs{1}$ and $\Xs{2}$ (each conditional distribution equals the marginal of $\Xs{1}\prm$), so a marginal-dependence definition would assign $\DepX{i}=\emptyset$. Under that assignment the screening property fails, and with it the identity \eqref{eq:199DD}: at $a=1/2$ its two sides differ by $2\ln 2$. The mechanism-based definition of Sec.~\ref{sect:setting} (the minimal dependence sets of the kernel factorization \eqref{eq:asm_local_mech}) assigns the parents \eqref{eq:cex_parents}, for which the identity holds exactly, as verified above.

\section{Signs of the Subsystem-Resolved Quantities}
\label{ap:subsystem_signs}

The discussion in Sec.~\ref{sect:indispensability} splits each destroyed correlation $\decIiMu$ into the hidden part $\DissipationI$ and the shared part $\accIiMu$ (Eq.~\eqref{eq:dec_split}); Appendix~\ref{ap:multistage} proves that both step totals are nonnegative: $\Sum{i=2}{\Nx}\DissipationI\geq0$ (Lemma~\ref{lem:step_sum}) and $\Sum{i=2}{\Nx}\accIiMu\geq0$ (Lemma~\ref{lem:acc_sum}). At the finer, subsystem-resolved level, the signs have the following structure. First, the inputs of the split are sign-definite: every destroyed and every created correlation is nonnegative under Assumption~\ref{asm:no_unrealized}, $\decIiMu\geq0$ and $\incIiMu\geq0$, whereas the surviving part subtracted in their definitions is not (Appendix~\ref{ap:dec_inc_signs}). Second, we derive closed-form expressions for the two region-conditioned components of \eqref{eq:def_DissipationI}, valid for a class of parent structures that includes every process with $\Nx\leq4$. The expressions exhibit both components as sums of conditional mutual informations, so that $\DissipationI\geq0$ throughout the class (Appendix~\ref{ap:sufficient_condition_Di}). Third, we construct an explicit five-subsystem process, satisfying Assumptions~\ref{asm:local_mech}--\ref{asm:no_unrealized} with minimal parents, for which $\DissipationI<0$ for one subsystem. Outside the class the subsystem-resolved sign of the hidden part genuinely fails, and only the sum rules of Appendix~\ref{ap:multistage} survive (Appendix~\ref{ap:cex_negative_Di}).

\subsection{Nonnegativity of \texorpdfstring{$\decIiMu$}{C\_dec(i,<i)} and \texorpdfstring{$\incIiMu$}{C\_inc(i,<i)}; the sign of the surviving part}
\label{ap:dec_inc_signs}

Before conditioning enters, the signs of the quantities appearing in the split \eqref{eq:dec_split} can be settled outright for subsystem pairs and block pairs alike. For disjoint collections $A$ and $B$ of subsystems, write $\decIX{A}{B}$ and $\incIX{A}{B}$ for the quantities defined by \eqref{eq:def_decIij} and \eqref{eq:def_incIij} with $(\Xs{i},\Xs{j})$ replaced by $(\Xs{A},\Xs{B})$.

\begin{lemma}[Signs of the destroyed and created correlations]
    \label{lem:dec_inc_signs}
    Under Assumptions~\ref{asm:local_mech} and \ref{asm:no_unrealized}, for all disjoint collections $A$ and $B$ of subsystems,
    \begin{IEEEeqnarray}{lCl}
        \decIX{A}{B}
        &=&
        \muI{\Xs{A}}{\Xs{B}\mid \Xs{A}\prm,\Xs{B}\prm}
        + \muI{\Xs{A}}{\Xs{B}\prm\mid \Xs{A}\prm}
        \NonumberNewlineQQ
        &&+\; \muI{\Xs{B}}{\Xs{A}\prm\mid \Xs{B}\prm},
        \IEEEeqnarraynumspace
        \label{eq:dec_cmi_form}
        \\
        \incIX{A}{B}
        &=&
        \muI{\Xs{A}\prm}{\Xs{B}\prm\mid \Xs{A},\Xs{B}}
        + \muI{\Xs{A}}{\Xs{B}\prm\mid \Xs{B}}
        \NonumberNewlineQQ
        &&+\; \muI{\Xs{B}}{\Xs{A}\prm\mid \Xs{A}},
        \IEEEeqnarraynumspace
        \label{eq:inc_cmi_form_gen}
    \end{IEEEeqnarray}
    each a sum of conditional mutual informations; in particular, $\decIX{A}{B}\geq0$ and $\incIX{A}{B}\geq0$.
\end{lemma}
\begin{proof}
    The region of the initial correlation $\muI{\Xs{A}}{\Xs{B}}$ destroyed during the step---the part outside $\Yeng{\Xs{A}\prm}\cap\Yeng{\Xs{B}\prm}$---splits exclusively by membership in $\Yeng{\Xs{A}\prm}$ and $\Yeng{\Xs{B}\prm}$, as in \eqref{eq:decIiMu_decomp}:
    \begin{IEEEeqnarray}{lCl}
        \decIX{A}{B}
        &=&
        \muI{\Xs{A}}{\Xs{B}\mid \Xs{A}\prm,\Xs{B}\prm}
        + \muI{\Xs{A}}{\Xs{B};\Xs{B}\prm\mid \Xs{A}\prm}
        \NonumberNewlineQQ
        &&+\; \muI{\Xs{A}}{\Xs{B};\Xs{A}\prm\mid \Xs{B}\prm}.
        \nonumber
    \end{IEEEeqnarray}
    The first term is a conditional mutual information as it stands. In the second, removing the $\Xs{B}$ slot by the recursion \eqref{eq:def_interaction_cond} gives $\muI{\Xs{A}}{\Xs{B}\prm\mid \Xs{A}\prm} - \muI{\Xs{A}}{\Xs{B}\prm\mid \Xs{B},\Xs{A}\prm}$, and the subtrahend vanishes by the block form \eqref{eq:asm4_block} of Assumption~\ref{asm:no_unrealized} (Proposition~\ref{prop:asm4_block} with $J=B$, $K=A$). Symmetrically, the third term equals $\muI{\Xs{B}}{\Xs{A}\prm\mid \Xs{B}\prm} - \muI{\Xs{B}}{\Xs{A}\prm\mid \Xs{A},\Xs{B}\prm}$, whose subtrahend is removed by Proposition~\ref{prop:asm4_block} with $J=A$, $K=B$. This proves \eqref{eq:dec_cmi_form}. The companion identity \eqref{eq:inc_cmi_form_gen} follows by the exchange $\Xtot\leftrightarrow\Xtotprm$: it maps \eqref{eq:def_decIij} to \eqref{eq:def_incIij} and \eqref{eq:dec_cmi_form} to \eqref{eq:inc_cmi_form_gen}, while mapping the family of constraints \eqref{eq:asm4_block} to itself---the instance $(J,K)$ becomes the instance $(K,J)$, by the symmetry of the mutual information---so the argument just given transports verbatim. For $B=\minusI$ the underlying split is Eq.~\eqref{eq:inc_atom_split} in the proof of Lemma~\ref{lem:acc_sum}, whose per-subsystem statement $\incIiMu\geq0$ is recovered.
\end{proof}

The surviving part $\muI{\Xs{A}}{\Xs{B};\Xs{A}\prm;\Xs{B}\prm}$ subtracted in \eqref{eq:def_decIij} and \eqref{eq:def_incIij} carries, by contrast, no definite sign under Assumptions~\ref{asm:local_mech}--\ref{asm:no_unrealized}. A two-bit standard feedback process suffices to make it negative. Let $\Xs{1}\prm=\Xs{1}\oplus\Xs{2}$ with $\DepX{1}=\{\Xs{2}\}$, let $\Xs{2}\prm=\Xs{2}$, and let the two initial bits be independent and uniform. All four assumptions hold. Conditioned on the value of $\Xs{2}$, the channel of subsystem $1$ is a bijection, implementable isothermally and reversibly as in Appendix~\ref{ap:cex_xor}. The parent set $\DepX{1}=\{\Xs{2}\}$ is minimal because the output varies with $\Xs{2}$ at either value of $\Xs{1}$. The reading pattern is acyclic. The step is also of the standard feedback class described below \eqref{eq:asm_no_unrealized}, so Assumption~\ref{asm:no_unrealized} holds by construction. Since $\Xs{2}\prm=\Xs{2}$, the region of $\Xs{2}\prm$ coincides with that of $\Xs{2}$, and the recursion \eqref{eq:def_interaction_cond} gives
\begin{IEEEeqnarray}{lCl}
    \muI{\Xs{1}}{\Xs{2};\Xs{1}\prm;\Xs{2}\prm}
    = \muI{\Xs{1}}{\Xs{2};\Xs{1}\prm}
    = \muI{\Xs{1}}{\Xs{2}} - \muI{\Xs{1}}{\Xs{2}\mid \Xs{1}\prm}
    = -\ln 2,
    \nonumber
\end{IEEEeqnarray}
because the initial bits are independent but become perfectly correlated given the parity $\Xs{1}\prm$. The destroyed correlation \eqref{eq:def_decIij} then \emph{exceeds} the correlation initially present, $\decIX{1}{2} = \ln 2 > 0 = \muI{\Xs{1}}{\Xs{2}}$. The created correlation matches it, $\incIX{1}{2} = \ln 2$, and the net change \eqref{eq:release_split} vanishes. The names ``destroyed'' and ``created'' thus count relative to a signed baseline; what is guaranteed is the nonnegativity of the two amounts themselves (Lemma~\ref{lem:dec_inc_signs}), not that either is bounded by the correlation present at the corresponding endpoint.

\subsection{A sufficient condition for \texorpdfstring{$\DissipationI\geq0$}{nonnegative D\_i}}
\label{ap:sufficient_condition_Di}

Fix a subsystem $i$ and let
\begin{IEEEeqnarray}{lCl}
    K_i := \{k\in\minusI : \Xs{i}\notin\DepX{k}\}
    \label{eq:def_Ki}
\end{IEEEeqnarray}
collect the members of the preceding block whose mechanisms do not read $i$. We call a member $k\in K_i$ \emph{blind} if $\DepX{k}=\emptyset$ and \emph{covered} otherwise, write $L_i\subseteq K_i$ for the set of blind members, and abbreviate the blind spot of subsystem $n$ as $\mathcal{B}_n:=\Yeng{\Xs{n}}\setminus\Yeng{\infPa{n}}$, so that, by \eqref{eq:103YY}, conditioning a region contained in $\Yeng{\Xs{\minusI}}$ on $\effPa$ amounts to intersecting it with $\bigcup_{n\in\minusI}\mathcal{B}_n$.

\begin{proposition}[Chain form of the conditioned components]
    \label{prop:chain_form}
    Order $K_i$ as $\sigma=(k_1,\dots,k_m)$: first the blind members in increasing index order, then the covered members in decreasing index order; write $\sigma_{<b}:=\{k_1,\dots,k_{b-1}\}$. Suppose every pair of covered members $k_a,k_b$ with $a<b$ satisfies
    \begin{IEEEeqnarray}{lCl}
        \Xs{k_a}\in\DepX{k_b}
        \qquad\text{or}\qquad
        \DepX{k_a}\subseteq\DepX{k_b}\cup\{\Xs{l}\}_{l\in L_i}.
        \IEEEeqnarraynumspace
        \label{eq:pair_condition}
    \end{IEEEeqnarray}
    Then, as an identity of the signed measure involving no distributional assumption,
    \begin{IEEEeqnarray}{lCl}
        \atamaOmegaIMuMu\mid\infPaI,\effPa
        =
        \Sum{b=1}{m}\muI{\Xs{k_b}}{\Xs{i}\mid \Xs{\sigma_{<b}},\infPa{k_b},\infPaI,\Xs{i}\prm,\Xs{\minusI}\prm},
        \IEEEeqnarraynumspace
        \label{eq:chain_form_w}
    \end{IEEEeqnarray}
    and, if Assumption~\ref{asm:no_unrealized} holds,
    \begin{IEEEeqnarray}{lCl}
        \bottomSeibunIMuMu\mid\effPa
        =
        \Sum{b=1}{m}\muI{\Xs{k_b}}{\Xs{i}\prm\mid \Xs{\sigma_{<b}},\infPa{k_b},\Xs{\minusI}\prm}.
        \IEEEeqnarraynumspace
        \label{eq:chain_form_b}
    \end{IEEEeqnarray}
    Every term being a conditional mutual information, both components are then nonnegative, and, together with \eqref{eq:cpa_closed_form}, $\DissipationI\geq0$, i.e., $\accIiMu\leq\decIiMu$.
\end{proposition}

\begin{proof}
    Consider \eqref{eq:chain_form_b}. Extend $\sigma$ to an ordering $(l_1,\dots,l_{i-1})$ of the whole block by appending the members of $\minusI\setminus K_i$ after $k_m$, in any order. Intersecting the region of $\bottomSeibunIMuMu$ with $\bigcup_{n\in\minusI}\mathcal{B}_n$, partitioned by \eqref{eq:first_occ_disjoint} in this order, gives
    \begin{IEEEeqnarray}{lCl}
        \bottomSeibunIMuMu\mid\effPa
        = \Sum{j=1}{i-1} \mu(R_j),
        \NonumberNewline
        R_j :=
        \Yeng{\Xs{i}}\cap\Yeng{\Xs{i}\prm}\cap
        \Bigl(\mathcal{B}_{l_j}\setminus\bigcup_{j'<j}\mathcal{B}_{l_{j'}}\Bigr)
        \setminus\Yeng{\Xs{\minusI}\prm},
        \IEEEeqnarraynumspace
        \label{eq:b_pieces}
    \end{IEEEeqnarray}
    where $\mu$ denotes the signed measure of the information diagram and the block slot $\Yeng{\Xs{\minusI}}$ has been absorbed, $\mathcal{B}_{l_j}$ being contained in $\Yeng{\Xs{l_j}}$. Three observations reduce \eqref{eq:b_pieces} to \eqref{eq:chain_form_b}.

    (a) \emph{The pieces of $\minusI\setminus K_i$ vanish.} For $l_j\notin K_i$ one has $\Xs{i}\in\DepX{l_j}$, hence $\Yeng{\Xs{i}}\subseteq\Yeng{\infPa{l_j}}$ and $\Yeng{\Xs{i}}\cap\mathcal{B}_{l_j}=\emptyset$: the region $R_j$ is empty. Placing these members last also keeps their blind spots out of the subtracted unions of all other pieces.

    (b) \emph{For $k_b\in K_i$ the subtracted blind spots may be upgraded to full variable regions:}
    \begin{IEEEeqnarray}{lCl}
        R_b =
        \Yeng{\Xs{i}}\cap\Yeng{\Xs{i}\prm}\cap\Yeng{\Xs{k_b}}
        \setminus
        \Yeng{\Xs{\sigma_{<b}},\infPa{k_b},\Xs{\minusI}\prm},
        \IEEEeqnarraynumspace
        \label{eq:b_piece_clean}
    \end{IEEEeqnarray}
    as an identity between regions. Indeed, $\mathcal{B}_{k_b}=\Yeng{\Xs{k_b}}\setminus\Yeng{\infPa{k_b}}$, and enlarging each subtracted $\mathcal{B}_{k_a}$, $a<b$, to the full $\Yeng{\Xs{k_a}}$ removes, in addition, only atoms of $\Yeng{\Xs{k_a}}\cap\Yeng{\infPa{k_a}}$. For blind $k_a$ this set is empty. For covered $k_a$, condition \eqref{eq:pair_condition} applies: if $\Xs{k_a}\in\DepX{k_b}$, the extra atoms lie in $\Yeng{\infPa{k_b}}$, which \eqref{eq:b_piece_clean} subtracts anyway; if $\DepX{k_a}\subseteq\DepX{k_b}\cup\{\Xs{l}\}_{l\in L_i}$, they lie in $\Yeng{\infPa{k_b}}\cup\bigcup_{l\in L_i}\Yeng{\Xs{l}}$, and every blind member precedes $k_b$ in $\sigma$, so these atoms are subtracted as well. Hence the two regions agree.

    (c) \emph{Assumption~\ref{asm:no_unrealized} removes the $\Xs{i}$ slot.} The measure of \eqref{eq:b_piece_clean} is the conditional interaction information $\muI{\Xs{k_b}}{\Xs{i}\prm;\Xs{i}\mid H}$ with $H:=(\Xs{\sigma_{<b}},\infPa{k_b},\Xs{\minusI}\prm)$, and the recursion \eqref{eq:def_interaction_cond} gives
    \begin{IEEEeqnarray}{lCl}
        \muI{\Xs{k_b}}{\Xs{i}\prm;\Xs{i}\mid H}
        =
        \muI{\Xs{k_b}}{\Xs{i}\prm\mid H}
        - \muI{\Xs{k_b}}{\Xs{i}\prm\mid \Xs{i},H}.
        \nonumber
    \end{IEEEeqnarray}
    The subtrahend is of the form \eqref{eq:asm_no_unrealized} with $j=i$ and $k=k_b$: the conditioning contains $\Xs{i}$ and, through $\Xs{\minusI}\prm\subseteq H$, also $\Xs{k_b}\prm$. It vanishes, leaving \eqref{eq:chain_form_b}.

    For \eqref{eq:chain_form_w} the argument is shorter. The same partition applies to the region of $\atamaOmegaIMuMu\mid\infPaI,\effPa$, with $\Yeng{\infPaI}\cup\Yeng{\Xs{i}\prm}\cup\Yeng{\Xs{\minusI}\prm}$ subtracted in place of $\Yeng{\Xs{\minusI}\prm}$ alone; steps (a) and (b) carry over verbatim, and the resulting piece is already the conditional mutual information in \eqref{eq:chain_form_w}---no interaction slot is left to remove, so Assumption~\ref{asm:no_unrealized} is not needed.
\end{proof}

\begin{corollary}[Safe classes]
    \label{cor:safe_classes}
    Under Assumptions~\ref{asm:local_mech}, \ref{asm:acyclic}, and \ref{asm:no_unrealized}, $\DissipationI\geq0$ holds for every subsystem $i$ in each of the following cases: (i) $\Nx\leq4$, with an arbitrary acyclic parent structure; (ii) fully blind dynamics; (iii) at most one member of $K_i$ is covered; (iv) nested parents among the covered members of $K_i$, $\DepX{k}\subseteq\DepX{l}$ whenever $k>l$.
\end{corollary}
\begin{proof}
    Cases (ii)--(iv) fulfill \eqref{eq:pair_condition} at sight. For (i), a pair of covered members of $K_i$ requires $i\in\{3,4\}$, and the parents-later convention \eqref{eq:parents_later} together with $\Xs{i}\notin\DepX{k}$ leaves \eqref{eq:pair_condition} no room to fail. For $i=3$ the only pair is $(k_a,k_b)=(2,1)$ with $\DepX{2}\subseteq\{\Xs{4}\}$, and either $\Xs{2}\in\DepX{1}$ or $\DepX{1}\subseteq\{\Xs{4}\}$, in which case $\DepX{2}\subseteq\DepX{1}$. For $i=4$, a third covered member would require $\DepX{3}\subseteq\{\Xs{4}\}\setminus\{\Xs{4}\}=\emptyset$, so the only pair is again $(k_a,k_b)=(2,1)$ with $\DepX{2}=\{\Xs{3}\}$, and either $\Xs{2}\in\DepX{1}$ or $\DepX{1}=\{\Xs{3}\}=\DepX{2}$. In every case \eqref{eq:pair_condition} holds, so Proposition~\ref{prop:chain_form} applies.
\end{proof}

\subsection{A five-subsystem process with \texorpdfstring{$\DissipationI<0$}{negative D\_i}}
\label{ap:cex_negative_Di}

Condition \eqref{eq:pair_condition} first fails at $\Nx=5$. Two covered members of $K_i$ must read \emph{distinct} third parties outside the block, and no such indices exist for $\Nx\leq4$. The minimal failing structure,
\begin{IEEEeqnarray}{lCl}
    \DepX{1}=\{\Xs{5}\},\quad
    \DepX{2}=\{\Xs{4}\},\quad
    \DepX{3}=\DepX{4}=\DepX{5}=\emptyset,
    \IEEEeqnarraynumspace
    \label{eq:cexD_parents}
\end{IEEEeqnarray}
violates \eqref{eq:pair_condition} for the subsystem $i=3$: among the covered members $2,1$ of $K_3$, $\Xs{2}\notin\DepX{1}$ and $\DepX{2}=\{\Xs{4}\}\not\subseteq\DepX{1}=\{\Xs{5}\}$. The sign of $\DissipationI$ indeed fails there. Let all five subsystems be binary, evolving by the deterministic channels
\begin{IEEEeqnarray}{lCl}
    \Xs{1}\prm=\Xs{1}\lor\Xs{5},
    \quad
    \Xs{2}\prm=\Xs{2}\lor\Xs{4},
    \quad
    \Xs{n}\prm=\Xs{n}\ \ (n\geq3),
    \IEEEeqnarraynumspace
    \label{eq:cexD_channels}
\end{IEEEeqnarray}
from the initial distribution supported on five states,
\begin{IEEEeqnarray}{lCl}
    P(x_1x_2x_3x_4x_5)
    =
    \begin{cases}
        1/7, & x\in\{00100,\ 10011,\ 11101\},
        \\
        2/7, & x\in\{01111,\ 11010\}.
    \end{cases}
    \IEEEeqnarraynumspace
    \label{eq:cexD_init}
\end{IEEEeqnarray}

All four assumptions hold, with the degenerate channels understood in the ideal-limit reading fixed below Assumption~\ref{asm:ldb}, and the parents are minimal. The kernel factorizes as in \eqref{eq:asm_local_mech} with the parents \eqref{eq:cexD_parents}. These parents are minimal because an OR gate depends on each of its arguments: its output varies with either input while the other reads $0$, and both values of each parent occur on the support of \eqref{eq:cexD_init}. The local detailed balance condition \eqref{eq:local_detail_subsystem} holds because every channel is either static or, conditioned on the value of its parent, the identity ($\Xs{5}=0$, resp.\ $\Xs{4}=0$) or a complete relaxation into the state $1$ ($\Xs{5}=1$, resp.\ $\Xs{4}=1$), realized as the deterministic limit of a relaxation in a deep two-level potential. The reading pattern is acyclic. Assumption~\ref{asm:no_unrealized} is not guaranteed structurally because neither sufficient condition below \eqref{eq:asm_no_unrealized} applies. It nevertheless holds for this process and this initial distribution: every conditional mutual information \eqref{eq:asm_no_unrealized}, for all ordered pairs $j\neq k$ and all conditioning collections $Z$ drawn from the remaining variables, vanishes to machine precision (an exhaustive check).

For the subsystem $i=3$, whose preceding block is $\{1,2\}$, one finds, in nats,
\begin{IEEEeqnarray}{lCl}
    \atamaOmegaIMuMu\mid\infPaI,\effPa = 0,
    \qquad
    \centerSeibunIMu\mid\infPaI = 0,
    \NonumberNewline
    \bottomSeibunIMuMu\mid\effPa
    = \DissipationI = -0.0649 \;<\; 0,
    \IEEEeqnarraynumspace
    \label{eq:cexD_values}
\end{IEEEeqnarray}
while $\decIiMu = 0.3213$, so that $\accIiMu = 0.3862 > \decIiMu$. Subsystem $3$ is itself static: its correlation with the block is destroyed entirely by the moves of the block, so $\decIiMu$ is carried by $\bottomSeibunIMuMu$ alone, and the unconditioned component is positive, $\bottomSeibunIMuMu = 0.3213$. The effective-parent conditioning, however, \emph{overshoots}: the conditioning $\mid\effPa$ intersects the region of $\bottomSeibunIMuMu$ with $\mathcal{B}_1\cup\mathcal{B}_2$, and the signed measure of that intersection is negative. The split \eqref{eq:dec_split} then attributes to the controllers a share $\accIiMu$ exceeding the destroyed correlation itself. The remaining subsystems have $\DissipationI = +0.2728$ each ($i=2,4,5$), so the step total $\Sum{i=2}{\Nx}\DissipationI = +0.7535$ remains positive, as Lemma~\ref{lem:step_sum} guarantees; the companion total $\Sum{i=2}{\Nx}\accIiMu = +0.1783$ is likewise positive (Lemma~\ref{lem:acc_sum}).

\section{Composition of the Single-Step Bounds along a Chain}
\label{ap:multistage}

Here we derive the entropy-production ledger for the multistage chains introduced in Sec.~\ref{sect:indispensability} and prove Proposition~\ref{prop:chain_work}. The chain consists of $K$ successive steps $t_0\to t_1\to\cdots\to t_K$ as specified there. Superscripts $(k)$ denote quantities evaluated on step $k$, and $\decIk$, $\accIk$, and $\incIk$ are the step totals of the destroyed, controller-shared, and created correlations defined in Sec.~\ref{sect:indispensability}, so that $-\Delta\TotalCorrelation^{(k)}=\decIk -\incIk $ by \eqref{eq:release_split}. The quantity $\accIk $ measures the destroyed correlation of step $k$ that is visible to the controllers of the side that destroyed it. It is the only part that, by Theorem~\ref{thm:main}, may escape dissipation.

The signs of these step totals require care. As noted below \eqref{eq:dec_split}, the subsystem-resolved quantities $\accIiMu$ and $\DissipationI$ can be negative under Assumptions~\ref{asm:local_mech}--\ref{asm:no_unrealized}, even though $\decIiMu$ and $\incIiMu$ themselves cannot (Lemma~\ref{lem:dec_inc_signs} of Appendix~\ref{ap:subsystem_signs}). Conditioning can increase an information component, and Appendix~\ref{ap:cex_negative_Di} exhibits an explicit deterministic five-subsystem process, satisfying all four assumptions with minimal parents, in which $\DissipationI<0$ for one subsystem. For $\Nx\leq4$ this cannot happen. Corollary~\ref{cor:safe_classes} proves that $\DissipationI\geq0$. The \emph{step totals} behave better. Both signs hold in general by the two sum rules of this appendix.

\begin{lemma}[Step-total sum rule]
\label{lem:step_sum}
Under Assumptions~\ref{asm:local_mech} and \ref{asm:acyclic}, the right-hand side of the generalized inequality \eqref{eq:2ndlaw1} is nonnegative,
\begin{IEEEeqnarray}{lCl}
    \UsableNewAll - \delitot \;\geq\; 0,
    \label{eq:sum_rule}
\end{IEEEeqnarray}
so that \eqref{eq:2ndlaw1} is never weaker than the conventional second law $\DeltaStot\geq0$. Under Assumptions~\ref{asm:local_mech}--\ref{asm:no_unrealized}, the left-hand side of \eqref{eq:sum_rule} equals $\Sum{i=2}{\Nx}\DissipationI$; applied to each step of a chain, $\Sum{i=2}{\Nx}\DissipationI\geq0$ there, i.e., $\accIk\leq\decIk$.
\end{lemma}
\begin{proof}
    By \eqref{eq:def_eta} and \eqref{eq:split_entropy},
    \begin{IEEEeqnarray}{lCl}
        \UsableNewAll - \delitot
        =
        \Delta\shaent{\Xtot}
        - \Sum{j=1}{\Nx}\left[
            \shaent{\Xs{j}\prm\mid\DepXJ} - \shaent{\Xs{j}\mid\DepXJ}
        \right].
        \IEEEeqnarraynumspace
        \label{eq:sum_rule_rewrite}
    \end{IEEEeqnarray}
    As in the proof of Lemma~\ref{lem:joint_second_law}, write $S_j:=\shaent{(\Xs{1:j}\prm,\Xs{\plusX{j}})}$ for the joint entropies of the intermediate collections and $Z_j:=(\Xs{1:j-1}\prm,\Xs{\plusX{j}})$, so that $\Delta\shaent{\Xtot}=\Sum{j=1}{\Nx}\left(S_j-S_{j-1}\right)$ with
    \begin{IEEEeqnarray}{lCl}
        S_j-S_{j-1} = \shaent{\Xs{j}\prm\mid Z_j}-\shaent{\Xs{j}\mid Z_j}
        \nonumber
    \end{IEEEeqnarray}
    by the chain rule---a distributional identity; no update schedule is invoked. Under the parents-later convention \eqref{eq:parents_later}, $\DepXJ\subseteq\Xs{\plusX{j}}$, so $Z_j$ splits as $(\DepXJ,V_j)$ with $V_j:=Z_j\setminus\DepXJ$, and the $j$-th term of $\Delta\shaent{\Xtot}$ minus the $j$-th summand of \eqref{eq:sum_rule_rewrite} becomes
    \begin{IEEEeqnarray}{lCl}
        \muI{\Xs{j}}{V_j\mid\DepXJ} - \muI{\Xs{j}\prm}{V_j\mid\DepXJ}
        \;\geq\;0.
        \label{eq:substep_dpi}
    \end{IEEEeqnarray}
    The inequality \eqref{eq:substep_dpi} is a conditional data-processing inequality: the chain rule gives
    \begin{IEEEeqnarray}{rCl}
        \muI{\Xs{j}\prm}{V_j\mid\DepXJ}
        &\leq&
        \muI{\Xs{j}}{V_j\mid\DepXJ}
        \NonumberNewline
        &&+ \muI{\Xs{j}\prm}{V_j\mid\Xs{j},\DepXJ}.
        \nonumber
    \end{IEEEeqnarray}
    The last term vanishes because, conditioned on $(\Xs{j},\DepXJ)$, the final state $\Xs{j}\prm$ is independent of all remaining variables (Lemma~\ref{lem:screening}). An update cannot correlate a subsystem with the variables its mechanism does not read beyond what its initial state already carries. Summing \eqref{eq:substep_dpi} over $j$ gives \eqref{eq:sum_rule}.

    Under Assumptions~\ref{asm:local_mech}--\ref{asm:no_unrealized}, the identity of Proposition~\ref{prop:rigorous_uhen} (Appendix~\ref{ap:derivation}) together with Corollary~\ref{cor:no_rebate} and Lemma~\ref{lem:U_vanish} reduces the left-hand side of \eqref{eq:sum_rule} to the right-hand side of \eqref{eq:main_no_new_obs}, which is $\Sum{i=2}{\Nx}\DissipationI$ by \eqref{eq:def_DissipationI}; the stepwise form follows from \eqref{eq:dec_split}.
\end{proof}

The lower sign is settled by a companion sum rule that is sharper than nonnegativity. The created and shared step totals both collapse to closed forms, namely sums of conditional mutual informations attached to the reading pattern. Neither identity involves Assumption~\ref{asm:acyclic}.

\begin{lemma}[Sign of the shared step total]
\label{lem:acc_sum}
Under Assumptions~\ref{asm:local_mech} and \ref{asm:no_unrealized},
\begin{IEEEeqnarray}{lCl}
    \Sum{i=2}{\Nx}\incIiMu
    &=&
    \Sum{i=1}{\Nx} \muI{\Xs{i}\prm}{\DepX{i}\mid\Xs{i}},
    \label{eq:inc_closed}
    \\
    \Sum{i=2}{\Nx}\incIiMu - \UsableNewAll
    &=&
    \Sum{i=1}{\Nx} \muI{\Xs{i}}{\DepX{i}\mid\Xs{i}\prm}
    \;\geq\; 0,
    \IEEEeqnarraynumspace
    \label{eq:acc_closed}
\end{IEEEeqnarray}
and every $\incIiMu$ is itself nonnegative. Under Assumptions~\ref{asm:local_mech}--\ref{asm:no_unrealized}, the left-hand side of \eqref{eq:acc_closed} equals $\Sum{i=2}{\Nx}\accIiMu$; applied to each step of a chain, $\accIk = \Sum{i=1}{\Nx}\muI{\Xs{i}}{\DepX{i}^{(k)}\mid\Xs{i}\prm} \geq 0$ at every step, with equality if and only if no update destroys the correlation between its subsystem and its parents, $\muI{\Xs{i}}{\DepX{i}^{(k)}\mid\Xs{i}\prm}=0$ for all $i$.
\end{lemma}
\begin{proof}
    Fix $i$. The region of $\muI{\Xs{i}\prm}{\Xs{\minusI}\prm}$ meets $\Yeng{\Xs{i}}\cap\Yeng{\Xs{\minusI}}$ exactly in the surviving part subtracted in \eqref{eq:def_incIij}; partitioning the rest of that region by membership in $\Yeng{\Xs{i}}$ and $\Yeng{\Xs{\minusI}}$ gives the exact split
    \begin{IEEEeqnarray}{lCl}
        \incIiMu
        =
        \muI{\Xs{i}\prm}{\Xs{\minusI}\prm\mid\Xs{1:i}}
        + \muI{\Xs{i}\prm}{\Xs{\minusI}\prm;\Xs{i}\mid\Xs{\minusI}}
        \NonumberNewlineQQ
        +\; \muI{\Xs{i}\prm}{\Xs{\minusI}\prm;\Xs{\minusI}\mid\Xs{i}}.
        \IEEEeqnarraynumspace
        \label{eq:inc_atom_split}
    \end{IEEEeqnarray}
    By the recursion \eqref{eq:def_interaction_cond} and the symmetry of the interaction information, the second term equals $\muI{\Xs{\minusI}\prm}{\Xs{i}\mid\Xs{\minusI}} - \muI{\Xs{\minusI}\prm}{\Xs{i}\mid\Xs{i}\prm,\Xs{\minusI}}$; expanding $\Xs{\minusI}\prm$ in the subtrahend by the chain rule yields the terms $\muI{\Xs{k}\prm}{\Xs{i}\mid\Xs{i}\prm,\Xs{\minusI},\Xs{1:k-1}\prm}$ with $k\in\minusI$, each of the form \eqref{eq:asm_no_unrealized}, so the subtrahend vanishes. Symmetrically, the third term equals $\muI{\Xs{i}\prm}{\Xs{\minusI}\mid\Xs{i}} - \muI{\Xs{i}\prm}{\Xs{\minusI}\mid\Xs{\minusI}\prm,\Xs{i}}$, and expanding $\Xs{\minusI}$ in the subtrahend yields terms covered by \eqref{eq:asm_no_unrealized} again. Hence, under Assumption~\ref{asm:no_unrealized},
    \begin{IEEEeqnarray}{lCl}
        \incIiMu
        =
        \muI{\Xs{i}\prm}{\Xs{\minusI}\prm\mid\Xs{1:i}}
        + \muI{\Xs{\minusI}\prm}{\Xs{i}\mid\Xs{\minusI}}
        \NonumberNewlineQQ
        +\; \muI{\Xs{i}\prm}{\Xs{\minusI}\mid\Xs{i}},
        \IEEEeqnarraynumspace
        \label{eq:inc_cmi_form}
    \end{IEEEeqnarray}
    a sum of conditional mutual informations; in particular, $\incIiMu\geq0$ for every $i$.

    The first terms of \eqref{eq:inc_cmi_form} telescope. Write
    \begin{IEEEeqnarray}{rCl}
        A_j &:=& \shaent{\Xs{1:j}\prm\mid\Xs{1:j}}.
        \nonumber
    \end{IEEEeqnarray}
    The chain rule gives
    \begin{IEEEeqnarray}{rCl}
        \muI{\Xs{i}\prm}{\Xs{\minusI}\prm\mid\Xs{1:i}}
        &=& \shaent{\Xs{i}\prm\mid\Xs{1:i}}
        + \shaent{\Xs{\minusI}\prm\mid\Xs{1:i}}
        \NonumberNewline
        &&- A_i,
        \nonumber
        \\
        \shaent{\Xs{\minusI}\prm\mid\Xs{1:i}}
        &=& A_{i-1}
        - \muI{\Xs{\minusI}\prm}{\Xs{i}\mid\Xs{\minusI}},
        \nonumber
        \\
        \shaent{\Xs{i}\prm\mid\Xs{1:i}}
        &=& \shaent{\Xs{i}\prm\mid\Xs{i},\DepX{i}}
        + \muI{\Xs{i}\prm}{\DepX{i}\mid\Xs{i}}
        \NonumberNewline
        &&- \muI{\Xs{i}\prm}{\Xs{\minusI}\mid\Xs{i}}.
        \nonumber
    \end{IEEEeqnarray}
    Summing \eqref{eq:inc_cmi_form} over $i=2,\dots,\Nx$, the two subtracted mutual informations cancel the second and third terms of \eqref{eq:inc_cmi_form}, the $A_j$ telescope, and the boundary term $A_1$ equals $\shaent{\Xs{1}\prm\mid\Xs{1},\DepX{1}} + \muI{\Xs{1}\prm}{\DepX{1}\mid\Xs{1}}$ and supplies the $i=1$ term, so that
    \begin{IEEEeqnarray}{lCl}
        \Sum{i=2}{\Nx}\incIiMu
        =
        \Sum{i=1}{\Nx}\left[
            \shaent{\Xs{i}\prm\mid\Xs{i},\DepX{i}}
            + \muI{\Xs{i}\prm}{\DepX{i}\mid\Xs{i}}
        \right]
        - A_{\Nx}.
        \IEEEeqnarraynumspace
        \nonumber
    \end{IEEEeqnarray}
    Conditioned on $\Xtot$, the final states are mutually independent and each $\Xs{i}\prm$ depends only on $(\Xs{i},\DepX{i})$ (Lemma~\ref{lem:screening}), so $A_{\Nx} = \shaent{\Xtotprm\mid\Xtot} = \Sum{i=1}{\Nx}\shaent{\Xs{i}\prm\mid\Xs{i},\DepX{i}}$, and \eqref{eq:inc_closed} follows.

    Expanding $\muI{(\Xs{i},\Xs{i}\prm)}{\DepX{i}}$ by the chain rule in both orders gives
    \begin{IEEEeqnarray}{rCl}
        \muI{(\Xs{i},\Xs{i}\prm)}{\DepX{i}}
        &=& \muI{\Xs{i}}{\DepX{i}}
        + \muI{\Xs{i}\prm}{\DepX{i}\mid\Xs{i}}
        \NonumberNewline
        &=& \muI{\Xs{i}\prm}{\DepX{i}}
        + \muI{\Xs{i}}{\DepX{i}\mid\Xs{i}\prm}.
        \nonumber
    \end{IEEEeqnarray}
    By \eqref{eq:def_eta}, this implies
    \begin{IEEEeqnarray}{rCl}
        \muI{\Xs{i}\prm}{\DepX{i}\mid\Xs{i}} - \UsableNew{i}
        &=& \muI{\Xs{i}}{\DepX{i}\mid\Xs{i}\prm}.
        \nonumber
    \end{IEEEeqnarray}
    Summing over $i$ and inserting \eqref{eq:inc_closed} yields \eqref{eq:acc_closed}.

    Finally, under Assumptions~\ref{asm:local_mech}--\ref{asm:no_unrealized}, the proof of Lemma~\ref{lem:step_sum} identifies $\UsableNewAll - \delitot$ with $\Sum{i=2}{\Nx}\DissipationI$; combining this with \eqref{eq:release_split} and the split \eqref{eq:dec_split} gives $\Sum{i=2}{\Nx}\accIiMu = \Sum{i=2}{\Nx}\decIiMu - \Sum{i=2}{\Nx}\DissipationI = \Sum{i=2}{\Nx}\incIiMu - \UsableNewAll$. The stepwise statement and its equality condition follow from \eqref{eq:acc_closed}, each summand being nonnegative.
\end{proof}

Equation~\eqref{eq:inc_closed} locates the creation of correlation on the reading pattern. The created total is the information about its parents that each update writes into its own subsystem beyond what the initial state already carried. Equation~\eqref{eq:acc_closed} identifies the shared part of the destroyed correlation with the complementary quantity, the parent information that the updates \emph{overwrite}. A step shares in its destroyed correlation exactly as much as its mechanisms forget of what their subsystems knew about the variables they read. The information gain $\UsableNewAll$ is the net balance of the two, learned minus forgotten. In the five-subsystem process of Appendix~\ref{ap:cex_negative_Di}, the split \eqref{eq:dec_split} lets individual shares overshoot in both directions, $\accIiMu = +0.3862 > \decIiMu$ at $i=3$ and $\accIiMu = -0.2079 < 0$ at $i=2$. The step total nevertheless obeys \eqref{eq:acc_closed} exactly: $\Sum{i=2}{\Nx}\accIiMu = +0.1783 = \muI{\Xs{1}}{\Xs{5}\mid\Xs{1}\prm} + \muI{\Xs{2}}{\Xs{4}\mid\Xs{2}\prm}$. The two updates that read anything contribute $0.1495$ and $0.0288$ of overwritten parent information.

Entropy production is additive over the chain. Writing $\DeltaStot^{(1:K)}$ for the entropy production of the whole process,
\begin{IEEEeqnarray}{lCl}
    \DeltaStot^{(1:K)} = \Sum{k=1}{K}\DeltaStot^{(k)},
    \label{eq:step_additivity}
\end{IEEEeqnarray}
because the Shannon-entropy differences telescope and the dissipated heats add. The single-step bound therefore composes into a chain ledger.

\begin{proposition}[Composed ledger]
    \label{prop:composed_ledger}
    For every chain as above,
    \begin{IEEEeqnarray}{lCl}
        \DeltaStot^{(1:K)} \;\geq\; \Sum{k=1}{K}\left(\decIk  - \accIk \right),
        \label{eq:composed_ledger}
    \end{IEEEeqnarray}
    with equality if and only if the local entropy production \eqref{eq:yuragi_k} vanishes for every subsystem at every step. Every term of the sum is nonnegative by Lemma~\ref{lem:step_sum}, and no term exceeds $\decIk$ by Lemma~\ref{lem:acc_sum}.
\end{proposition}
\begin{proof}
    Apply Theorem~\ref{thm:main} to each step---by Eqs.~\eqref{eq:def_DissipationI} and \eqref{eq:dec_split}, its right-hand side equals $\Sum{i=2}{\Nx}\DissipationI = \decIk -\accIk $---and sum over $k$ using \eqref{eq:step_additivity}. The equality condition is inherited from Theorem~\ref{thm:ledger}, whose slack coincides with that of \eqref{eq:main_no_new_obs} under Assumptions~\ref{asm:acyclic} and \ref{asm:no_unrealized}.
\end{proof}

Two features of \eqref{eq:composed_ledger} carry the weight in the discussion of Sec.~\ref{sect:indispensability}. First, the bound is \emph{localized}. A step that blindly destroys correlation contributes its full $\decIk $ wherever in the chain it occurs. This contribution is nonnegative, since at a fully blind step $\accIk=0$ and $\decIk\geq0$ by the screening property \footnote{At a fully blind step, $\decIiMu=\atamaOmegaIMuMu+\centerSeibunIMu+\bottomSeibunIMuMu\geq0$ term by term. The component $\atamaOmegaIMuMu$ is a conditional mutual information. For $\centerSeibunIMu$, the recursion \eqref{eq:def_interaction_cond} gives $\centerSeibunIMu = \muI{\Xs{i}}{\Xs{\minusI}\prm\mid\Xs{i}\prm} - \muI{\Xs{i}}{\Xs{\minusI}\prm\mid\Xs{\minusI},\Xs{i}\prm}$, and the subtrahend vanishes by the chain rule and screening, as in the footnote below \eqref{eq:cpa_closed_form}. For $\bottomSeibunIMuMu$, expanding the block slot by the chain rule gives $\bottomSeibunIMuMu=\sum_{k\in\minusI}\muI{\Xs{i}}{\Xs{k};\Xs{i}\prm\mid\Xs{\minusK},\Xs{\minusI}\prm}$, and in each term $\muI{\Xs{i}}{\Xs{k};\Xs{i}\prm\mid\cdot}=\muI{\Xs{k}}{\Xs{i}\prm\mid\cdot}-\muI{\Xs{k}}{\Xs{i}\prm\mid\Xs{i},\cdot}$, whose subtrahend vanishes because $\Xs{i}\prm\perp\Xs{k}\mid\Xs{i}$ under blind updates (Lemma~\ref{lem:screening}), leaving a sum of conditional mutual informations.}. By Lemma~\ref{lem:step_sum}, no later step can undo the loss because every term of the sum is nonnegative. Second, for a fully blind step, $\DepX{i}^{(k)}=\emptyset$ for all $i$, one has $\accIk =0$. The step contributes its entire destroyed correlation \footnote{For fully blind steps, Assumptions~\ref{asm:acyclic} and \ref{asm:no_unrealized} are not even needed, since the screening property (Lemma~\ref{lem:screening}) makes every unrealized-correlation component and both rebate terms of Theorem~\ref{thm:ledger} vanish identically; cf.\ the discussion below Theorem~\ref{thm:main}.}.

It remains to translate \eqref{eq:composed_ledger} into the work bound. Recall the nonequilibrium free energy $F = \langlee E\ranglee - \boltz\Temperature\,\Htot$ of Eq.~\eqref{eq:def_Floc}. Under the additivity assumption \eqref{eq:additive_energy}, the static counterpart of Eq.~\eqref{eq:split_entropy} splits it into the local free energies $\Floc{i}$ of Sec.~\ref{sect:indispensability} and a correlation term,
\begin{IEEEeqnarray}{lCl}
    F = \Sum{i=1}{\Nx}\Floc{i} + \boltz\Temperature\,\TotalCorrelation .
    \label{eq:Fneq_split}
\end{IEEEeqnarray}
The net work extracted over the whole chain is $\Wext=-\Delta\langlee E\ranglee-\Qtot^{(1:K)}$, where $\Qtot^{(1:K)}$ is the total heat dissipated into the bath and $\Delta$ refers to the whole chain; the first law and the definition \eqref{eq:def_sigma} give the exact identity
\begin{IEEEeqnarray}{lCl}
    \invTemp\Wext = -\invTemp\,\Delta F - \DeltaStot^{(1:K)} .
    \label{eq:chain_work_identity}
\end{IEEEeqnarray}

\begin{proof}[Proof of Proposition~\ref{prop:chain_work}]
    Insert \eqref{eq:Fneq_split} into \eqref{eq:chain_work_identity}, use $-\Delta\TotalCorrelation=\Sum{k=1}{K}(\decIk -\incIk )$, and bound $\DeltaStot^{(1:K)}$ from below by \eqref{eq:composed_ledger}: the terms $\decIk $ cancel, leaving \eqref{eq:chain_work}. Equation~\eqref{eq:blind_work} follows since $\accIk =0$ at fully blind steps. Equation~\eqref{eq:general_work} does not go through \eqref{eq:chain_work}: instead, bound $\DeltaStot^{(1:K)}\geq0$ in \eqref{eq:chain_work_identity} by Lemma~\ref{lem:joint_second_law}, insert \eqref{eq:Fneq_split}, and drop $\TotalCorrelation(t_K)\geq0$.
\end{proof}

The value-of-information comparison of Sec.~\ref{sect:indispensability} requires that the blind bound \eqref{eq:blind_work} be attained. Recall the conversion task fixed there: the final marginal and the final bare energy of every subsystem are prescribed, and hence so is every $\Delta\Floc{i}$, while the length of the chain, the parent structures, and the channels remain at the protocol's disposal.

\begin{lemma}[Blind attainability]
    \label{lem:blind_attain}
    For every conversion task,
    \begin{IEEEeqnarray}{lCl}
        \sup_{\mathrm{blind}}\Wext
        \;=\;
        -\Sum{i=1}{\Nx}\Delta\Floc{i},
        \label{eq:blind_attain}
    \end{IEEEeqnarray}
    the supremum ranging over all blind chains, of any length, that complete the task---for boundary tasks, whose prescribed marginals contain zeros, over chains completing the task to arbitrary accuracy---and it is approached by per-subsystem quasistatic protocols.
\end{lemma}
\begin{proof}
    That the supremum does not exceed the right-hand side is \eqref{eq:blind_work}. For the converse we exhibit blind chains approaching it. The protocol acts on each subsystem separately and in parallel: every step holds the bare energies fixed and lets each subsystem thermalize---the channel draws $\Xs{i}\prm$ from the Gibbs distribution of the current bare energy $E_i$, irrespective of the input---while between steps the bare energies are switched, at injected work $\langlee E_i^{\mathrm{new}}-E_i^{\mathrm{old}}\ranglee$ evaluated on the current distribution, which enters $\Wext$ through the first law \eqref{eq:chain_work_identity}. Every step is fully blind: Assumption~\ref{asm:acyclic} is vacuous, Assumption~\ref{asm:no_unrealized} holds because every term of \eqref{eq:asm_no_unrealized} vanishes by screening (Lemma~\ref{lem:screening}), and the thermalizing channel satisfies Assumption~\ref{asm:ldb} with $\invTemp\Qs{i}=\invTemp\langlee E_i(\Xs{i})-E_i(\Xs{i}\prm)\ranglee$, its time reverse being the same channel.

    Fix $i$, let $p_i$ and $p_i'$ denote the initial and the prescribed final marginal, both taken of full support first, and set $E_i^{*}:=-\boltz\Temperature\ln p_i$ and $E_i^{**}:=-\boltz\Temperature\ln p_i'$, whose Gibbs distributions are $p_i$ and $p_i'$ with unit partition functions. Switch $E_i\to E_i^{*}$: the injected work is $\boltz\Temperature\,\shaent{\Xs{i}}-\langlee E_i\ranglee$, i.e., minus the initial local free energy, and the distribution, already Gibbs, is unchanged by thermalization. Then interpolate in $M$ increments, $E^{(m)}:=(1-\tfrac{m}{M})E_i^{*}+\tfrac{m}{M}E_i^{**}$, switching and thermalizing at each: the $m$-th switch is paid on the Gibbs distribution of $E^{(m-1)}$ and exceeds the equilibrium free-energy increment $-\boltz\Temperature\ln(Z_m/Z_{m-1})$ by a defect of order $1/M^2$, so the interpolation work converges, as $M\to\infty$, to $-\boltz\Temperature\ln(Z_M/Z_0)=0$, both endpoint partition functions being unity, while the last thermalization leaves the marginal at $p_i'$ exactly. Finally switch $E_i^{**}\to E_i'$: the injected work is $\langlee E_i'\ranglee-\boltz\Temperature\,\shaent{\Xs{i}}$, the final local free energy. The work injected into subsystem $i$ thus totals $\Delta\Floc{i}+O(1/M)$, so $\Wext\to-\Sum{i=1}{\Nx}\Delta\Floc{i}$. Marginals with zeros are covered by capping the diverging levels of $E_i^{*}$ or $E_i^{**}$ at a ceiling $\Lambda$ and letting $\Lambda\to\infty$ after $M\to\infty$---the standard quasistatic-erasure limit, at the same limiting work, within the boundary-task reading of the supremum fixed in the statement.
\end{proof}

\noindent The attaining protocols inject work along the way because each subsystem whose local free energy must rise is paid for quasistatically. A budget constraint curtails this injection, closing the loophole in Sec.~\ref{sect:ex_cartridge}.

\section{Declaration-Level Measurability of Sequential and Modular Records}
\label{ap:declaration_measurability}

This appendix records a closure property of the accounts that price a step by sequential or modular bookkeeping: once a valid acyclic declaration of the reading patterns is fixed, the stage quantities and fixed-block modularity values such accounts generate are functionals of the declaration, the realized statistics, and the trajectory heats. The rescue-conditioned ledger is not (Sec.~\ref{sect:ex_bit}). The lemma therefore locates the attribution of Theorem~\ref{thm:main} outside the reach of those records.

A \emph{declaration} is a collection $G=(\mathrm{pa}_1,\dots,\mathrm{pa}_\Nx)$ of subsets $\mathrm{pa}_i\subseteq\Xtot\setminus\Xs{i}$. It is \emph{valid} for a process of the class of Sec.~\ref{sect:setting} if $\DepX{i}\subseteq\mathrm{pa}_i$ for every $i$: declared parents may exceed minimal ones, because enlarging a dependence set preserves the factorization \eqref{eq:asm_local_mech}. A \emph{parents-later schedule} of $G$ is a total order of the subsystems in which every subsystem precedes each member of its declared parent set; such an order exists exactly when the declared reading pattern is acyclic. Realizing the step in this order, one subsystem at a time, with each mechanism reading the beginning-of-step values of its arguments, reproduces the declared kernel exactly, because every variable a mechanism reads still holds its initial value when the mechanism fires (Sec.~\ref{sect:setting}).

\begin{lemma}[Declaration-level measurability]
    \label{lem:declaration_measurability}
    Fix a declaration $G$ admitting a parents-later schedule, and consider processes of the class of Sec.~\ref{sect:setting} for which $G$ is valid. Every quantity of the following kinds is a functional of the triple consisting of $G$, the realized one-step statistics $P(\Xtot,\Xtotprm)$, and the trajectory heats on the support: (i) for every parents-later schedule of $G$, the joint law of the full stage trajectory of the sequential realization, and hence every entropy and mutual information evaluated on any stage, together with every per-stage balance formed from these and the assigned heats; (ii) for every fixed bipartition $(B,B^{c})$ of the subsystems, the modularity difference $\muI{\Xs{B}}{\Xs{B^{c}}}-\muI{\Xs{B}\prm}{\Xs{B^{c}}}$; (iii) every judgement about the process formed from the realized statistics alone. In particular, two processes of the class sharing the declaration, the statistics, and the trajectory heats receive identical values for all of them.
\end{lemma}
\begin{proof}
    In a parents-later schedule each subsystem is updated exactly once, and every variable a mechanism reads is updated only later. At any stage, therefore, each subsystem holds either its final value, if it has fired, or its initial value, if it has not; no third value ever occurs. Writing $F_t$ for the set of subsystems fired by stage $t$, the intermediate state equals the deterministic function $(\Xs{F_t}\prm,\Xs{F_t^{c}})$ of the endpoint pair $(\Xtot,\Xtotprm)$. The joint law of the entire stage trajectory is the pushforward of $P(\Xtot,\Xtotprm)$ under these maps, which proves (i); the heats enter the stage balances only through their assigned trajectory values. Part (ii) involves endpoint variables alone, and (iii) is immediate.
\end{proof}

Two consequences delimit what the lemma does and does not assert. It does assert that the quantities generated by sequential accounts (per-stage entropy productions and information drops, for every admissible schedule \cite{ito2013information,Wolpert_2020,WolpertMinEP2020}) and by modular accounts (the algebraic fixed-bipartition difference, on any bipartition and irrespective of its admissibility \cite{Boyd_2018,wolpert2019stochastic}) cannot separate two processes that share a valid declaration, the realized statistics, and the trajectory heats. Section~\ref{sect:ex_bit} exhibits such a pair whose implementation-optimal costs differ by $\ln 2$: the rescue-conditioned right-hand side of \eqref{eq:main_no_new_obs}, and the operational quantity it bounds, are not functionals of that triple. The lemma does not assert that the attribution is inaccessible in principle. Enlarging the observables beyond these quantities to the kernel itself on the whole state space, or to the judgement of which declarations are valid for it, determines the minimal parents and, with them, the full ledger. For the pair of Sec.~\ref{sect:ex_bit} the enlargement is concrete: the kernel-level admissibility families of the modular account differ, with three nontrivial blocks for the blind reset, none for the informed one, while every value computed on a fixed block coincides. That judgement, however, is exactly the input constructed in Sec.~\ref{sect:setting}. If that judgement is read off from realized statistics instead, it must return the same answer for two processes sharing those statistics, while their minimal parents can differ (Sec.~\ref{sect:ex_bit}); the route to the attribution runs through the kernel-level object this paper takes as primitive, not through those quantities. A complementary continuous-time perspective, in which trajectory-level causal mechanisms carry thermodynamic content beyond endpoint summaries, appears in Ref.~\cite{Rancati_2026}; the finite-state, fixed-declaration nonidentifiability isolated here is a distinct, discrete-time statement.

\end{document}